\documentclass[twocolumn,fleqn,usenatbib]{mnras}

\usepackage{newtxtext,newtxmath}

\usepackage[T1]{fontenc}
\usepackage[T4, T1]{fontenc}
\DeclareRobustCommand{\VAN}[3]{#2}
\let\VANthebibliography\thebibliography
\def\thebibliography{\DeclareRobustCommand{\VAN}[3]{##3}\VANthebibliography}
    \makeatletter
\def\@fnsymbol#1{\ensuremath{\ifcase#1\or \dagger\or \ddagger\or
   \mathsection\or \mathparagraph\or \|\or **\or \dagger\dagger
   \or \ddagger\ddagger \else\@ctrerr\fi}}
    \makeatother

\usepackage{graphicx}	
\usepackage{amsmath}	

\usepackage{mathtools, cuted}
\usepackage{lipsum}
\usepackage{widetext}
\usepackage{orcidlink}

\title[A theoretical perspective of the nano-Hertz GW hosts]{Shining Light on the Hosts of the Nano-Hertz Gravitational Wave Sources: A Theoretical Perspective}

\author[Saeedzadeh et al. (2023)]{
Vida Saeedzadeh,$^{1}$\orcidlink{0009-0000-7559-7962}
Suvodip Mukherjee,$^{2}$\thanks{E-mail:suvodip@tifr.res.in}\orcidlink{0000-0002-3373-5236}
Arif Babul$^{1,3}$\orcidlink{0000-0003-1746-9529}
Michael Tremmel$^{4}$\orcidlink{0000-0002-4353-0306}
and Thomas R. Quinn$^{5}$
\\
$^{1}$Department of Physics and Astronomy, University of Victoria, 3800 Finnerty Road, Victoria, BC, V8P 1A1, Canada\\
$^{2}$Department of Astronomy \& Astrophysics, Tata Institute of Fundamental Research, 1, Homi Bhabha Road, Colaba, Mumbai 400005, India\\
$^{3}$Infosys Visiting Chair Professor, Indian Institute of Science, Bangalore 560012, India\\
$^{4}$School of Physics, University College Cork, College Road, Cork T12 K8AF, Ireland\\
$^{5}$Astronomy Department, University of Washington, Box 351580, Seattle, WA, 98195-1580, USA
}

\date{Accepted XXX. Received YYY; in original form ZZZ}

\pubyear{2023}

\begin{document}
\label{firstpage}
\pagerange{\pageref{firstpage}--\pageref{lastpage}}
\maketitle

\begin{abstract}
The formation of supermassive black holes (SMBHs) in the Universe and its role in the properties of the galaxies is one of the open questions in astrophysics and cosmology. Though, traditionally, electromagnetic waves have been instrumental in direct measurements of SMBHs, significantly influencing our comprehension of galaxy formation, gravitational waves (GW) bring an independent avenue to detect numerous binary SMBHs in the observable Universe in the nano-Hertz range using the pulsar timing array observation. This brings a new way to understand the connection between the formation of binary SMBHs and galaxy formation if we can connect theoretical models with multi-messenger observations namely GW data and galaxy surveys. Along these lines, we present here the first paper on this series based on {\sc Romulus25} cosmological simulation on the properties of the host galaxies of SMBHs and propose on how this can be used to connect with observations of nano-Hertz GW signal and galaxy surveys. We show that the most dominant contribution to the background will arise from sources with high chirp masses which are likely to reside in low redshift early-type galaxies with high stellar mass, largely old stellar population, and low star formation rate, and that reside at centers of galaxy groups and manifest evidence of recent mergers. The masses of the sources show a correlation with the halo mass and stellar mass of the host galaxies. This theoretical study will help in understanding the host properties of the GW sources and can help in establishing a connection with observations.
\end{abstract}




\section{Introduction}\label{bbhIntro}

The discovery of Gravitational Waves (GWs) by the LIGO-Virgo-KAGRA (LVK) Collaboration from coalescing compact object binaries of a few tens of solar masses inaugurated the era of gravitational-wave astronomy, enabling the observations of previously inaccessible astrophysical phenomena 
\citep{LIGOScientific:2014pky,Martynov:2016fzi,Acernese_2014,PhysRevLett.123.231108,KAGRA:2013pob,Akutsu:2018axf,KAGRA:2020tym}. Following this initial discovery, several more binary objects have been detected, one of which (GW170817) also had an electromagnetic (EM) counterpart and stands as the first multi-messenger measurement involving GW signal \citep{PhysRevLett.116.061102,TheLIGOScientific:2017qsa,LIGOScientific:2020kqk,Abbott_2021, LIGOScientific:2021psn,LIGOScientific:2021djp,Virgo:2021bbr}. With the aid of ongoing and upcoming networks of GW detectors, several more detections of coalescing black hole binaries are likely over the frequency range of $10$ Hz and above. 

Along with the high-frequency GW signal, coalescing supermassive black holes (SMBHs) can also produce GW signals detectable at low-frequency  bands, ranging from a few nano-Hertz to milli-Hertz range. In the milli-Hertz frequency range, upcoming GW detectors -- such as Laser Interferometer Space Antenna \citep[LISA;][]{2017arXiv170200786A, Baker:2019pnp} -- can probe signal from the SMBHs of masses in the range approximately $10^4$- $10^7$ M$_\odot$. The nano-Hertz GW signal from sources with masses above $10^8$ M$_\odot$ can be detected and characterized using the timing data from several extremely well-studied millisecond pulsars \citep{sesana2008pta,1990ApJ...361..300F}.  These signals are the target of the International Pulsar Timing Array (IPTA) collaboration \citep{Antoniadis:2022pcn}, comprising the European Pulsar Timing Array \citep[EPTA;][]{2016MNRAS.458.3341D}, the North American Nanohertz Observatory for Gravitational Waves \citep[NANOGrav;][]{2013CQGra..30v4008M}, the Indian Pulsar Timing Array Project \citep[InPTA;][]{2018JApA...39...51J} and the Parkes Pulsar Timing Array \citep[PPTA;][]{2013PASA...30...17M}. {Along with the IPTA Collaboration, the Chinese Pulsar Timing Array \citep[CPTA;][]{Xu_2023} are also making measurements in this band.} In the future with the operation of the Square Kilometer Array (SKA) \citep{sesana2008pta,Terzian:2006hnm,janssen2014gravitational} more accurate measurement of this signal will be possible \citep{Burke-Spolaor:2018bvk}.  The recent {evidence} of the stochastic GW background (SGWB) in the nano-Hertz range by CPTA \citep{Xu:2023wog}, EPTA+InPTA \citep{Antoniadis:2023ott},  NANOGrav \citep{NANOGrav:2023gor} and PPTA 
{\citep{reardon2023}} promises to open an exciting new window onto the evolving population of binary supermassive black holes (SMBBHs) in the Universe. 

The presence of nano-Hertz GW signal leads to several interesting questions such as: What are the astrophysical properties of the host galaxies of the source SMBBHs, and can one use conventional galaxy surveys to identify (if not uniquely detect) these host galaxies?  
Since SMBHs reside in the centers of galaxies, SMBBHs are expected to be byproducts of galaxy mergers.  Consequently, SMBBHs-host galaxy identifications can potentially shed light on the pathways leading to the formation of the SMBBHs, including their dynamical evolution from the time of first encounter, and more generally on the astrophysics of galaxy mergers.  They can also potentially provide insights into the growth of SMBHs and the implications of SMBH-SMBH mergers on galaxy formation \citep{2009Natur.460..213C}, including conditions leading to the  transition between radiative versus kinetic feedback modes \citep[eg][]{Narayan-Quataert-2005,MerloniHeinz2008Radiative-RIAF,Benson2009,Babul2013}. (See also 
\citealt{OSullivan2012CL0910}, \citealt{Reynolds2014}, \citealt{prasad-phoenix2020} and \citealt{2021MNRAS.508.3796O}  for rare examples of observed SMBHs in unexpected state.)  This would be an important step towards a new paradigm of multi-messenger science capable of addressing a broad spectrum of questions related to astrophysics and cosmology.  

However, typical PTA localizations in the near term are expected to encompass several thousand (if not more) galaxies. {Theoretical and computational modeling offers an opportunity to not only explore environments where SMBH - SMBH mergers take place but also a way to narrow the field of candidate host galaxies for more detailed observational scrutiny \citep{rosado2014,goldstein2019, volonteri2003,volonteri2020,Kozhikkal:2023gkt}}.  In this study, we use results from a high-resolution {\sc Romulus} cosmological simulation \citep{tremmel2017romulus,tremmel2019introducing} to explore this possibility.  As we discuss in \S \ref{sec:romulus}, {\sc Romulus} 
suite of simulations is especially suited for investigating SMBH/SMBBH-galaxy connections because of its unique approach to seeding, accretion, and especially the dynamics of supermassive black holes. Consequently, the simulations have previously been used to explore a variety of related topics, including the timescale for the formation of close SMBH pairs following galaxy mergers \citep{Tremmel_2018}, the galaxy-SMBH coevolution \citep{Ricarte_2019}, the origin and demographics of wandering black holes \citep{Ricarte_2021}, and the demographics of dual active galactic nuclei \citep{saeedzadeh2023dual}. 

The paper is organized as follows:  In \S\ref{sec:motv}, we briefly discuss the motivation behind the present study and in \S\ref{sec:romulus}, we discuss the {\sc Romulus} simulation.  The expected SGWB based on the simulation and the astrophysical properties of the galaxies hosting SGWB sources are discussed in \S\ref{sec:sgwb} and \S\ref{sec:host}.  Among the properties we consider are: gas density ($\rho_{\rm gas}$), star formation rate ($\dot{M}_*$ or SFR), stellar mass ($\rm M_*$), galaxy morphology, galaxy color, specific star formation rate (sSFR $\equiv \dot{M}_*/M_*$) and halo mass ($\rm M_h$, which we take to be $\rm M_\mathrm{500}$; see \S \ref{selection} for definition of $\rm M_\mathrm{500}$). Then, we discuss possible techniques to validate the connection between the SMBBHs and their host galaxies that we find in \S\ref{sec:connectiontoobs}. Finally, we summarise our findings and discuss the future outlook in \S\ref{sec:conc}.  

\section{Motivation}\label{sec:motv}
On one hand, we have the recently detected SGWB in the nano-Hertz range from coalescing SMBHs of mass $\rm M> 10^7$ M$_{\odot}$. On the other hand, we have spectroscopic/photometric galaxy surveys that are capable of detecting faint galaxies up to high redshifts, some of which will be hosts of the SMBBHs that are contributing to the nano-Hertz SGWB. The combination of these two opens up the prospect of a new multi-messenger science that can shed light on several key questions in astrophysics and cosmology. A  limited list of these key questions are: (i) How do the 
SMBHs grows with time? (ii) How do SMBBHs form and is there a relationship between their formation and one or more properties of the host galaxies? (iii) Do the astrophysical properties of the host galaxies play a role in coalescing of the SMBBHs?  (iv) What is the occupation number of the SMBHs in galaxies (or halos) of different masses?

We are interested in understanding the theoretical dimensions of these questions and {in identifying whether the} key astrophysical properties of the host galaxies can be predicted based on our current understanding of galaxy formation. In this paper, we explore the astrophysical ``tells'' of galaxies that host SMBBHs in the {\sc Romulus} simulation volume.  We also investigate the properties of the halos of these galaxies.  Although the {\sc Romulus} simulations can track black holes across nearly three orders of magnitude in mass ($10^6$- $10^9$ M$_\odot$), in the present paper we focus primarily on coalescing binaries black holes that can contribute to the stochastic gravitational wave background in the frequency band accessible to PTA. We perform a simulation-based study of the correlations between the SMBBHs and their host galaxies.  The specific galaxy properties we focus on include their morphology, star formation rate, galaxy color, stellar mass, gas density, and halo mass. Uncovering a theoretical connection between the properties of the host galaxy and its SMBBH will help motivate observational and data analysis strategies aimed at identifying the host galaxies of the GW sources from the photometric/spectroscopic galaxy catalogs.  This, in turn, can contribute to building a data-driven understanding of the evolution of SMBBHs in galaxies. 

In future papers in this series, we will consider black holes accessible to LISA, examine possible connections between these and SMBBHs detectable with PTA, and its implementation on the latest nano-Hertz observations \citep{Xu:2023wog,antoniadis2023astrophysics,NANOGrav:2023gor,Zic:2023gta} to identity the possible host candidates. For completeness, we note that there are several analytical and numerical simulation-based studies estimating SGWB signal in the PTA frequency range \citep{1995ApJ...446..543R, 2003ApJ...583..616J,2011MNRAS.411.1467K, Chen:2016zyo, DeGraf:2020yzt, 2022MNRAS.509.3488I, Izquierdo-Villalba2023}. {Additional studies are referenced throughout the text.}

\section{The {\sc Romulus} Simulations}\label{sec:romulus}

In this work, we present results from the analysis of the {\sc Romulus25} simulation, which is a $\rm (25\; cMpc)^3$ cosmological volume simulation from the Romulus suite \citep{tremmel2017romulus,tremmel2019introducing,butsky2019ultraviolet,jung2022massive,saeedzadeh2023cgm}.   

{The simulation was run using the Tree+Smoothed Particle Hydrodynamics (Tree+SPH) code CHaNGa \citep{menon2015adaptive,wadsley2017gasoline2}. CHaNGa incorporates the standard physics models previously employed in the simulation codes GASOLINE/GASOLINE2 and has been extensively tested \citep{Stinson_2006}. Including modules for star formation, stellar feedback, turbulent diffusion \citep{Shen_2010}, the UV background, low-temperature metal cooling, and an improved treatment of both weak and strong shocks. However, the models for SMBH formation, dynamics, growth, and feedback are novel \citep{tremmel2015off,tremmel2017romulus}. We will discuss them in more detail in the following subsections.}

{The {\sc Romulus25} simulation was run assuming a flat $\Lambda$CDM universe with cosmological parameters consistent with the Planck 2016 results \citep{Planck_2016}: $\Omega_{\rm m} = 0.309$, $\Omega_{\rm \Lambda} = 0.691$, $\Omega_{\rm b} = 0.0486$, $\rm H_{\rm 0} = 67.8\, {\rm km}\,{\rm s}^{-1} \rm Mpc^{-1}$, and $\sigma_{\rm 8} = 0.82$. The simulation has a Plummer equivalent gravitational force softening of 250 pc (or 350 pc spline kernel) and a maximum SPH resolution of 70 pc.  Differing from many similar cosmological runs, the dark matter distribution in {\sc Romulus25} is oversampled with 3.375 times more dark matter particles than gas particles. This results in a dark matter particle mass of $3.39 \times 10^5 M_\odot$ and a gas particle mass of $2.12\times 10^5 M_\odot$. This deviation from the standard approach of simulating equal numbers of gas and dark matter particles minimizes numerical noise and allows for more precise black hole dynamics \citep{tremmel2015off}.} 

{More} details about the {\sc Romulus25} simulation have been described in a number of published papers. {We} refer interested readers to \citet{tremmel2015off,tremmel2017romulus,tremmel2019introducing,tremmel2020formation,sanchez2019not,butsky2019ultraviolet,chadayammuri2020fountains,jung2022massive}; and \citet{saeedzadeh2023cgm}. The latter two especially offer a concise yet complete summary.  {Additionally, for comparisons with galaxy formation models employed in other cosmological simulations, interested readers are referred to \citet{somerville2014,vogelsberger2020,Oppenheimer2021GroupsSimReview}}

There are, however, a few aspects of the {\sc Romulus25} simulation that are important to highlight as these are relevant to the present discussion. These pertain to the treatment of SMBH seeding, growth, and dynamical evolution in the   
{\sc Romulus25} \citep{tremmel2017romulus}.

\subsection{SMBH Seeding}

{Unlike many other cosmological simulations \citep[e.g.,][]{schaye2015eagle,weinberger2017tng,pillepich2018tng,nelson2019tng,dave2019simba}, the {\sc Romulus25} SMBH seed model does not depend on a halo or a galaxy to exceed a certain mass threshold for a SMBH to form. Rather, the seeding depends only on the local gas properties \citep{tremmel2017romulus}. As a result, the SMBHs in {\sc Romulus25} can form in low mass halos and tend to form much earlier \citep[z $>$ 5, ][]{tremmel2017romulus}.
Additionally, one can also have multiple SMBHs arising in the same halo.}

The criteria for converting gas particle into a SMBH seed in {\sc Romulus25} are as follows: (i) The gas particle must be \emph{both} eligible and selected to form a star.   The latter is a probabilistic process. (ii) The gas particle must have very low metallicity ($ Z < 3 \times 10^{-4}$); (iii) its density must be very high; i.e.~at least $3\;m_p/{\rm cc}$ or greater.  And, (iv) its temperature is within the range of $9500$ - $10000\;$K.  This seeding prescription resembles the direct collapse black hole scenario, where high temperatures and low metallicities suppress fragmentation and allow sizeable gas clouds to collapse directly into an SMBH seed \citep{lodato2007mass,alexander2014rapid,natarajan2021new}. 

{In the {\sc Romulus25} simulation, SMBHs are seeded with a mass of $\rm 10^6 \ M_\odot$. This seeding mass differs slightly from other simulations like TNG50/100/300, EAGLE, Horizon AGN, and SIMBA-C where the initial SMBH seed masses are set at $\sim 8 \times 10^5 \rm M_\odot$, $10^8 \rm M_\odot$, $10^5 \rm M_\odot$ and $10^4 \rm M_\odot$respectively \citep{nelson2019tng,kaviraj2017horizonagn,crain2015eagle,hough2023simbac}. The choice of the SMBH seed mass in {\sc Romulus25} is constrained primarily by two factors: (i) the resolution of the simulation, with the dark matter and gas mass resolutions in {\sc Romulus25} being $3.39 \times 10^5 M_\odot$ and $2.12 \times 10^5 M_\odot$ respectively, and (ii) the necessity to keep SMBHs more massive than dark matter and star particles. This latter requirement is crucial for reducing the occurrence of spurious scattering events \citep{tremmel2015off}.}

The resulting SMBH occupation fraction at z=0 is consistent with current observations even on the scale of dwarf galaxies \citep{ricarte2019tracing}. {Additionally, the SMBH masses correlate with the stellar masses of their host galaxies, following the observed SMBH mass - stellar mass relation \citep{tremmel2017romulus,ricarte2019}.}  

\subsection{SMBH Dynamics and Mergers}

{The other difference in the SMBH model between {\sc Romulus25} and other cosmological simulations is SMBH dynamics. Unlike many other cosmological simulations, where the SMBHs are artificially placed at the gravitational potential minimum of their host galaxies \citep[e.g.][]{Crain.2009,Sijacki_2015MNRAS,dave2019simba},} the {\sc Romulus25} simulation accurately tracks the dynamical evolution of SMBHs down to sub-kpc scales, which is highly advantageous for the present study. To achieve this, a sub-grid correction is employed that accounts for the unresolved dynamical friction from stars and dark matter that the SMBHs ought to be experiencing\citep{tremmel2015off}. For each SMBH in the simulation, this force is estimated by assuming a locally isotropic velocity distribution and integrating Chandrasekhar's equation \citep{chandrasekhar1943dynamical} from the 90-degree deflection radius ($\rm r_{90}$) to the SMBH's gravitational softening length ($\epsilon_g$).  The resulting acceleration is

\begin{equation}
{\mathbf a}_{DF} = -4\pi G^2\; M_\bullet\; \rho(v < v_{BH})\; {\rm ln}\Lambda\;\frac{{\mathbf v}_{BH}}{v^3_{BH}},
\end{equation}

In order for two SMBHs to merge, they must be within a distance of two gravitational softening lengths ($0.7$  kpc) and possess a low enough relative velocity to be mutually bound; ~i.e.~$\frac{1}{2} \; \Delta {\mathbf v}^2 < \Delta {\mathbf a} \cdot \Delta {\mathbf r}$, where $\Delta {\mathbf v}$ and $\Delta {\mathbf a}$ are the differences in velocity and acceleration of the two black holes, and $\Delta {\mathbf r}$ is the distance between them \citep{bellovary2011,tremmel2017romulus}\footnote{Note that there is a typographical error in the criterion for boundedness in \citet{tremmel2017romulus}.}.
The separation limit of two gravitational softening lengths is deemed appropriate because once the separation drops below this limit, the simulation's ability to accurately track the SMBH pair's dynamics becomes less reliable. 

When a merger takes place, the resulting SMBH is assigned a velocity that conserves momentum, and its mass is the sum of the masses of its progenitors.   Mergers are one of the two processes driving the growth of SMBHs.

\subsection{SMBH Growth and Feedback}

The other process by which SMBHs grow is via the accretion of gas. In {\sc Romulus25}, this accretion rate is estimated via a modified Bondi-Hoyle {(\citealt{bondi1952spherically},  for modifications see \citealt{tremmel2017romulus})} prescription applied to the smoothed properties of the 32 nearest gas particles:

\begin{equation}
    \dot{M}_\bullet = \alpha \times
    \begin{cases}
    \frac{\pi (G M_\bullet)^2 \rho_{\rm gas}}{(v_{bulk}^2 + c_s^2)^{3/2}} ~~~~~ \textrm{if} ~ v_{bulk} > v_{\theta} \\
    \\
    \frac{\pi (G M_\bullet)^2 \rho_{\rm gas} c_s}{(v_{\theta}^2 + c_s^2)^{2}} ~~~~ \textrm{if} ~ v_{bulk} < v_{\theta}
    \end{cases}, 
\end{equation}
where $\rho_{\rm gas}$ is the ambient gas density, $c_s$ is the ambient sound speed, $v_\theta$ is the local rotational velocity of surrounding gas, and $v_{bulk}$ is the bulk velocity relative to the SMBH.  All ambient quantities are calculated using the 32 nearest gas particles.   The introduction of $v_\theta$ and $v_{bulk}$ terms in the above aims to remedy the neglect of gas bulk motion and angular momentum in the original Bondi-Hoyle formulation.  Finally, 
the coefficient $\alpha$ is introduced to correct for the suppression of the black hole accretion rate due to resolution effects.  It is defined as 

\begin{equation}
    \alpha = 
    \begin{cases}
    (\frac{n}{n_{th,*}})^2 ~~~ \textrm{if} ~~ n \geq n_{th,*}\\
    \\
    1 ~~~~~~~~~~~~ \textrm{if} ~~ n \leq n_{th,*}
    \end{cases},
\end{equation}
where $n_{th,*}$ is the star formation number density threshold ($0.2\; m_p/cc$).

Gas accretion onto a SMBH results in energy release into the environment around the black hole.  In {\sc Romulus25}, it is assumed that this energy is electromagnetic and that a fraction of it will couple to the ambient gas and contribute to its internal energy.  The thermal energy deposition rate is given by $\dot{E}_{\bullet,th} = \epsilon_r \epsilon_f \dot{M}_\bullet c^2,$ where $\epsilon_r$ is the radiative efficiency (assumed to be 10\%) and $\epsilon_f$ is gas coupling efficiency (set to 2\%).   The thermal energy is imparted isotropically to the 32 nearest gas particles, with the energy being distributed among these gas particles according to the smoothing kernel.  We refer readers to \citet{tremmel2017romulus} for further details.

\subsection{Selection of Halos and Binary SMBHs}\label{selection}

The halos in {\sc Romulus} simulations are extracted and processed using  the Amiga Halo Finder \citep[hereafter, AHF;][]{knebe2008relation,knollmann2009ahf}, and tracked across time with TANGOS \citep{pontzen2018tangos}.  

The halos and subhalos exist in a nested hierarchy, where the halos are the primary structures and the subhalos are incorporated within them. To identify these structures, AHF first locates density peaks in an adaptively smoothed density field and identifies all the particles (dark matter, gas, stars, and black holes) that are gravitationally bound to these peak.  This process is repeated on successively larger scales until all the structures in the hierarchy have been found. Once the halos are identified, their centers are found by applying the shrinking sphere approach \citep{power2003inner} to the distribution of bound particles associated with each of the halos.

The masses of the halos ($M_\Delta$) are determined by creating a sphere with a radius of $R_\Delta$ around each halo center. This sphere is constructed so that the average density within it, $\left\langle{\rho_\mathrm{m,\Delta}}(z)\right\rangle$, is equal to $\Delta$ times the critical cosmological density, $\rho_\mathrm{crit}(z) = \rm 3H^2(z)/8\pi G$
 \citep[see, for example,][]{babul2002physical}.  In this study, we reference ($M_\mathrm{200}, R_\mathrm{200}$) and ($M_\mathrm{500}, R_\mathrm{500}$), which correspond to $\Delta = 200$ and $\Delta = 500$, respectively. For our assumed cosmology, $M_\mathrm{500}/M_\mathrm{200} \approx 0.7$ and $R_\mathrm{500}/R_\mathrm{200} \approx 0.68$.

In the case of subhaloes, AHF tracks the  local density profile from the peak center outward. At some point, the external gravitational field starts to dominate, altering the shape of the density profile.  The distance from the peak to where this happens is taken to be the size of the subhalo, and the mass enclosed is recorded as the subhalo's mass.

We also track all the SMBHs in the {\sc Romulus25} simulation volume.  We use the resulting information to construct merger trees for all the black holes.  At each redshift, we then identify black holes that have experienced a merger during the immediately preceeding {output} and flag the about-to-merge SMBH pairs as candidate sources of nano-Hertz SGWB. {The time resolution ($\Delta t$) and redshift resolutions ($\Delta z$) for the saved output files within our redshifts of interest (see \S \ref{sec:bhp1}) vary in the ranges of $\rm 10 \ Myr < \Delta t < 400 \ Myr$ and $\rm 0.002 < \Delta z < 0.1$, getting to smaller values as approaching z = 0.} The typical separation of merging SMBH pairs is $\sim 1$ kpc and their maximum separation is 2.8 kpc.   For completeness, we also identify all black hole pairs separated by $\leq$ 1.4 kpc and which are not flagged as merging in the next {output}. We will refer to these as proximate pairs.

We emphasize that the SGWB from flagged SMBBHs with separation scale of $\sim 1\;$kpc cannot contribute to the nano-Hertz frequency band unless they coalescence down to sub-parsec ($10^{-5}$ pc) scale.   This journey of the SMBBHs from the scale of $\sim 1$ kpc to $\leq 10^{-5}$ pc is governed not only by GW emission but also by  environmental effects such as dynamical friction, stellar loss cone and viscous gas drag.  These processes are not resolvable in {\sc Romulus25} or, for that matter, in any other cosmological simulation.  We therefore need to model this coalescence separately. 

\section{Estimation of SGWB in the nano-Hertz}\label{sec:sgwb}
\subsection{Modeling SGWB signal from coalescing SMBBHs}\label{sec:sgwb-smbbhs}

In order to calculate the contribution to the SGWB signal from the coalescing SMBBHs, we start with the expression for the characteristic strain of the GW signal $h_c$ at frequency $f$ for a source emitting at a rest-frame frequency $f_r= (1+z)f$  \citep{1995ApJ...446..543R, Phinney:2001di, Sesana:2008mz}:

\begin{align}\label{GW-strain}
h_c^2(f) = \frac{4G}{c^2\pi f^2} 
 \iiint dz\, dm_1\, dm_2\, &\frac{d^3n_{GW} (m_1, m_2, z)}{dm_1dm_2dz} 
 \\ \nonumber & \times 
\frac{1}{1+z}\frac{dE_{\rm GW}(m_1, m_2, z)}{d\ln f_r},
\end{align}
where the distribution function, $\frac{d^3n_{\rm GW}(m_1, m_2, z)}{dm_1 dm_2 dz}$, is the number density of SMBBH GW sources with black hole masses in the range $[m_1,\; m_1+dm_1]$ and $[m_2,\;m_2+dm_2]$ at redshift $[z,\; z+dz]$ and determines the amplitude and spectral shape of the SGWB signal. The second term, $\frac{dE_{\rm GW}(m_1, m_2, z)}{d\ln f_r}$, quantifies the amount of GW energy released per logarithmic rest-frame frequency by a binary of source masses $m_1$ and $m_2$ at redshift $z$.  The latter is the product of the GW energy emission rate ($\frac{dE_{\rm GW}(m_1, m_2, z)}{dt_r}$), and the residence time (i.e.~the amount of time a source spends at a frequency: $\frac{dt_r}{d\ln f_r}$).  Following  
\citet{Kelley:2016gse, Kelley:2017lek}, we write the energy released as
\begin{align}\label{eq:gw1}
    \frac{dE_{\rm GW}(m_1, m_2, z)}{d\ln f_r}
    =& \frac{dE_{\rm GW}(m_1, m_2, z)}{d\ln f_r}\bigg|_{\rm GW} \frac{\tau_h}{\tau_{\rm GW}}(f),\\ \nonumber
    \\ \nonumber
    =&  f_r \frac{dE_{\rm GW}(m_1, m_2, z)}{df_r}\bigg|_{\rm GW}\frac{\tau_h}{\tau_{\rm GW}}(f),
\end{align}
where 
\begin{equation}
    \frac{dE_{\rm GW}(m_1, m_2, z)}{df_r}\bigg|_{\rm GW}= \frac{(\pi G)^{2/3} M_c^{5/3}}{3(1+z)f_r^{1/3}},
\end{equation}
for circular orbits emitting signals up to the innermost circular stable orbit (ISCO).   Here, $M_c= (m_1m_2)^{3/5}/(m_1+m_2)^{1/5}$ is the binary's chirp mass, and $f$ is the frequency, which at ISCO is given by $f_{r, \rm ISCO}= c^3/(6^{3/2}\pi G M_{\rm tot})$ in terms of total mass of the binary $M_{\rm tot}= m_1+m_2$. In the presence of higher harmonics, this equation modifies to a sum over the higher harmonics \citep{Enoki:2006kj}. 

As for the second term in Eq. \eqref{eq:gw1}, the ratio $\frac{\tau_h}{\tau_{\rm GW}}(f)$ captures the residence time of the GW signal at a particular frequency.  The numerator ($\tau_h \equiv \frac{a}{da/dt_t}$) is the binary hardening time expressed in terms of the semi-major axis of the binary $a$.  Initially, this timescale depends on the environmental effects arising due to the interaction between the binaries and their local environment.  These effects include (i) dynamical friction, (ii) stellar loss-cone scattering, and (iii) viscous drag.   The impact of these environmental effects is among the major sources of  uncertainty in the spectral shape of the signal but 
typically these environmental effects reduce the residence time of the GW signal at a particular frequency and the ratio will be less than one.    
 
 As we have noted, the above environmental effects cannot be directly computed from the {\sc Romulus25} simulation.  Moreover, from an EM observations point of view, resolving galaxies on sub-parsec scales at a cosmological distance is not possible with currently ongoing and upcoming surveys.  However, we can determine the average astrophysical properties of a galaxy --- like gas density, stellar mass, halo mass, and other properties --- on kpc scales from cosmological simulations as well as observations.  We therefore model the ratio, $\frac{\tau_{h}}{\tau_{\rm GW}}$, in terms of the average astrophysical properties of host SMBBH galaxies: 
\begin{align}\label{hardening}
    \frac{\tau_{h}}{\tau_{\rm GW}}(f)= \mathcal{E}(f, \dot{M}_*, M_*, & M_h, \rho_{\rm gas}, z).
\end{align}
In effect, we want to construct a framework that can relate the nano-Hertz (nHz) GW signal detectable from PTA with the observable quantities of galaxies.

\subsection{Modelling the environmental effect}\label{sec:env}

In this subsection, we discuss the model for $\mathcal{E}(f, \dot{M}_*, M_*, M_h, \rho_{\rm gas}, z)$ in greater detail.  But first we note that the impact of the environmental effects is greatest when the binaries are further away from each other and are radiating at lower frequencies 
of GW \citep{volonteri2003,volonteri2020,2011MNRAS.411.1467K, PhysRevD.91.084055, Chen:2016zyo, Kelley:2016gse, Kelley:2017lek}. As the SMBBHs inspiral and their separation decreases, they emits GW signals at increasingly higher frequencies. 
At frequencies of around 1 $\rm yr^{-1}$ (or about a few $\times 10^{-8}$ Hz), the environmental effects are no longer dominant.  The SMBBHs' evolution is dominantly through GW emission, and the frequency-dependent part of the ratio $\frac{\tau_h}{\tau_{\rm GW}}(f)$ approaches unity. Impact of these effects on the GW strain are often modelled using parametric forms \citep{PhysRevD.91.084055, Chen:2016zyo}.

{In the present case, \textsc{Romulus25} allows us to track the evolution of the black holes down to a few hundred parsecs} but none
of the current generation of cosmological simulations have the resolution to follow {their evolution due to the above processes to smaller scales.} 
Moreover, one also needs very high-resolution observations to determine the density profile of stars and gas at these scales.  We therefore use a parametric equation to capture the environmental effect $\mathcal{E}$. {In effect, $\mathcal{E}$ can be thought of as a subgrid model that uses accessible galaxy properties to estimate the overall number of SMBBH sources that will contribute to the nHz signal as well as the amount of their orbital energy that goes into GWs.}

{As noted in the last section, one of the key aspects of the problem where the astrophysical properties of the galaxies play an important role is in determining the fraction of SMBBHs that can contribute to the SGWB in the nHz range. These SMBBHs can successfully  reach from the orbital separation $\sim$ kpc scales to about $10^{-5}$ pc (i.e. the GW emission-dominated regime) from within the age of the Universe.} We model this via a dimensionless parameter $\alpha$, which quantifies how efficiently the SMBBHs identified in the simulations on a $\sim$ kpc scale will overcome the last parsec problem.  
We expect that $\alpha$ will depend on the various astrophysical properties of the host galaxy and given the specific nature of the processes involved, we make an ansatz that $\alpha$ will primarily depend on the galaxy’s gas density ($\rho_{\rm gas}$), stellar mass ($M_*$), and star formation rate ($\dot{M}_*$), and can be written as
\begin{align}\label{scaling-1}
    \alpha= \,&\alpha_\rho \bigg(\frac{\log(\rho_{\rm gas}= 10^7 M_\odot/\rm kpc^3)}{\log(\rho_{\rm gas})}\bigg)  \nonumber\\ & + \alpha_{M_*} \bigg(\frac{\log(M_*= 10^{10} M_\odot)}{\log(M_*)}\bigg) \nonumber \\ & + \alpha_{\dot{M}_*} \bigg(\frac{\log(\dot{M}_* = 10^{8} M_\odot/\rm Gyr)}{\log(\dot{M}_*)}\bigg),
\end{align}
where $\alpha_\rho$, $\alpha_{M_*}$, and $\alpha_{\dot M_*}$ govern how gas density, stellar mass and star formation rate, respectively, are in driving the hardening of the black hole binaries.

{The other aspect, which plays a crucial role in controlling the shape of the stochastic GW power spectrum, is the amount of orbital energy that is lost via environmental processes. This energy loss is modeled by the factor $(1+ \beta(\frac{f}{f_t})^{-\kappa})^{-\gamma}$.}  {Here the dimensionless factor $\beta$ captures the frequency-dependent loss of GW signal due to processes like dynamical friction, stellar hardening, and viscous drag, relative to the case where these environmental effects are absent and the hardening of the binary is driven only by GW emission.
We model this as}
\begin{align}\label{scaling-2}
    \beta =\, &\beta_\rho \bigg(\frac{\log(\rho_{\rm gas}= 10^7 M_\odot/\rm kpc^3)}{\log(\rho_{\rm gas})}\bigg) + \beta_{M_*} \bigg(\frac{\log(M_*= 10^{10} M_\odot)}{\log(M_*)}\bigg).
\end{align}

The term $\kappa$ controls the spectral behavior of the environmental effects \citep{PhysRevD.91.084055}. The value for some of the effects such as stellar scattering is $10/3$. However, the combination of various effects can lead to a different spectral index. Finally, the parameter $\gamma$ controls the overall tilt of the environmental effects. For a fiducial case with a GW-emission-only scenario, the value of $\gamma=1$ can be considered as a fiducial. However, there can be deviations due to astrophysical effects. For the spectral shape of the signal, we use three parameters, namely $\gamma, \kappa$, and $f_t$ where $f_t$ is the transition frequency at which the GW dominant effects become important over the environmental effects. The transition frequency
can be expressed in terms of the stellar density $\rho_\ast$ (in units of M$_\odot$ pc$^{-3}$) and velocity dispersion $\sigma_\ast$ (in units of km/s), eccentricity $e$, and chirp mass $M_c$ (in units of solar mass) as 

\begin{align}
    f_t= f_0\bigg(\frac{\rho_{\ast}}{F(e)\sigma_\ast}\bigg)^{3/10}M_{c}^{-2/5},
\end{align}
where $F(e)= (1+ (72/24) e^2 + (37/96)e^4)/((1-e^2)^{7/2})$. $f_0$ is a correction factor incorporating any effect that may not be captured by this simplistic approximate formula (such as mass ratio). For $\rho_\ast= 100 M_\odot$ pc$^{-3}$, $\sigma_\ast= 200$ km/s, $M_c=10^9$ M$_\odot$, $f_0=1$, the value of $f_t$ is around 0.4 nHz \citep{Chen:2016zyo}. {Combining all the together, we can write the total contribution from the host galaxy properties as}  
\begin{align}\label{hardening-1}
    \mathcal{E}(f, \dot{M}_*, M_*& , \rho_{\rm gas}, z)= \alpha \bigg[ 1+ \beta \bigg(\frac{f}{f_t}\bigg)^{-\kappa}\bigg]^{-\gamma},
\end{align}
{where $\alpha$, $\beta$ are defined above in Eq. \eqref{scaling-1} and Eq. \eqref{scaling-2}.}

\subsection{SGWB estimation from SMBBHs in \textsc{Romulus} simulation }\label{sec:sgwb-sims}

\begin{figure*}
    \centering
    \includegraphics[width= 1.\textwidth]{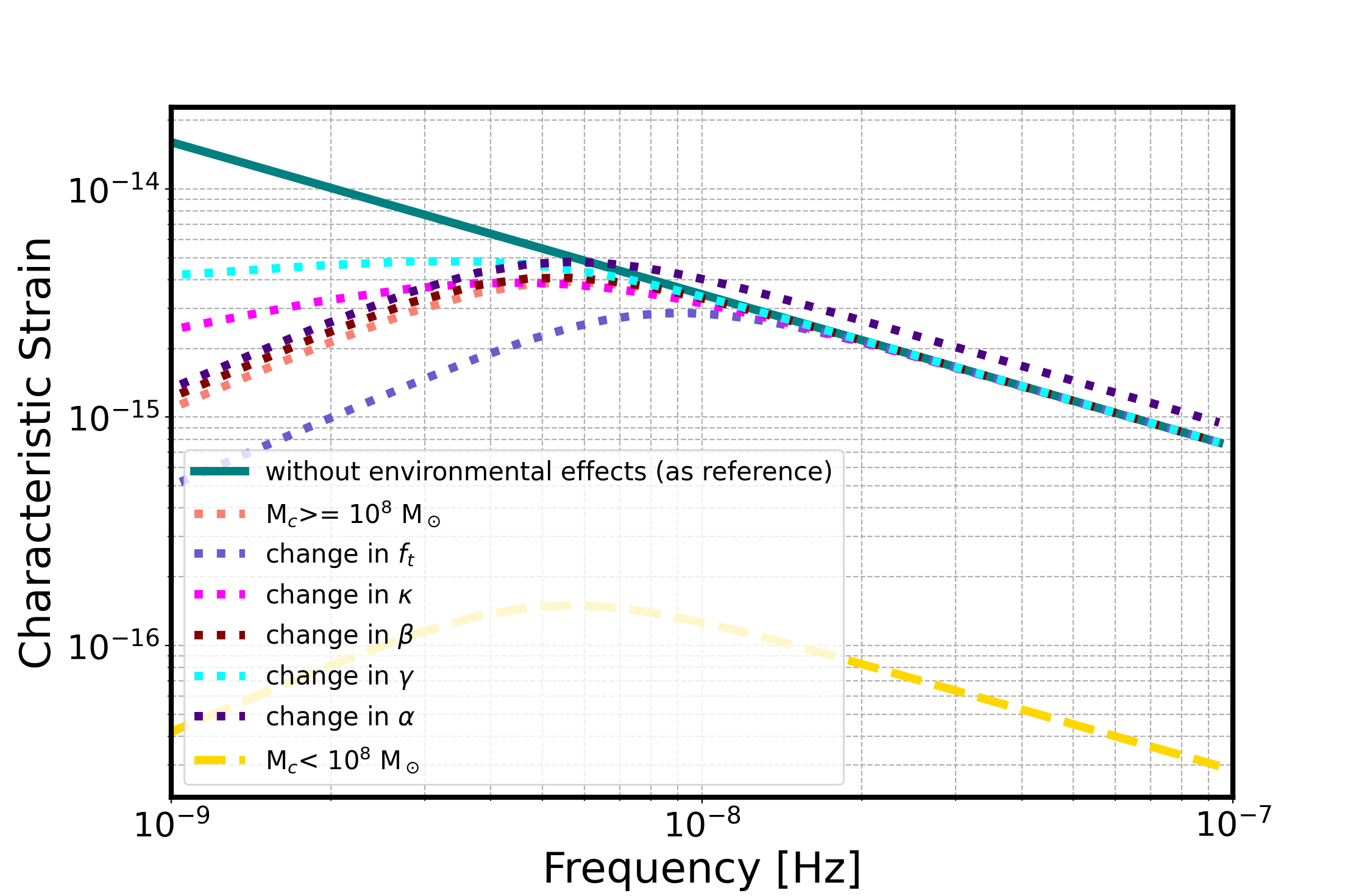}
    \caption{We show the SGWB strain as a function of the frequency with change in the fiducial values of the parameters $\alpha= 0.2$, $f_t= 5$ nano-Hertz, $\beta=1$, $\kappa=10/3$, and $\gamma=1$, and also for SMBBHs with chirp mass $\rm M_c < 10 ^ 8 M_\odot$ and $\rm M_c >= 10 ^ 8 M_\odot$. }
    \label{fig:sgwb}
\end{figure*}

Having put in place the above elements, we can use Eq. \eqref{GW-strain}, Eq. \eqref{eq:gw1}, and Eq. \eqref{hardening} to estimate the SGWB signal from discrete sources in a cosmological simulation as follows

\begin{align}\label{eq:conn2}
        h_c^2&(f) =  \frac{4G}{c^2\pi f^2V_c} \sum_{i} \frac{1}{(1+z^i)}\left[\frac{dE_{\rm GW}(m^i_1, m^i_2, z^i)}{d\ln f_r}\bigg|_{GW}\frac{\tau_{h}}{\tau_{\rm GW}}(f)\right]_i, 
\end{align}
where the sum runs over all the source SMBHHs (or equivalently, 
host galaxies in which coalescing SMBHs are present) in a simulation box of comoving volume $V_c= (25 \rm {Mpc})^3$ that contributes to GW background.

{We use the GW sources identified in the {\sc Romulus25} simulation (see \S \ref{sec:bhp1} for details about how these sources are identified) and the above relationships to model the SGWB. As described in \S \ref{sec:env}, the environmental effects --- the $\frac{\tau_{h}}{\tau_{\rm GW}}(f)$ part of Eq.\ref{eq:conn2} --- are modelled by $\mathcal{E}$ via the $\alpha$ and $\beta$ parameters defined in 
Eq. \eqref{scaling-1} and Eq. \eqref{scaling-2}. 
For $\rho_{gas}$, $M_*$, and $\dot{M}_*$ values that enter into these equations, we use the median values of these properties from the central 1 kpc region of the host galaxies of the 
{\sc Romulus 25} sources.  For additional details, see \S \ref{sec:bhp1}. As for our procedure, after inputting the median values of $\rho_{gas}$, $M_*$, and $\dot{M}_*$ into Eq. \eqref{scaling-1} and Eq. \eqref{scaling-2}, the free parameters of $\alpha_{\rho}$, $\alpha{M_*}$, $\alpha_{\dot{M}*}$, $\beta_{\rho}$, and $\beta_{{M}_*}$ are chosen such that that the amplitude of the signal matches with the observed nHz signal at frequency $f=1 yr^{-1}$ \citep{Xu:2023wog,Antoniadis:2023ott,NANOGrav:2023gor,Zic:2023gta}. This results in $\alpha = 0.2$ and $\beta = 1$.}

{The results of modeling the SGWB are shown in Fig. \ref{fig:sgwb}. For SMBHs with chirp mass $M_c \ge 10^8$ M$_\odot$, the results for the case where environmental effects are \emph{not considered} is shown as solid teal line and denoted as ``reference''.  The dotted and dashed lines show the results for SMBBHs with chirp mass $M_c \ge 10^8$ M$_\odot$ and  $M_c \le 10^8$ M$_\odot$, respectively, when environmental effects are taken into account.  The salmon dotted line corresponds to  $M_c \ge 10^8$ M$_\odot$ and parameters $\alpha = 0.2$, $\beta=1$, $f_t= 5$ nHz, $\kappa=10/3$ and $\gamma=1$.
The values of $\alpha$ and $\beta$ are computed as described above and the values of 
$f_t$, $\kappa$ and $\gamma$, which control the shape of the signal, are fiducial values from physical models discussed in the previous subsection (cf \S \ref{sec:env}). For the same parameters, the result for SMBBHs with $M_c \le 10^8$ M$_\odot$ is shown by a yellow dashed line.  An important point to note is that the contribution to the total signal from SMBBHs with chirp mass $M_c <10^8$ M$_{\odot}$ is not significant; they only contribute about $10\%$ of the signal arising from sources with $M_c \ge 10^8$ M$_\odot$.}

{To illustrate the impact of variations in environmental properties, we present results for SMBBHs with $M_c \ge 10^8$ M$_\odot$ where we vary the values of $\alpha$, $\beta$, $f_t$, $\kappa$ and $\gamma$.   Specifically, we examine the effects of changing $f_t$ from 5 to 8 nHz (shown in purple), $\kappa$ from 10/3 to 7/3 (in magenta), $\beta$ from 1.0 to 0.8 (in brown), $\gamma$ from 1. to 0.5 (in cyan), and $\alpha$ from 0.2 to 0.3 (in indigo). As each parameter value is changed, the spectral shape of the signal exhibits only moderate changes. We will explore the parameter estimation using the galaxy catalog in future work (in preparation). If the properties of the underlying host galaxy can be inferred, then the values of the parameters that control the environmental effects can be measured.}

\subsection{Connecting SGWB signal with galaxy properties}\label{sec:sgwb-gal}

In the case of {\sc Romulus25} simulation, we have firsthand knowledge of  $\frac{d^3n_{\rm GW}(m_1, m_2, z)}{dm_1 dm_2 dz}$, the number density of GW sources.   However, we aim to find a map between the GW sources and the EM observations -- specifically, the galaxies in a complete galaxy catalog.  

To connect the EM and GW observational sectors, we assert that the total number of GW sources should equal the total number of GW source host galaxies
\begin{widetext}
\begin{align}\label{eq:num}
   \int dz\, \frac{dV}{dz}  \iint d m_1\, dm_2 \frac{d^3n_{\rm GW}(m_1, m_2, z)}{dm_1 dm_2 dz} = \int dz\, \frac{dV}{dz}  \iiint d(\dot{M}_*)\,  dM_{*}\, d M_{h}\, d \rho_{\rm gas}\, \frac{d^4n_{\rm EM}(\dot{M}_*, M_*, M_h, \rho_{\rm gas},z)}{d \dot{M}_*  d M_{*} d M_{h} d \rho_{\rm gas}}, 
\end{align}
\end{widetext}

\noindent where $\frac{d^3n_{\rm GW}(m_1, m_2, z)}{dm_1 dm_2 dz}$ is as described previously and $\frac{d^4n_{\rm EM}(\dot{M}_*, M_*, M_h, \rho_{\rm gas}, z)}{d \dot{M}_*  d M_{*} d M_{h} d \rho_{\rm gas}}$ is the number density of EM sources (that are hosts of a GW source) in a bin of star formation rate ($\rm \dot{M}_*$), stellar mass ($\rm M_*$), halo mass ($\rm M_h$), and gas density ($\rho_{\rm gas}$). This is just the conservation of numbers; we are, however, implicitly assuming that there is only one SMBBH per host galaxy which can emit in the PTA frequency range.

Given a complete galaxy catalog, we can relate the GW source host galaxies to the galaxies in the catalog via $\eta(z, m_1, m_2,\dot{M}_*, M_*, M_h, \rho_{\rm gas})$, the occupancy fraction of PTA sources of masses $m_1$ and $m_2$ in galaxies at redshift $z$ with star formation rate ($\rm \dot{M}_*$), stellar mass ($\rm M_*$), halo mass ($\rm M_h$), and gas density ($\rho_{\rm gas}$).  Incorporating this and the number conservation (Eq. \ref{eq:num}), we can connect the SGWB signal to the distribution of galaxies in a catalog:

\begin{widetext}
\begin{align}\label{eq:conn}
        h_c^2(f)= & \frac{4G}{c^2\pi f^2} \iint\iiint dz\, d\dot{M}_*\, dM_*\, dM_h\, d\rho_{\rm gas} 
        \eta(z, m_1, m_2,\dot{M}_*, M_*, M_h, \rho_{\rm gas}) 
        \frac{d^4n_{\rm gal} (\dot{M}_*, M_*, M_h, \rho_{\rm gas}, z)}{d\dot{M}_* dM_*dM_hd\rho_{\rm gas}} 
        \frac{1}{1+z}\frac{dE_{\rm GW}(m_1, m_2, z)}{d\ln f_r}\bigg|_{\rm GW} \frac{\tau_{h}}{\tau_{\rm GW}}(f).
\end{align}
\end{widetext}

The term $\alpha$, coming from $\frac{\tau_{h}}{\tau_{\rm GW}}(f)$ (cf Eq. \ref{hardening-1}), and the occupation fraction $\eta$ appear in the above as a multiplicative factor ($\alpha\times \eta$), which now takes care of the overall occupation of the number of the SMBBHs contributing to the SGWB in terms of both black hole masses and also the astrophysical properties of the galaxies. One can compare this with the observations and make an inference of this quantity.

Finally, we can also write the GW source distribution function,  $\frac{d^3n_{\rm GW}(m_1, m_2, z)}{dm_1 dm_2 dz}$, 
in terms of the halo mass function 

\begin{align}\label{rel-bh-halo}
    &\frac{d^3n_{\rm GW}(m_1, m_2, z)}{dm_1 dm_2 dz} = \int \frac{d^4n_{\rm GW}(m_1, m_2, z)}{dm_1 dm_2 dz dn_{\rm halo}(M_h,z)} \frac{dn_{\rm halo}(M_h,z)}{dM_h} dM_h,
\end{align}
where $\frac{d^4n_{\rm GW}(m_1, m_2, z)}{dm_1 dm_2 dz dn_{\rm halo}(M_h,z)}$ is the SMBBHs occupation number density in a halo of mass $\rm M_h$ and $\frac{dn_{\rm halo}(M_h,z)}{dM_h}$ is the halo mass function, i.e.~the number density of halos in halo mass bin $\rm M_h$ (We remind the reader that in the present study, we identify $\rm M_h$ with $\rm M_\mathrm{500}$.).

From a simulation, we can estimate the population of SMBBHs which can contribute to the SGWB in the PTA frequency range, and also identify the mass and redshift of the host halo. This gives us a connection between the SMBHs and halo mass $M_h$ written in Eq. \eqref{rel-bh-halo}. Similarly we can identify the astrophysical properties of the host galaxies of the coalescing binaries from simulations.   These would also be accessible from EM observations. This gives us an avenue to connect the astrophysical properties of the host galaxies with the SMBH properties. 

\section{Properties of the host galaxies of the nano-Hertz GW sources}\label{sec:host}

The dynamics of the SMBBHs and their contribution to the nHz frequency depends on the local astrophysical properties as discussed in the previous section. However, to identify the key astrophysical properties of the host galaxies that can be identified from observations, we need to explore the large-scale properties of the host galaxy. In this section, { we explore both local astrophysical properties in the vicinity of SMBBHs and their host galaxy properties}. The properties we consider include gas density, stellar mass, and star formation rate.  While discussing the host galaxies, we also comment on the properties of the halos in which these galaxies reside.

\begin{figure}
    \centering
    \includegraphics[width= 0.5\textwidth]{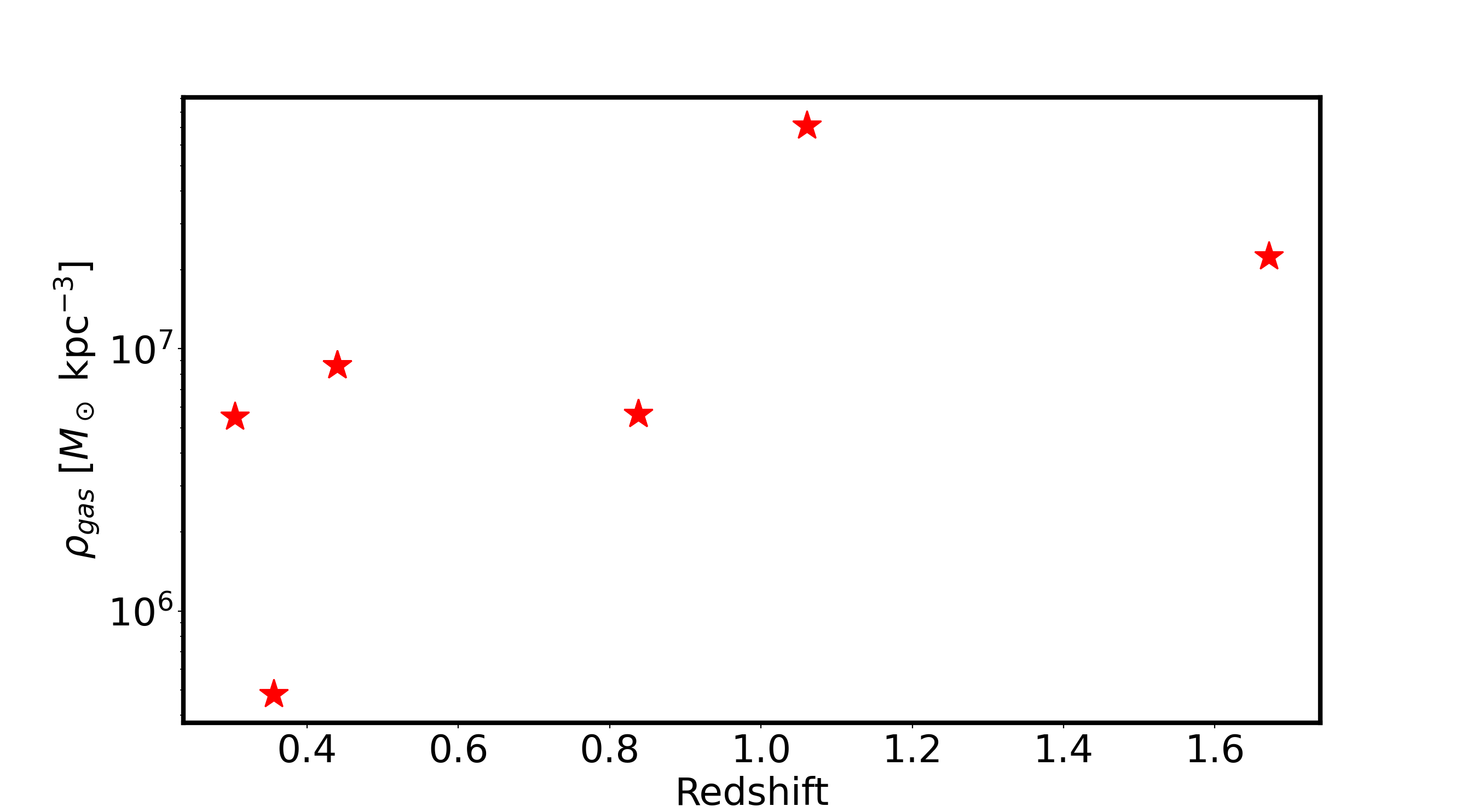}
    \includegraphics[width= 0.5\textwidth]{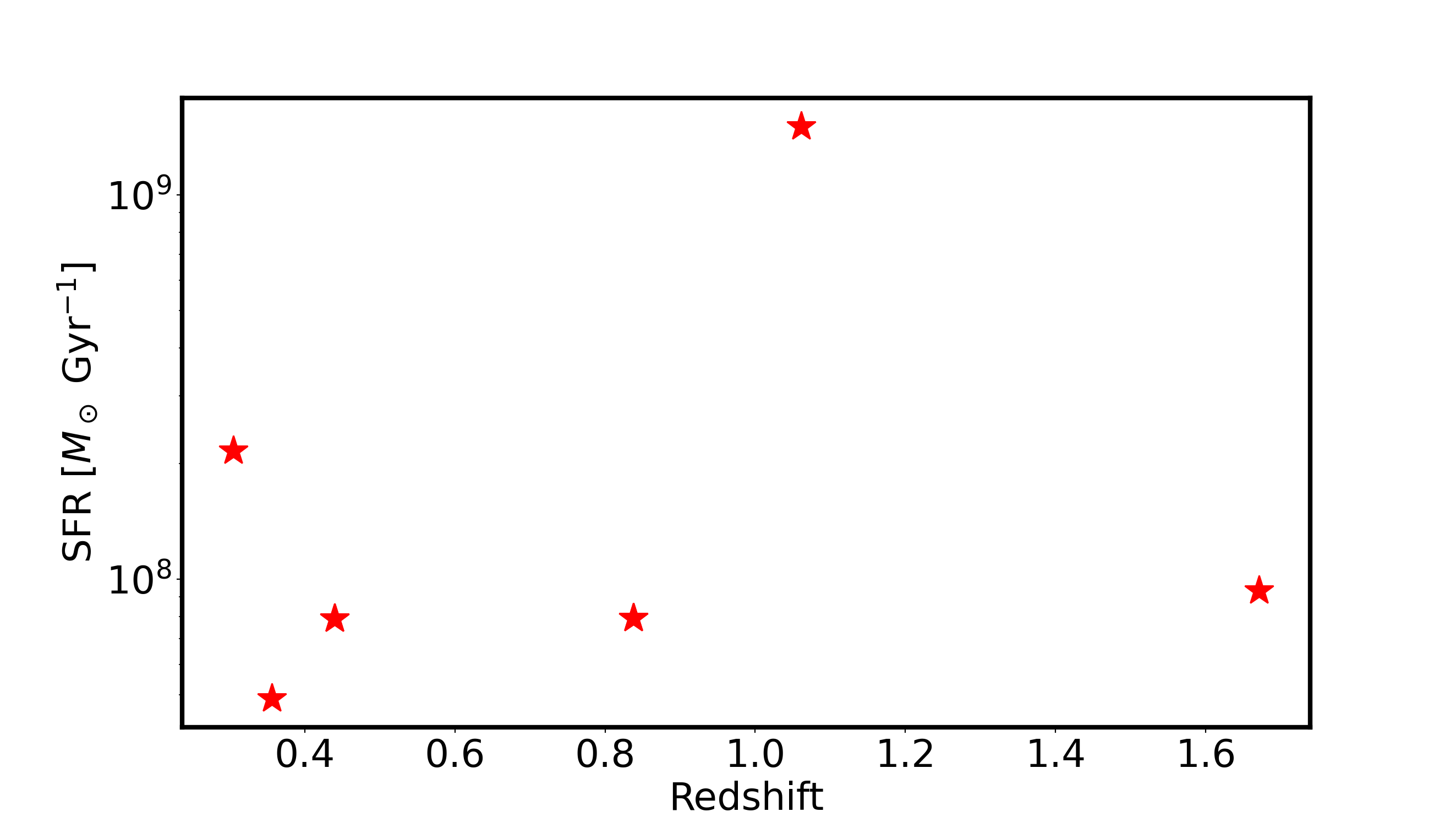}
     \includegraphics[width= 0.5\textwidth]{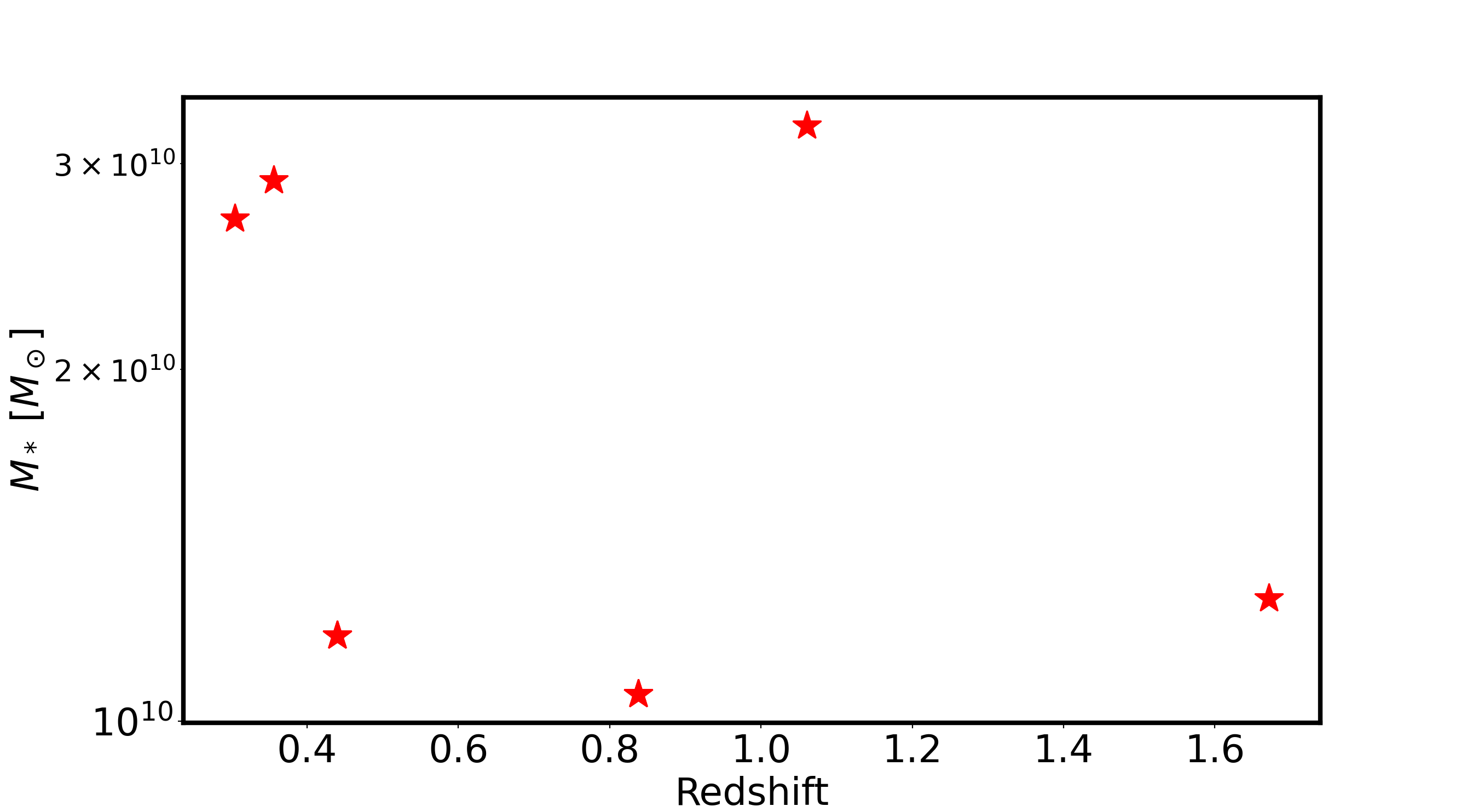}
    \caption{The local astrophysical properties of the galaxies: gas density (top), SFR (middle), and stellar mass (bottom) within 1 kpc of the more massive SMBH in merging SMBBHs {detected in {\sc Romulus25} and } contributing  most of the SGWB signal
    (i.e.~with chirp masses $M_c \geq 10^8$ M$_\odot$),  shown as a function of the host redshift.}
    \label{fig:BHs-z}
\end{figure}

\begin{figure*}
    \centering
    \includegraphics[width= 0.32\textwidth]{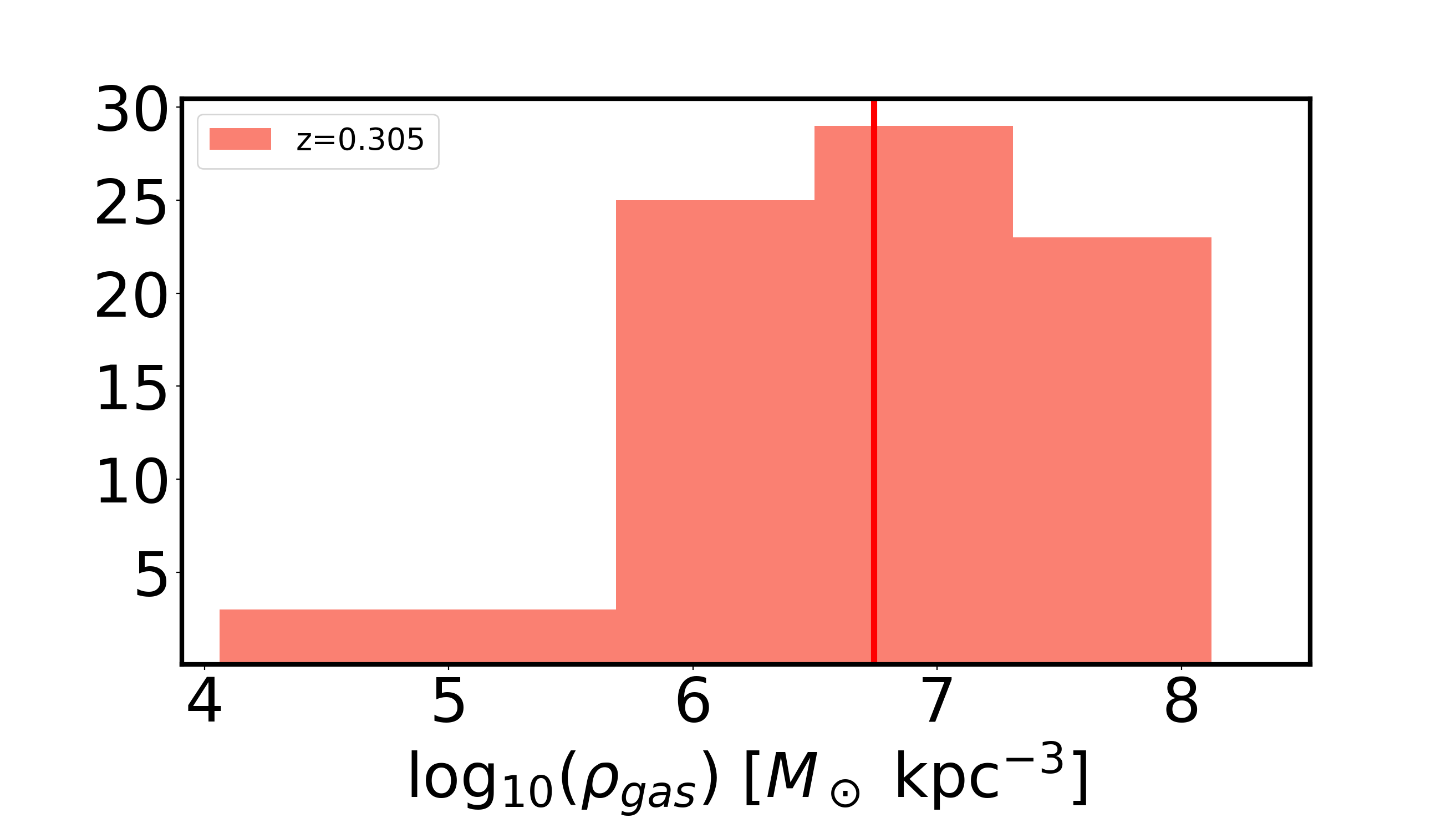}
    \includegraphics[width= 0.32\textwidth]{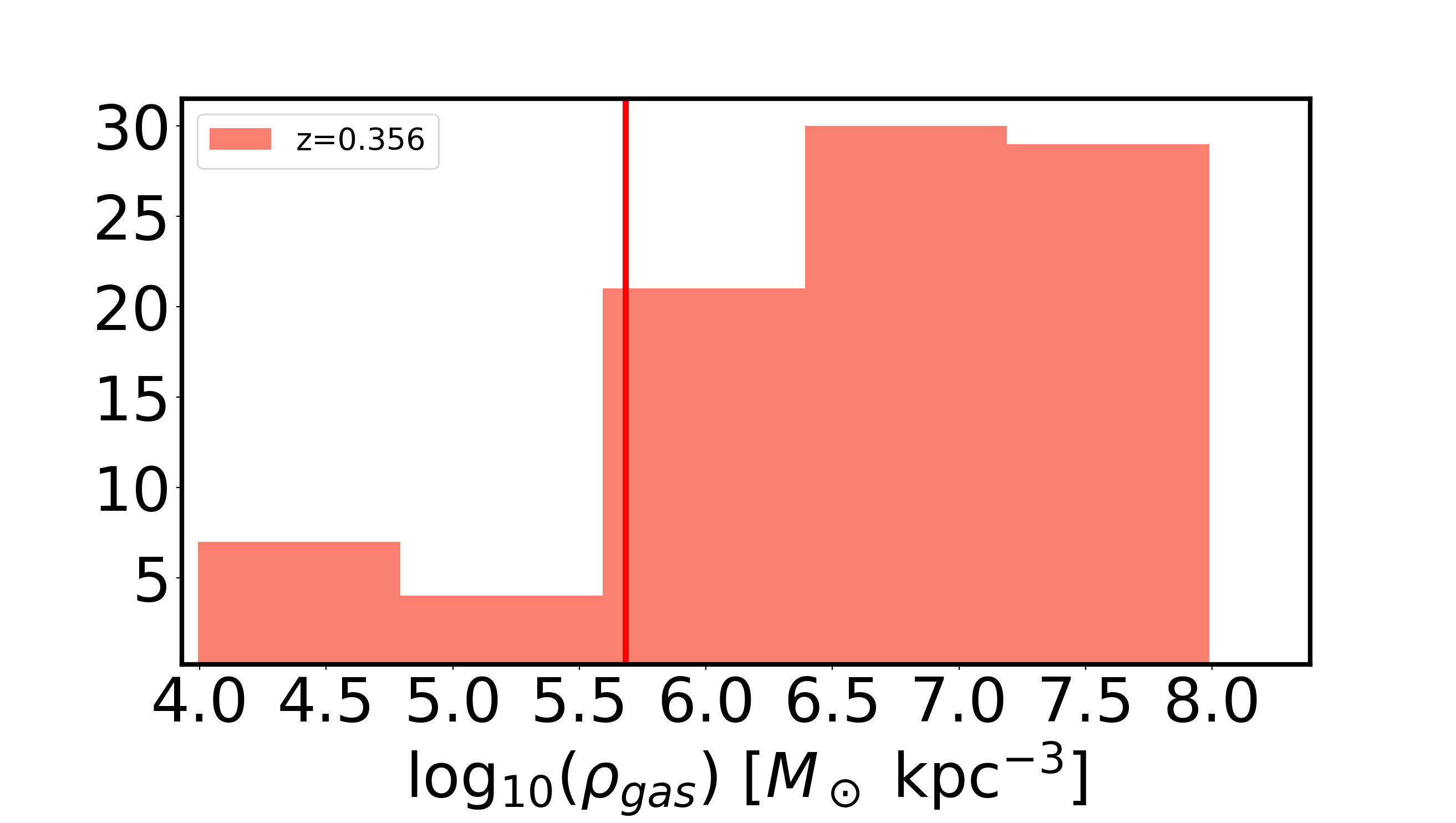}
    \includegraphics[width= 0.32\textwidth]{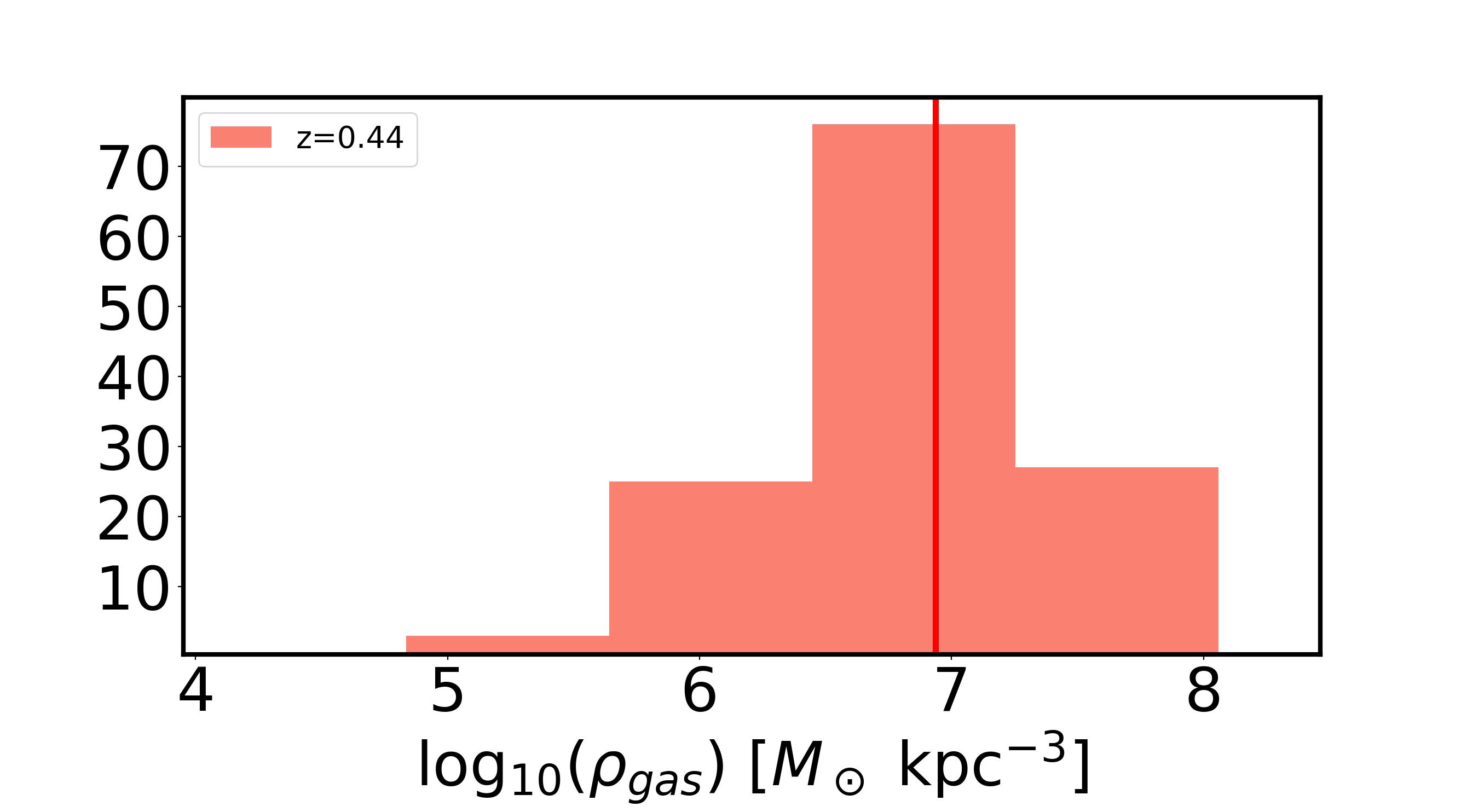}
    \includegraphics[width= 0.32\textwidth]{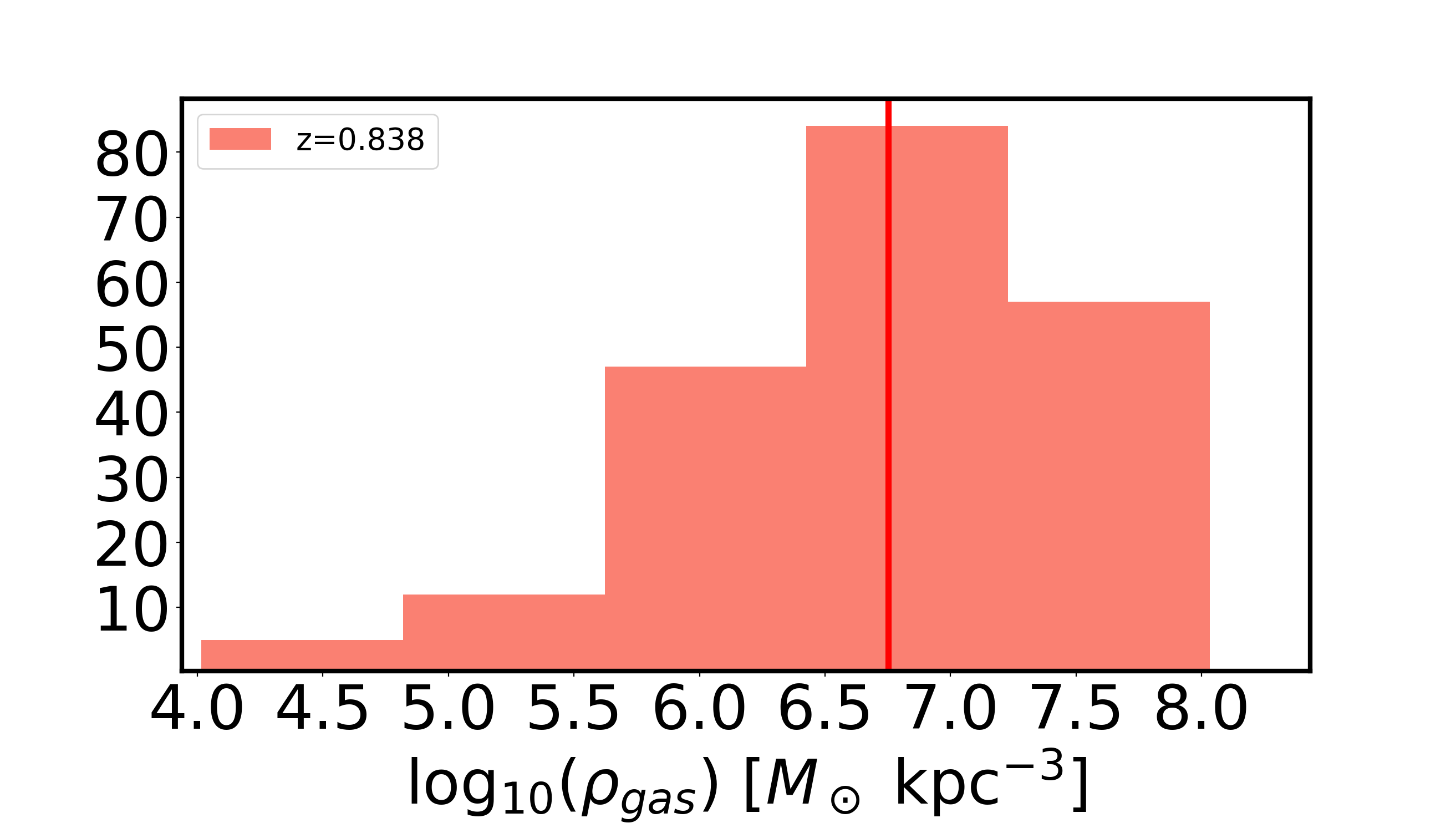}
    \includegraphics[width= 0.32\textwidth]{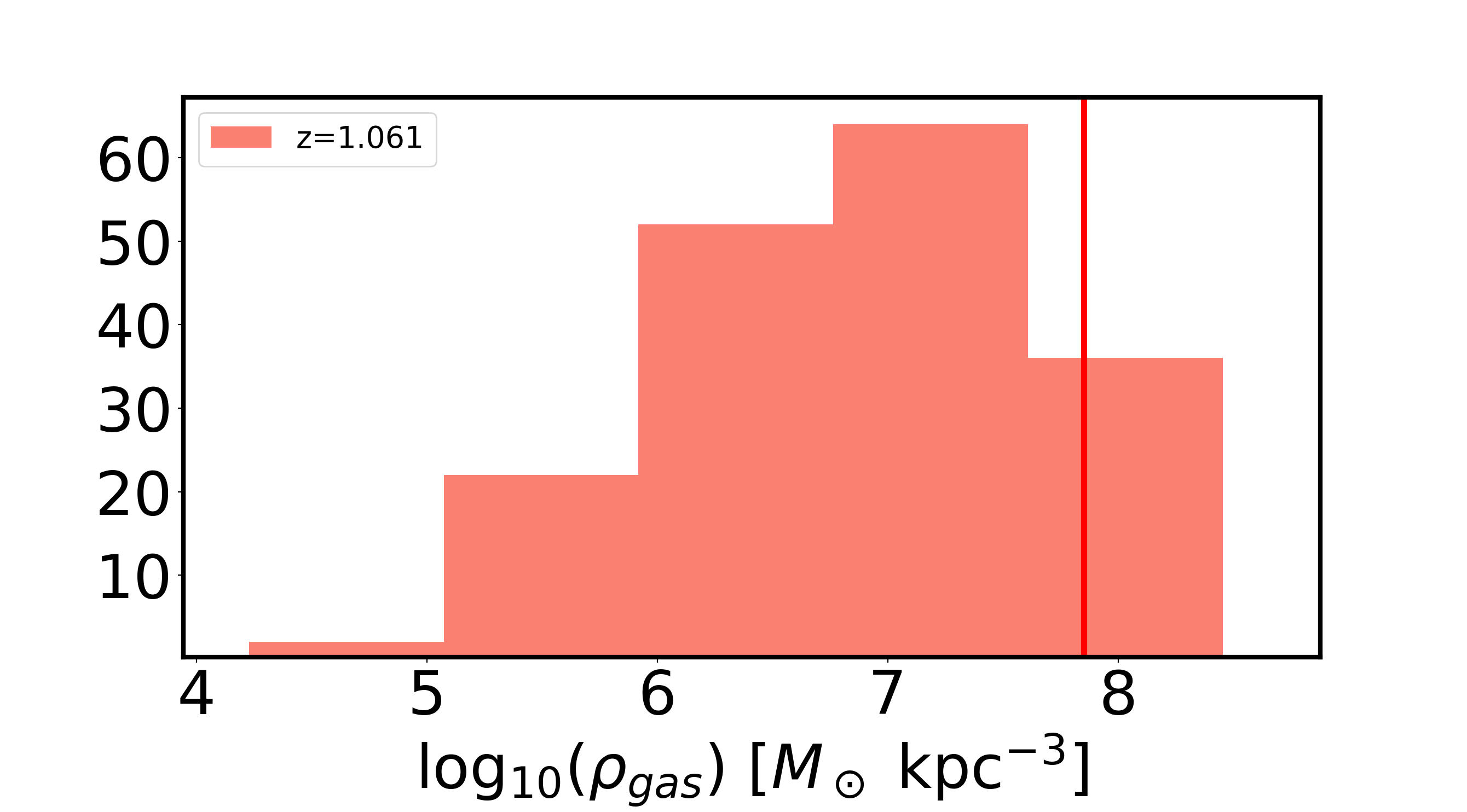}
    \includegraphics[width= 0.32\textwidth]{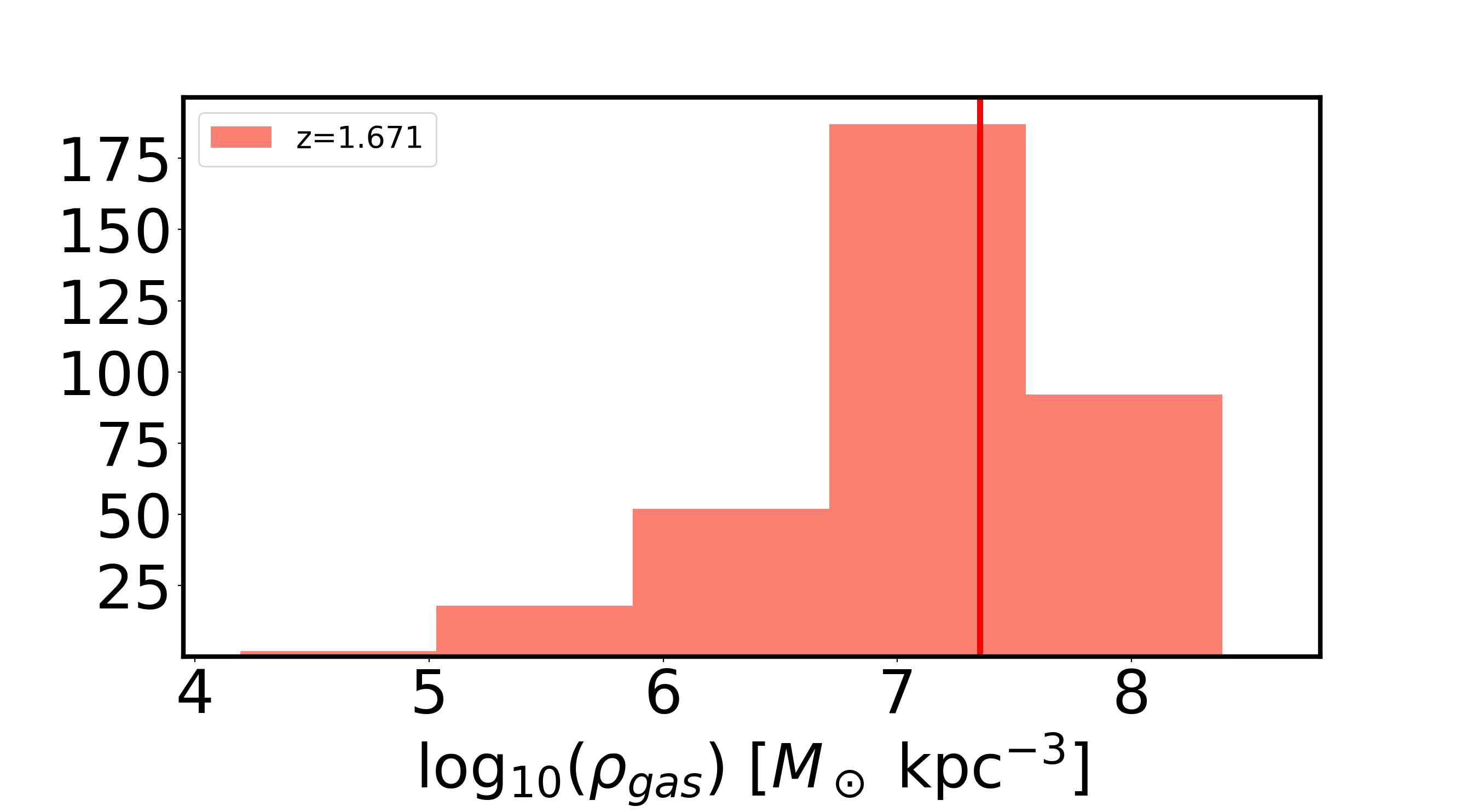}

    \hrule 

    \includegraphics[width= 0.32\textwidth]{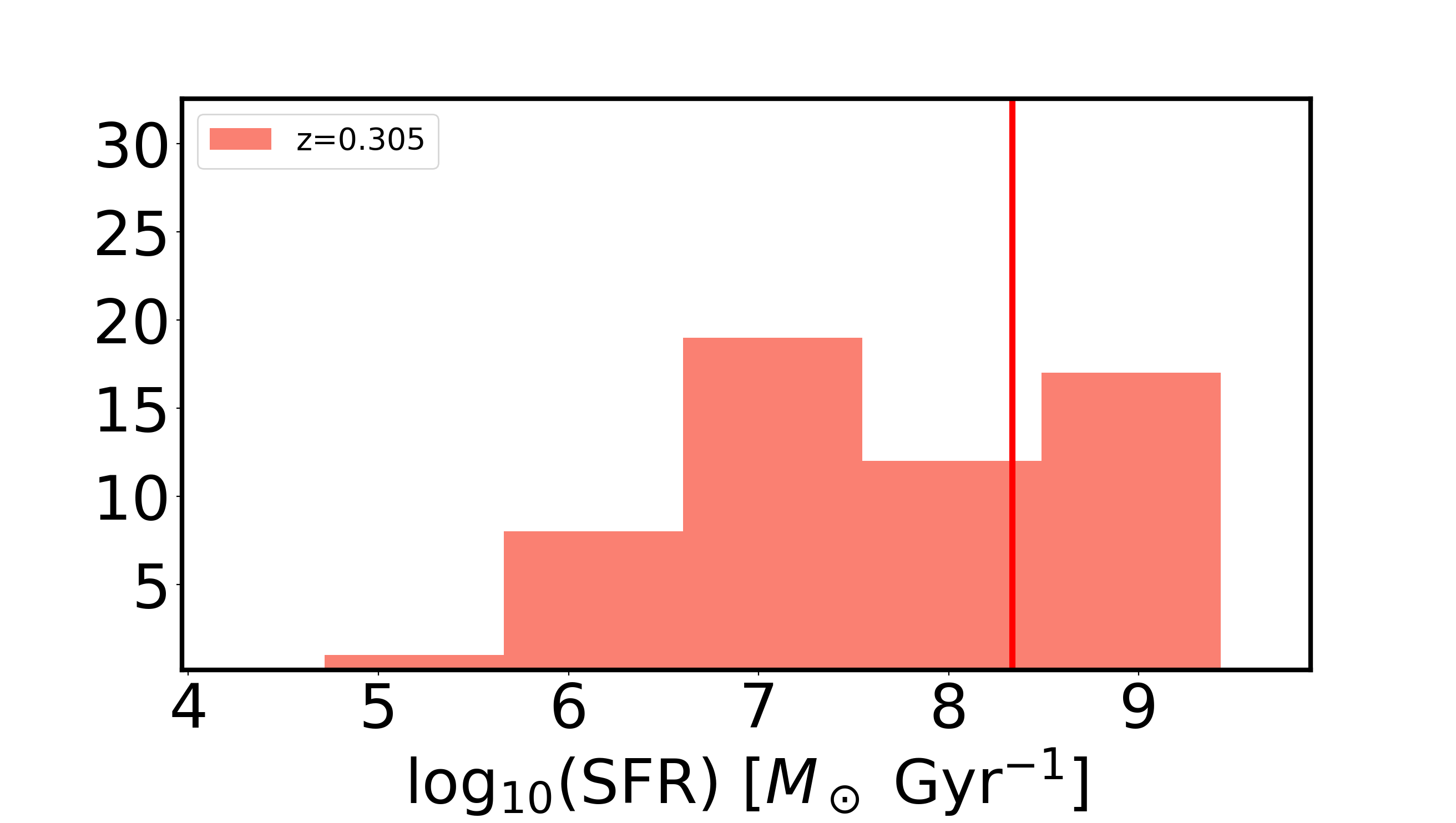}
    \includegraphics[width= 0.32\textwidth]{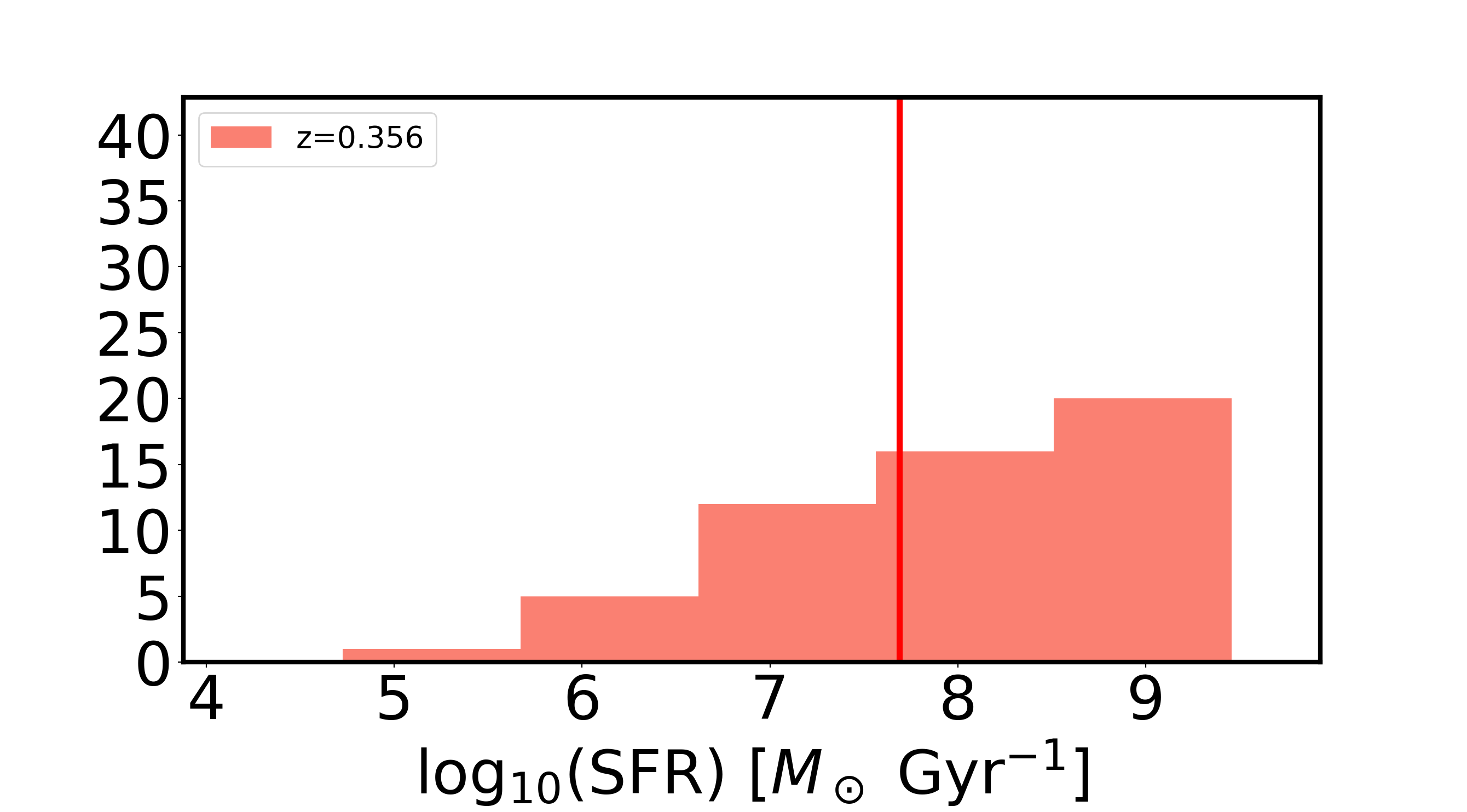}
        \includegraphics[width= 0.32\textwidth]{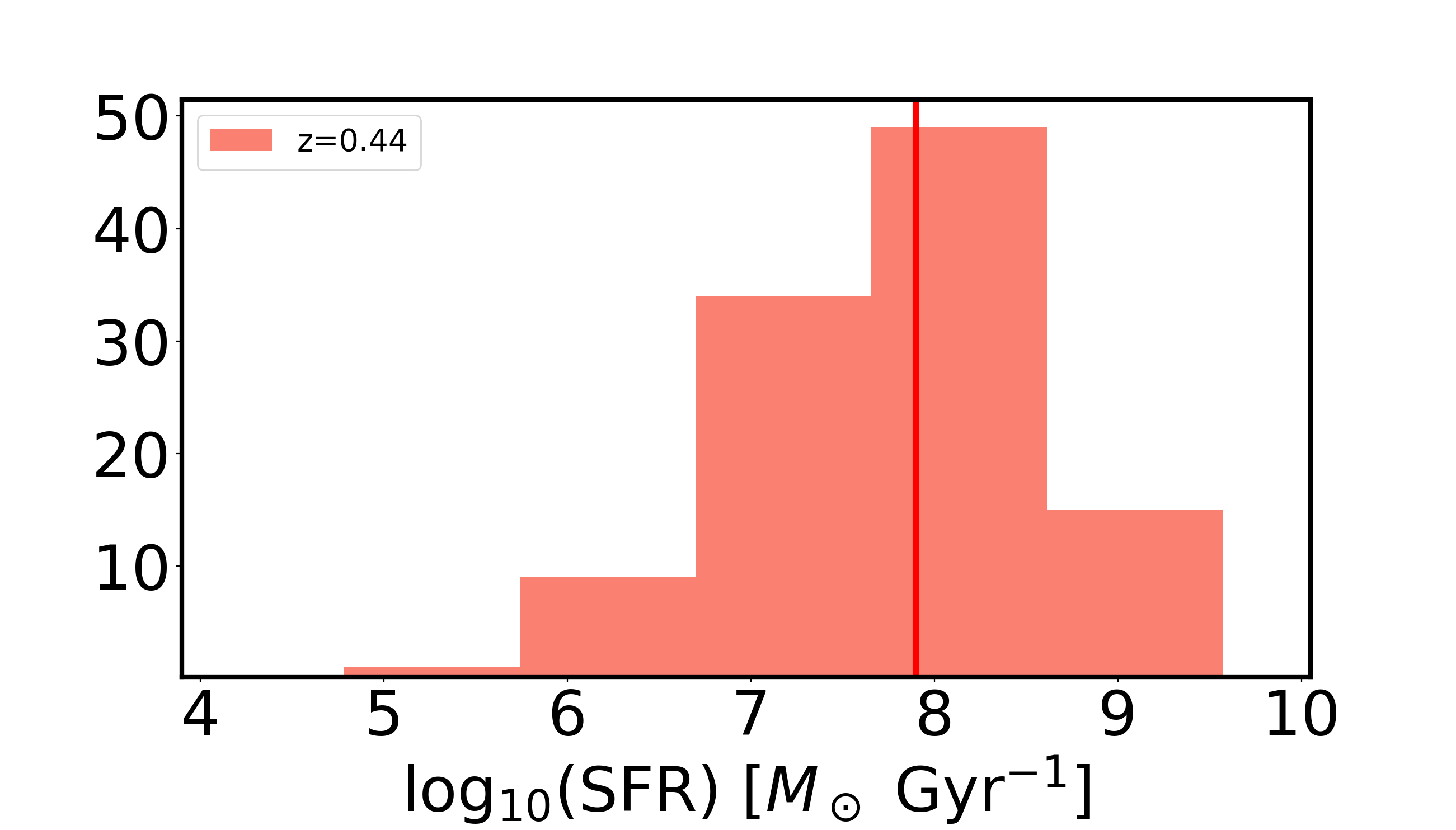}
    \includegraphics[width= 0.32\textwidth]{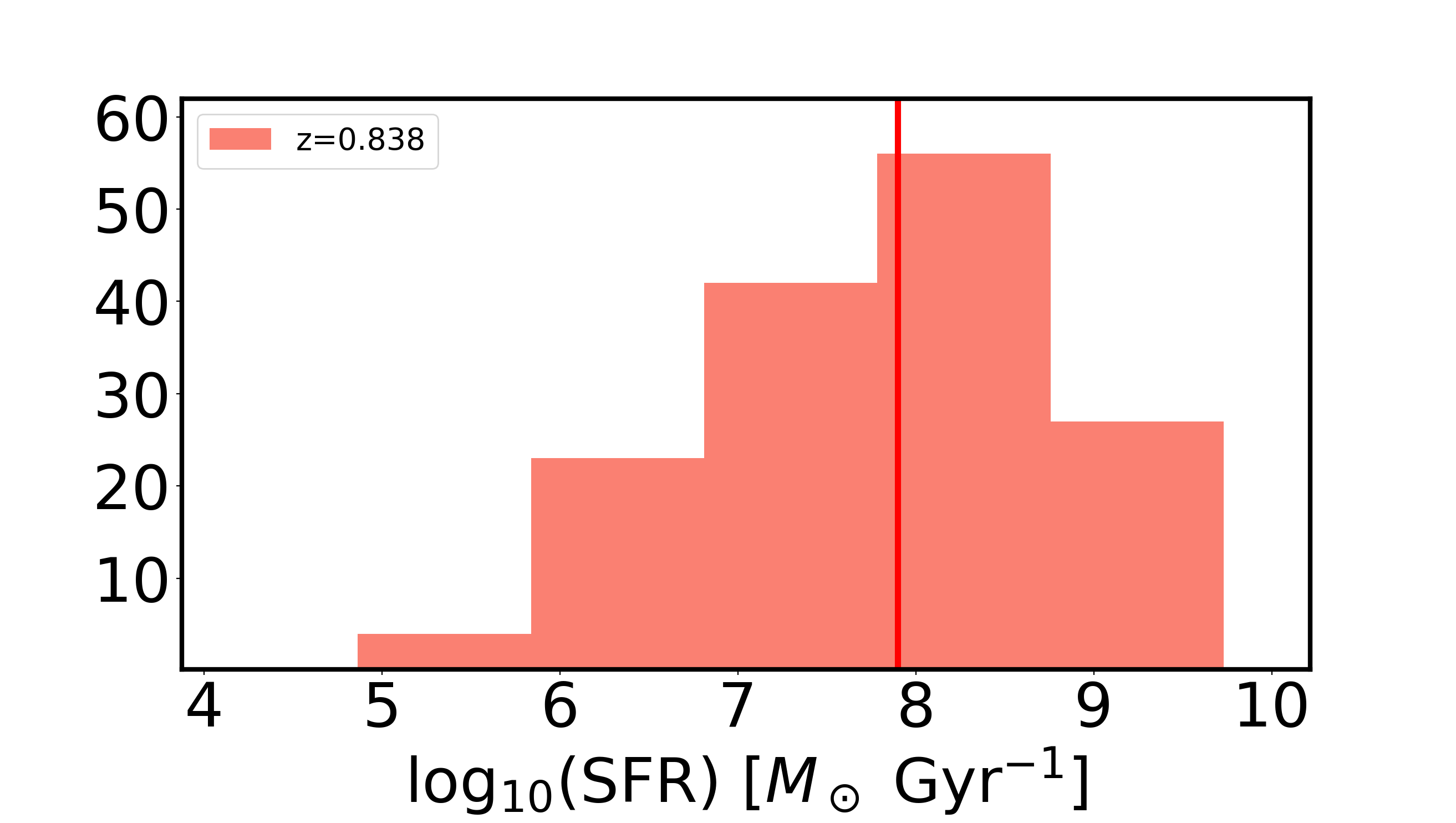}
        \includegraphics[width= 0.32\textwidth]{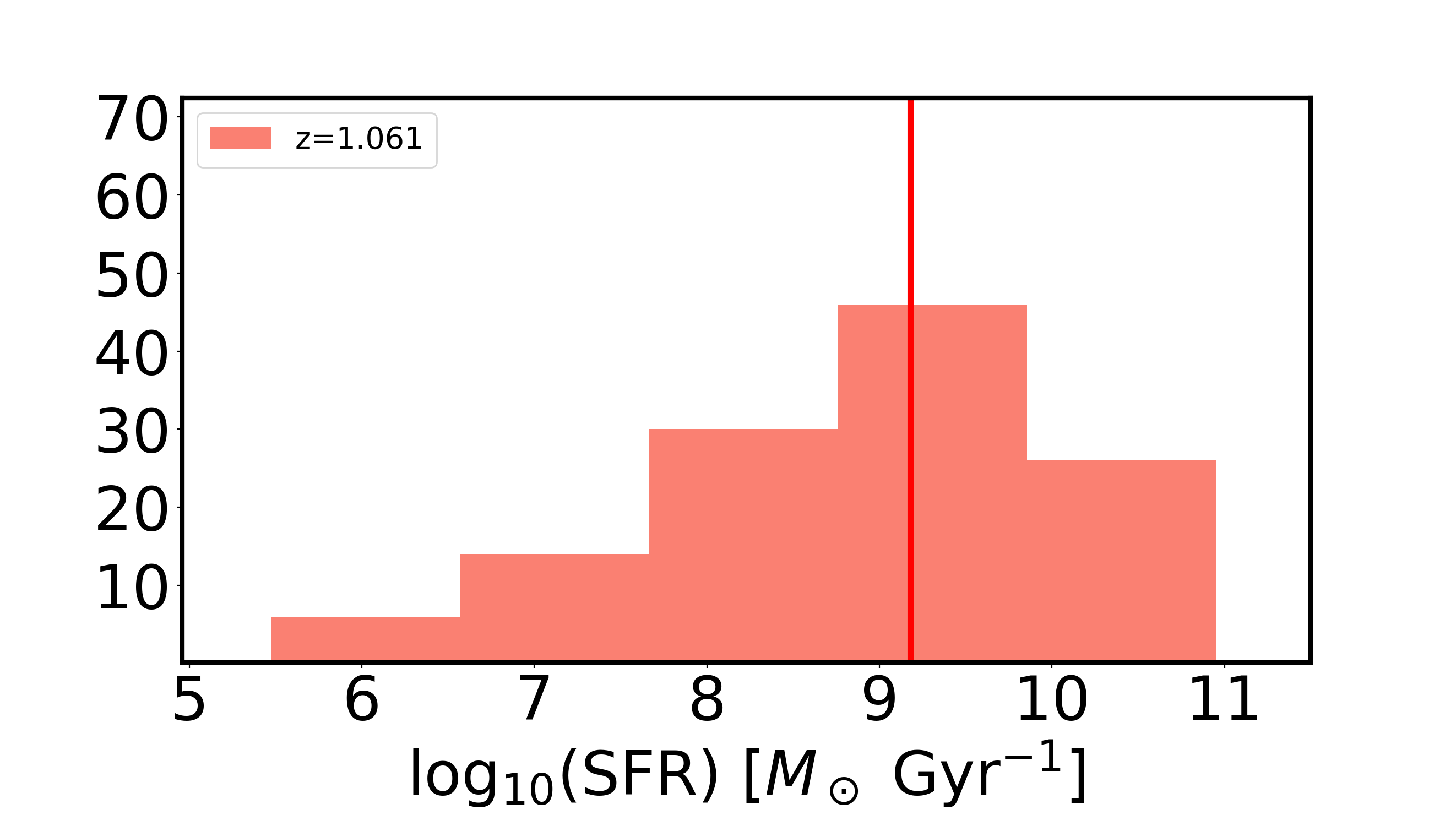}
    \includegraphics[width= 0.32\textwidth]{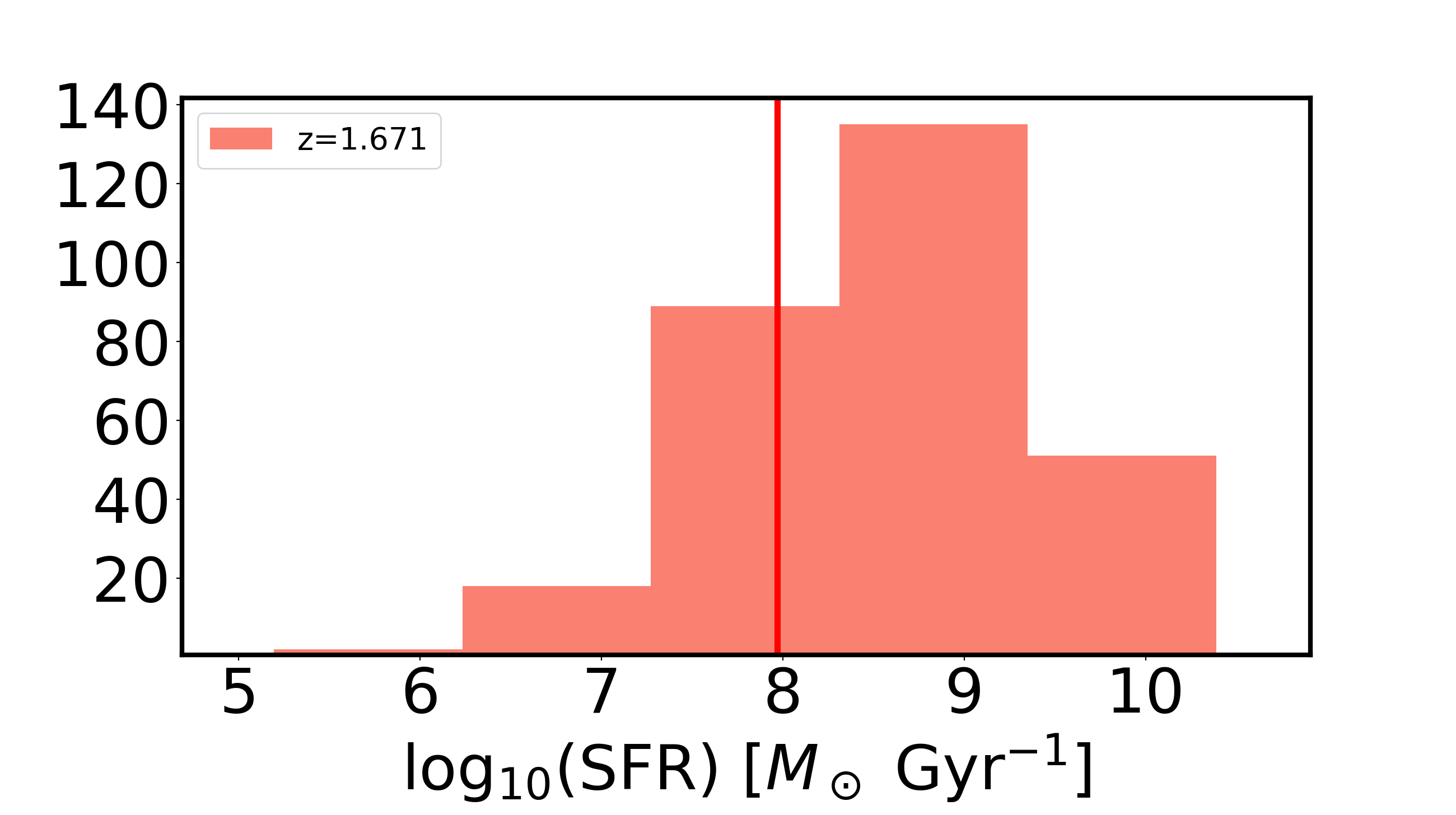}

    \hrule 
    
    \includegraphics[width= 0.32\textwidth]{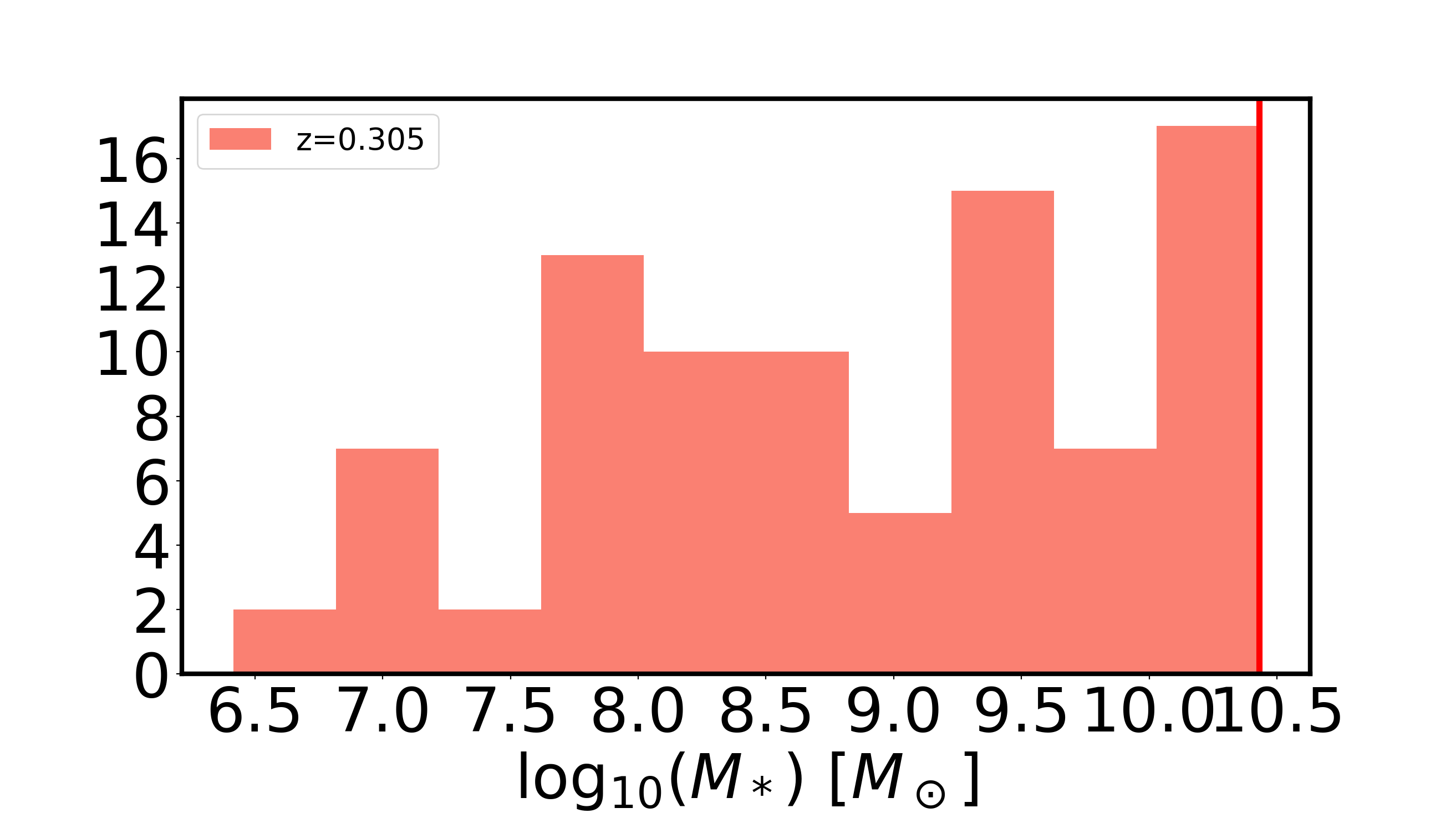}
    \includegraphics[width= 0.32\textwidth]{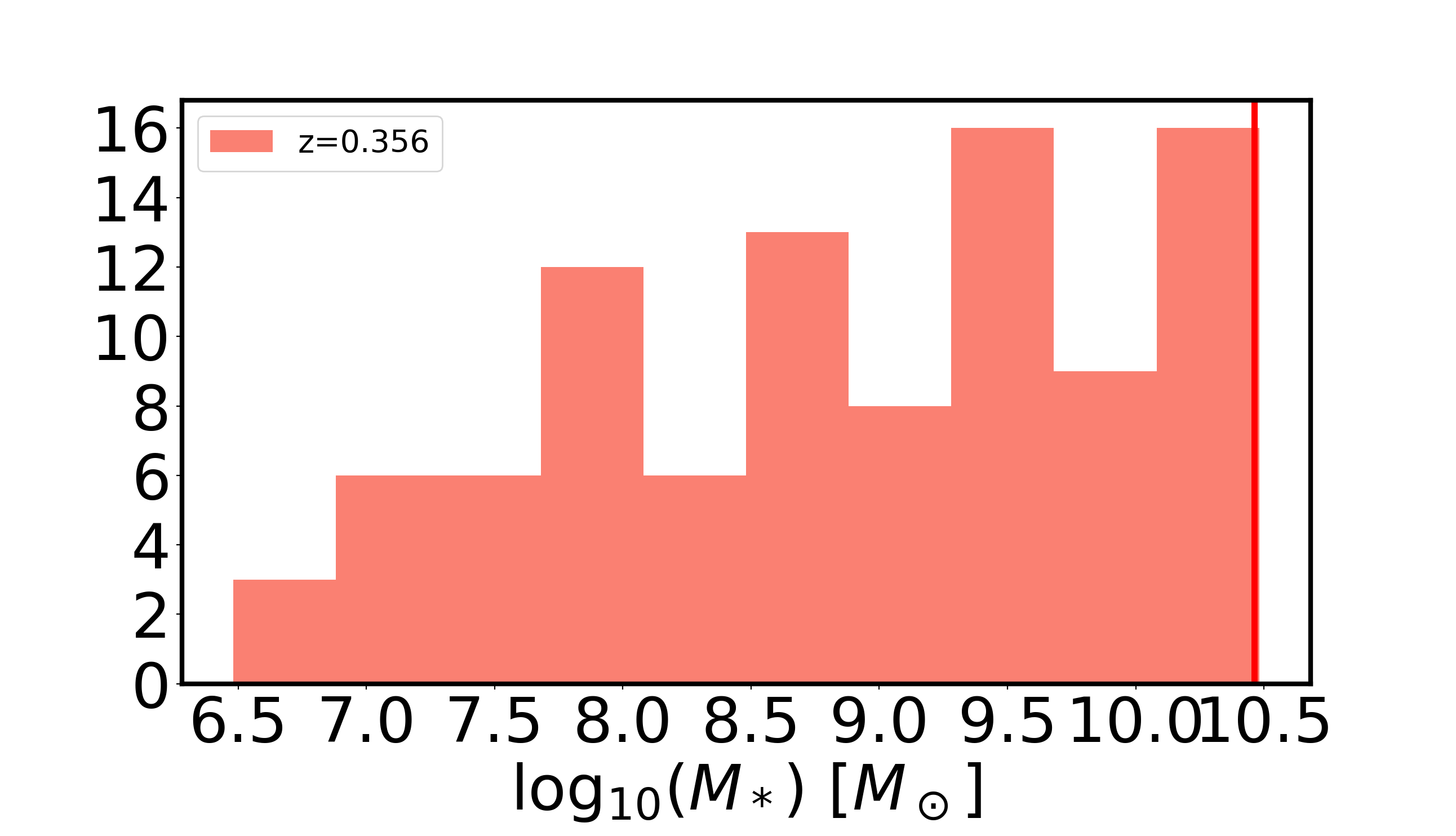}
      \includegraphics[width= 0.32\textwidth]{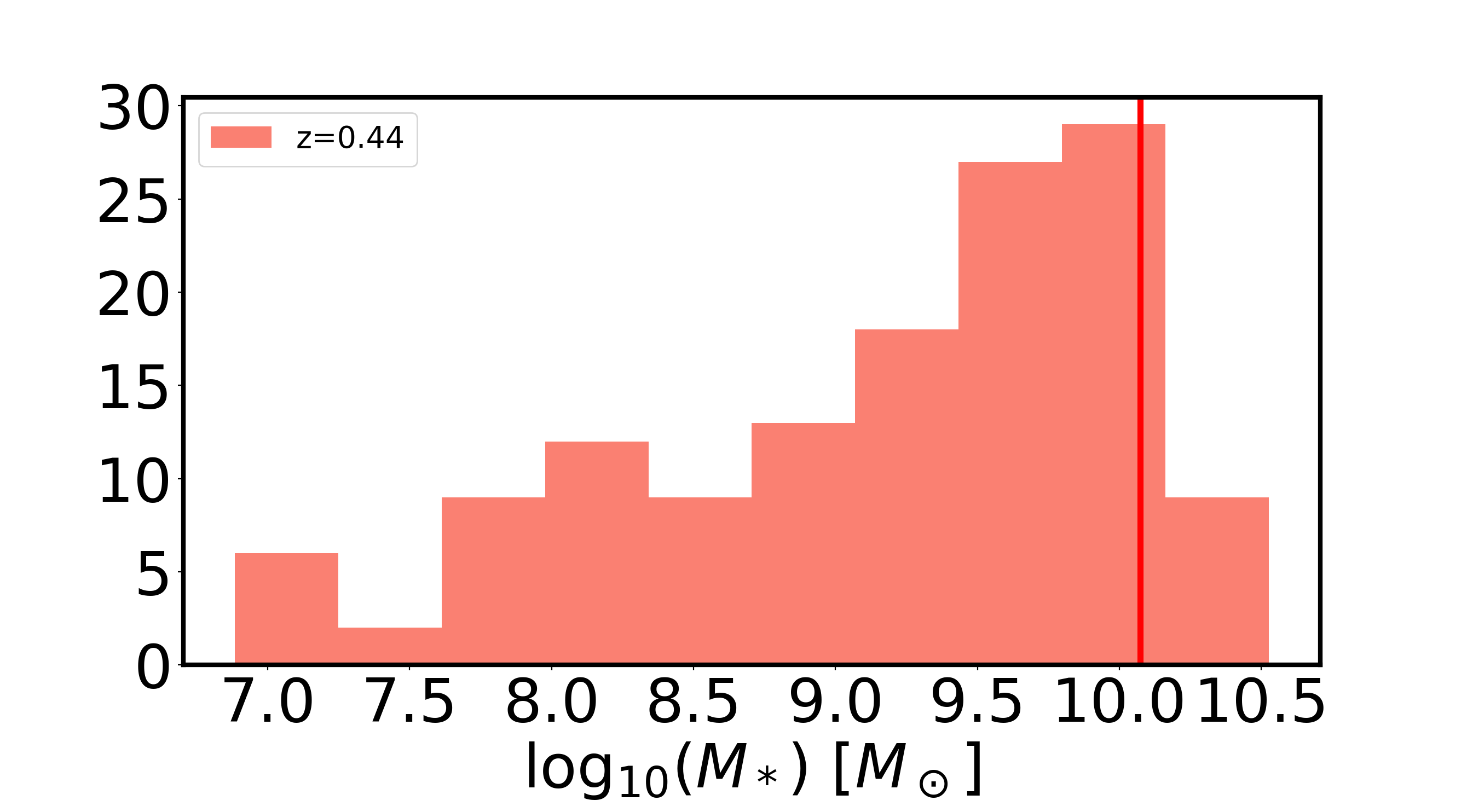}
    \includegraphics[width= 0.32\textwidth]{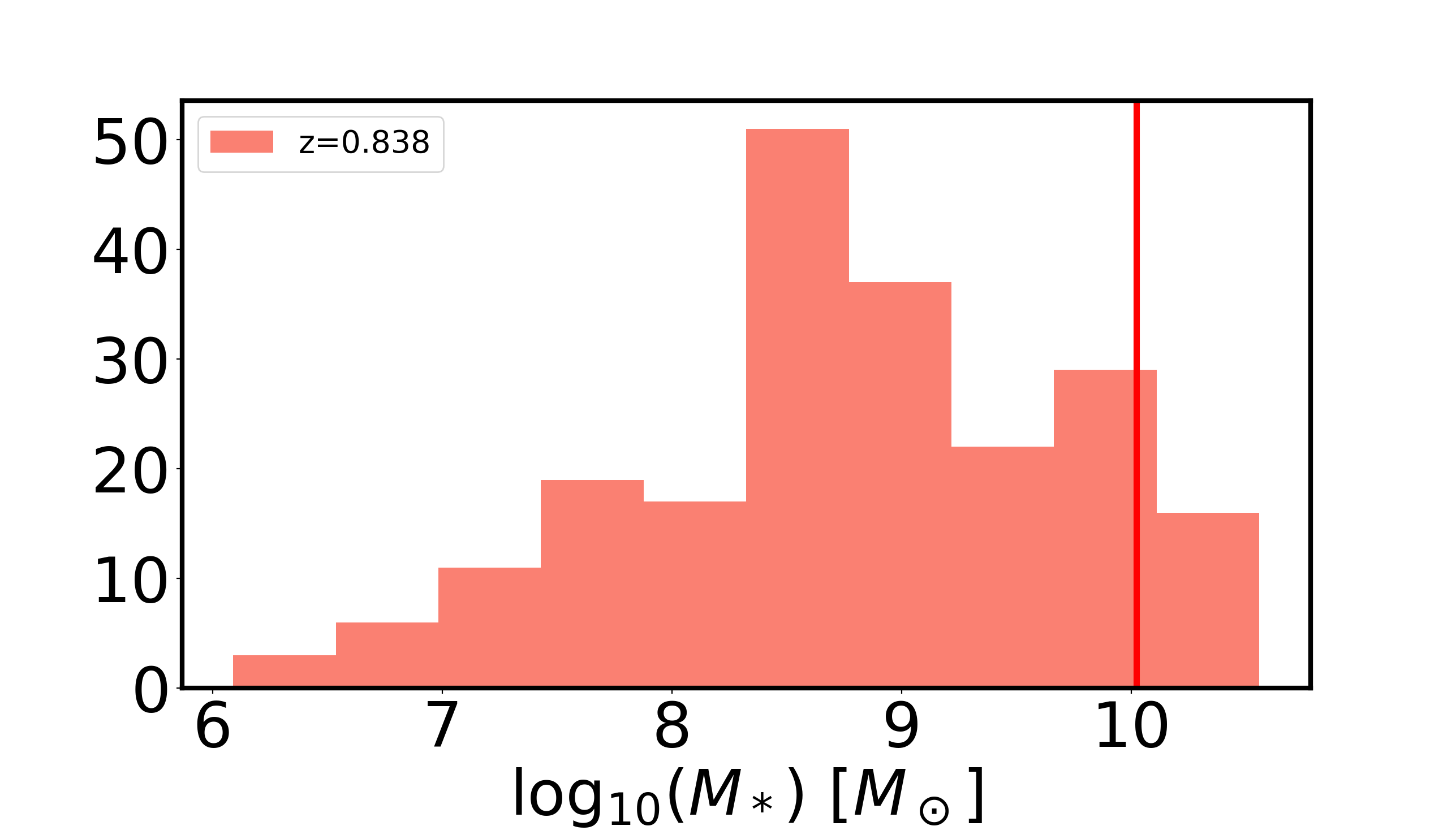}
       \includegraphics[width= 0.32\textwidth]{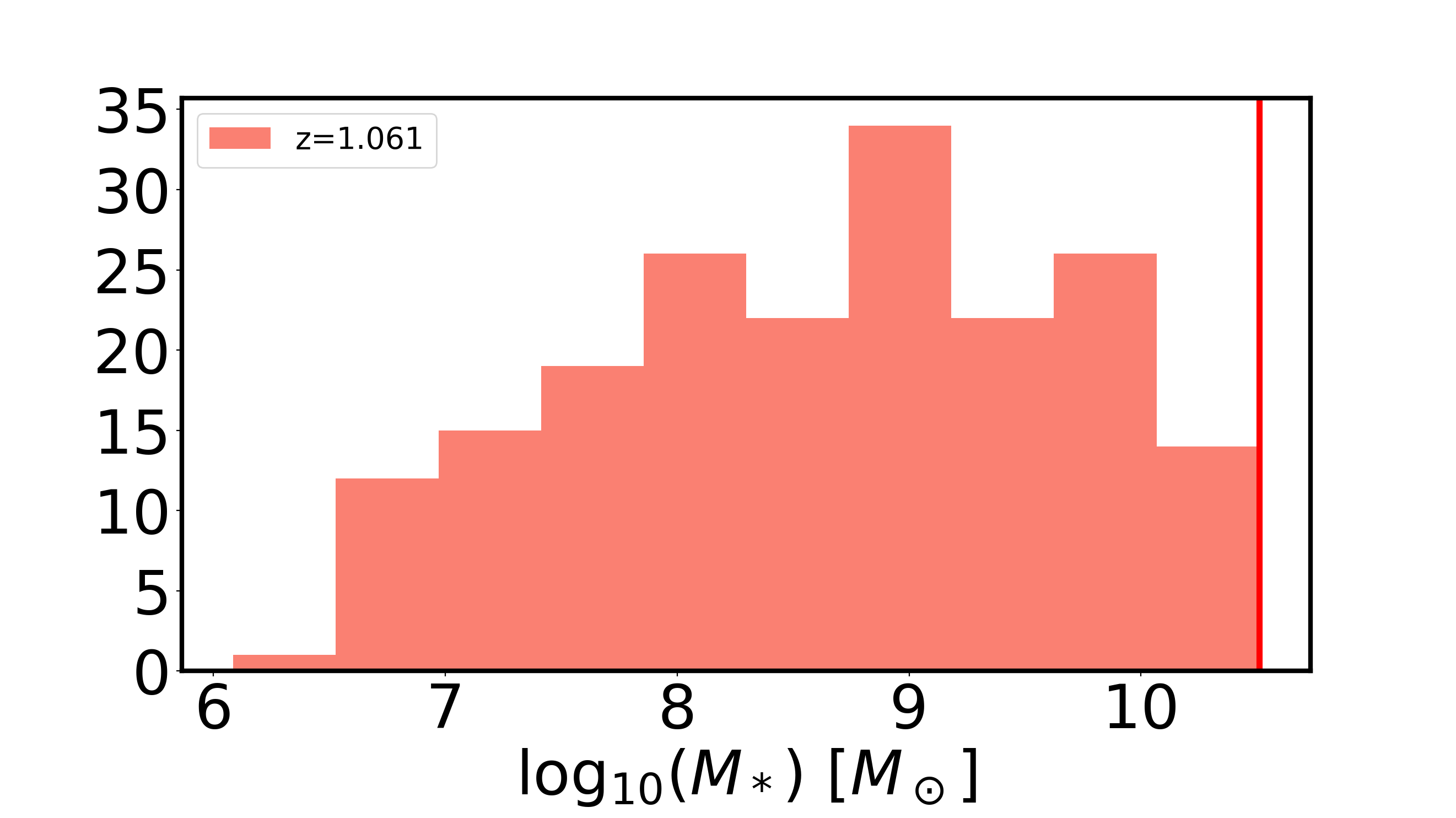}
    \includegraphics[width= 0.32\textwidth]{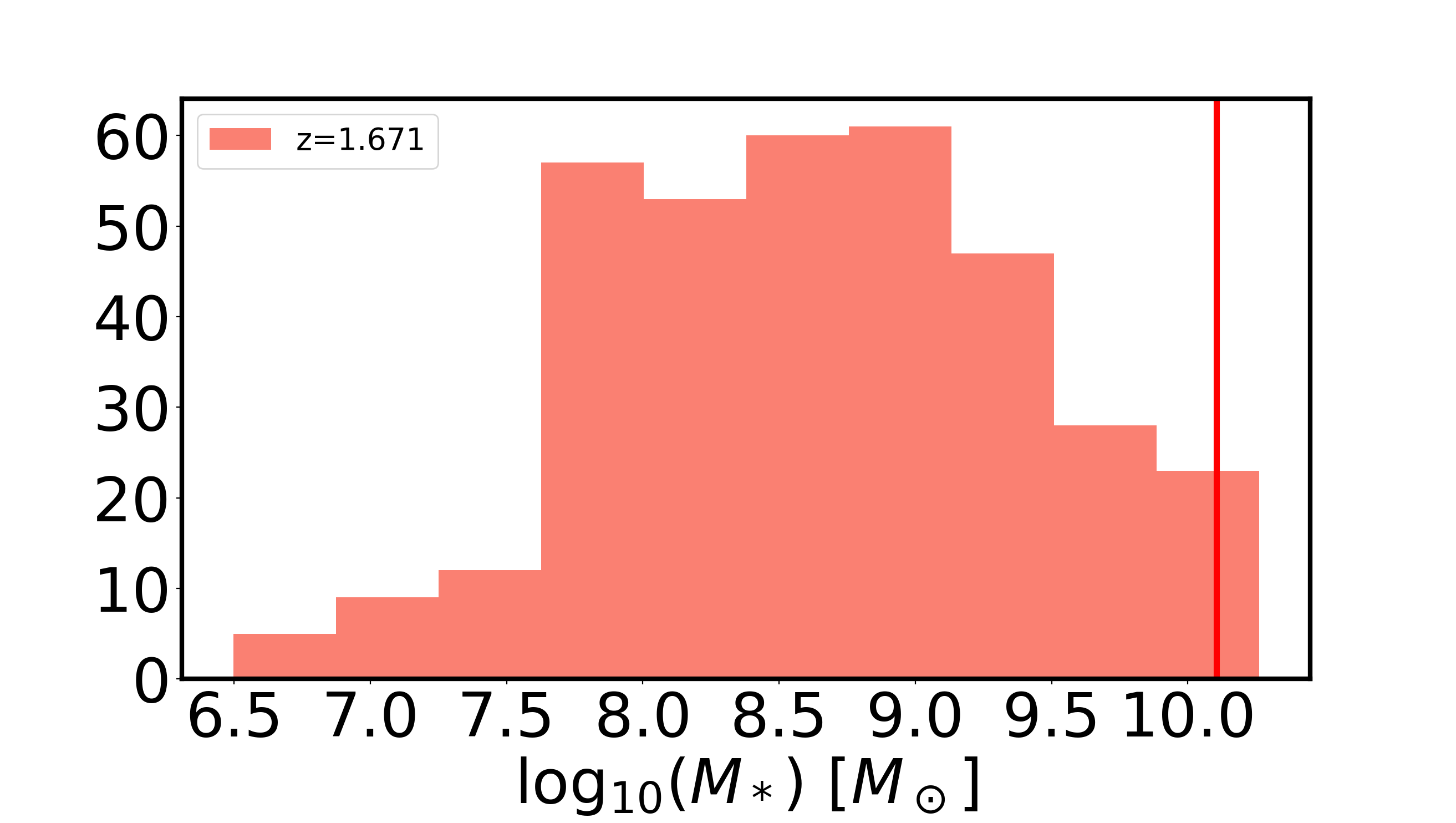}

    \caption{The vertical line in each panel shows the gas density (first and second row), star formation rate (third and fourth row) stellar mass (fifth and sixth row) within 1 kpc of the more massive SMBH in the merging SMBBH pairs identified as nano-Herz GW source (i.e.~with chirp masses $M_c \geq 10^8$ M$_\odot$) at the host galaxy's redshift.   The histogram shows the corresponding quantities around the more massive black holes in all  SMBH pairs (merging and proximate) in the {\sc Romulus25} simulation at the same redshift.}
    \label{fig:mstar-sfr-gas-bbh}
\end{figure*}

\subsection{Local galactic properties}\label{sec:bhp1}

In this section, we focus on the characteristics of the gas and stars 
in the vicinity of the SMBBHs, and specifically,
within a 1 kpc radius around the more massive black hole in SMBBHs, as the most massive BH drives the chirp mass and hence the strength of the signal. We also restrict ourselves to SMBBHs with a chirp mass $M_c \geq 10^8$ M$_\odot$\footnote{
$M_c \geq 10^8$ M$_\odot$ implies that the most massive SMBH in the pair is \emph{at least} $8.7\times 10^7$ M$_\odot$
if both SMBHs are equal mass, or more realistically $> 2.5\times 10^8$ M$_\odot$ since the BH mass ratios are typically $< 0.2$.}  and $z \leq 2$. These SMBBHs account for $\sim$ 90\% of the total SGWB power spectrum (see Fig. \ref{fig:sgwb}). Our analysis of the {\sc Romulus25} simulation reveals six nHz GW sources. {Considering the redshift range in which these sources are detected and our simulation volume, we calculate the number density of nHz GW sources to be $7.7 \times 10^{-6} \rm cMpc^{-3}$. This value closely aligns with the number densities estimated in \citet{mingarelli2017} and \citet{casey2022}  and that derived by \citet{antoniadis2023second} from PTA data release, which are $1.6 \times 10^{-6} \rm cMpc^{-3}$, $6.6 \times 10^{-6} \rm cMpc^{-3}$ and $1.5 \times 10^{-5} \rm cMpc^{-3}$ respectively.}

{In Fig. \ref{fig:sgwb} we have shown the GW background for various different cases
based on the median value of the gas density, star formation rate and stellar mass
within a 1 kpc radius around the more massive black hole in nHz GW SMBBH sources in the {\sc Romulus25} simulation.} {Fig. \ref{fig:BHs-z} shows, from top to bottom, the gas density ($\rho_{\rm gas}$), star formation rate (SFR), and stellar mass ($\rm M_*$), vs redshift, of the individual host galaxies of the nHz GW SMBBH sources.  No significant evolution with redshift is found for the local SFR and the local stellar mass. However, the top panel shows an increasing trend of $\rho_{\rm gas}$ with redshift.} {Nonetheless, examining the properties around black holes solely in nHz GW sources doesn't provide a comprehensive view. Therefore, we compare these to the same properties surrounding {the most massive of the two  black holes in our full set of proximate and merging (cf \S \ref{selection})  pairs of black holes} in the {\sc Romulus25} simulation. 
 {The resulting distributions for gas density, SFR, and stellar mass are shown as salmon histograms in} Fig. \ref{fig:mstar-sfr-gas-bbh}.
 The corresponding quantities for nHz GW sources are denoted by a vertical line. 
{We find that the local $\rho_{\rm gas}$ and SFR around our nHz GW sources are typical.  Specifically, the increasing gas density with redshift around the nHz GW sources simply reflects the fact that all systems have more gas at earlier epochs.}
However, the stellar mass in the vicinity of our nHz GW sources falls in the high mass tail of the corresponding distribution, as illustrated in Fig. \ref{fig:mstar-sfr-gas-bbh}. This indicates that nHz GW sources are more likely to be found in environments with high local stellar mass or equivalently, stellar density.} Though it is important to note that this conclusion is based on only 6 events detectable in a simulation box of (25 cMpc)$^3$. A study from a bigger box simulation will help in better understanding the statistical properties. 

\begin{figure}
    \centering
    \includegraphics[width= 0.5\textwidth]{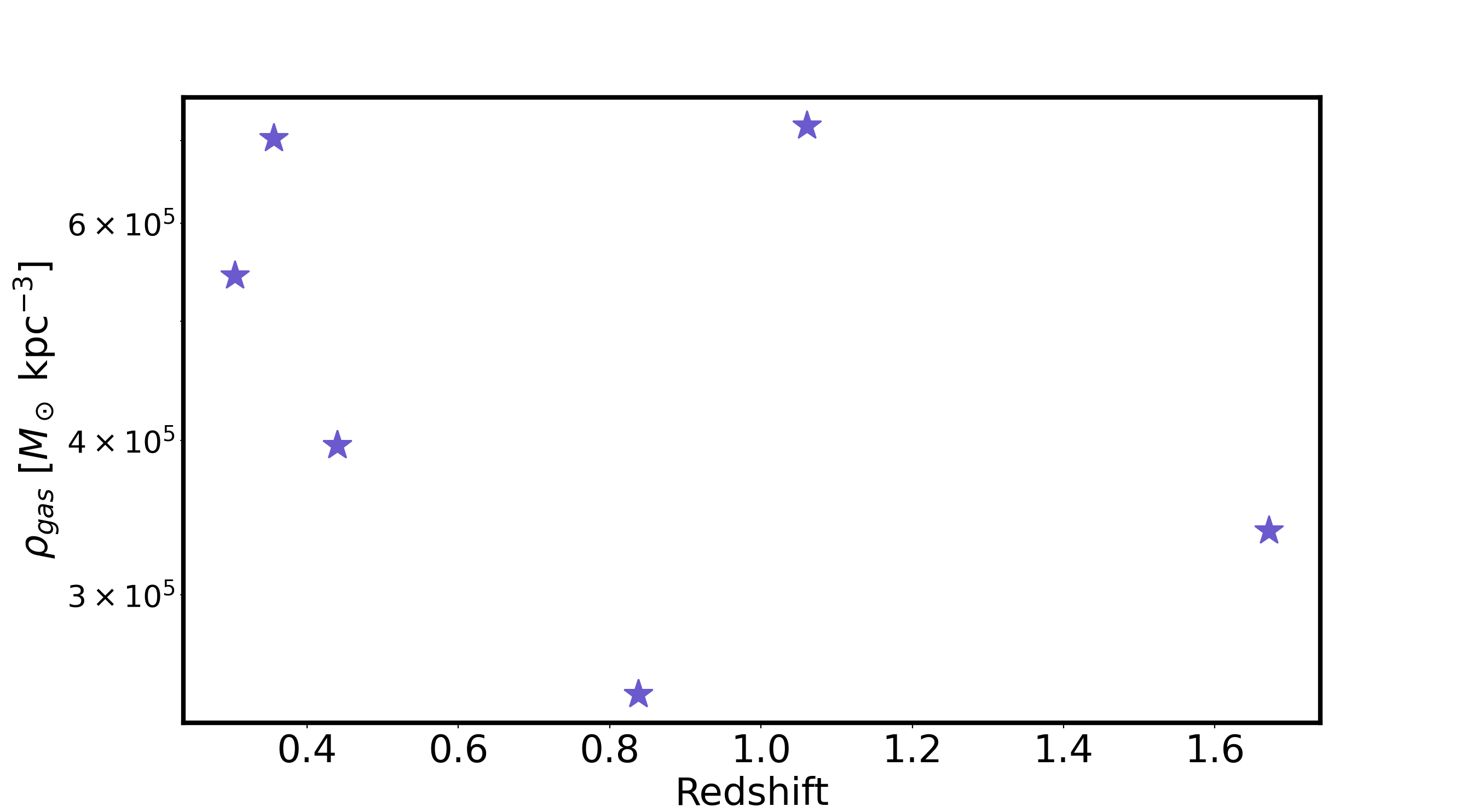}
     \includegraphics[width= 0.5\textwidth]{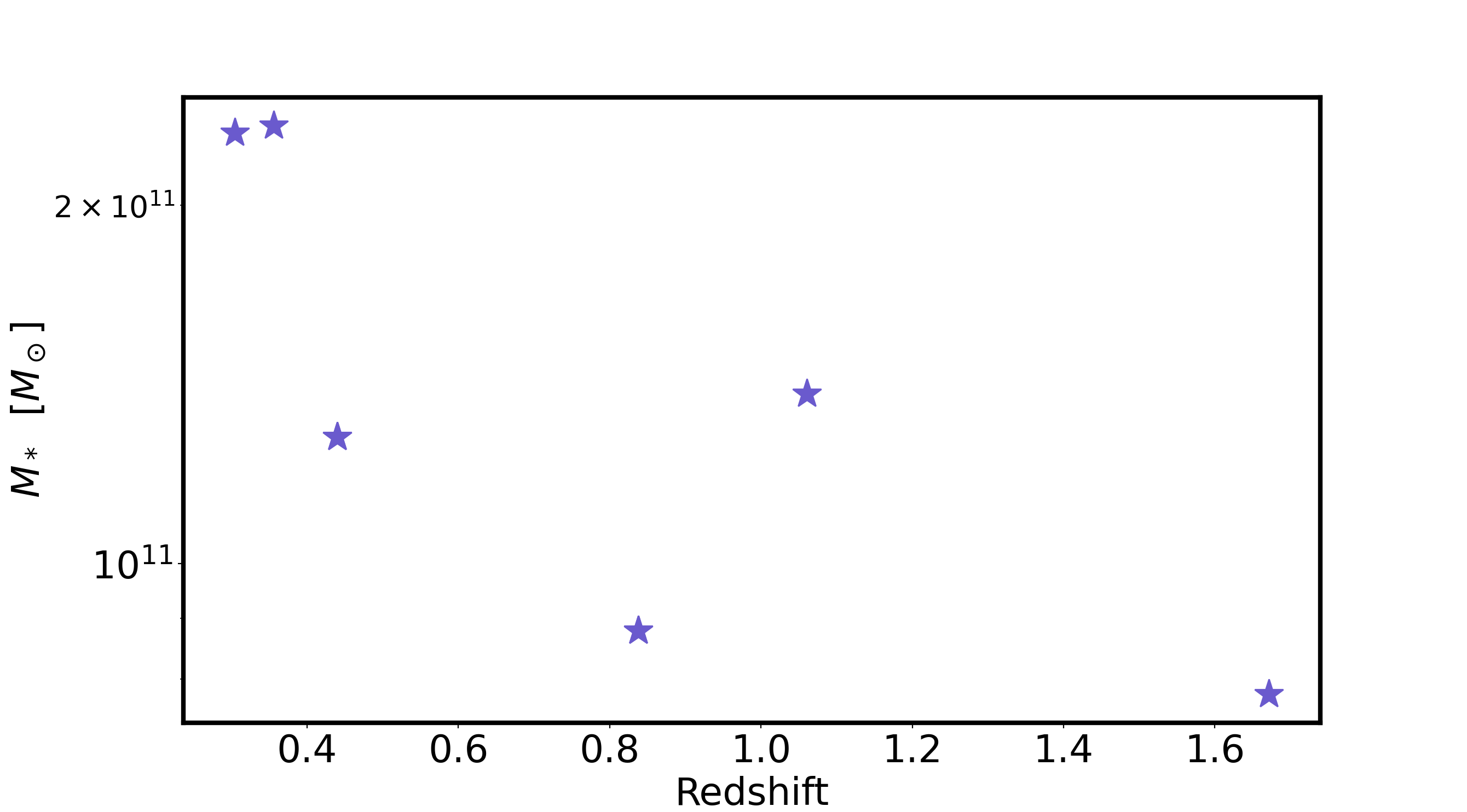}
     \includegraphics[width= 0.5\textwidth]{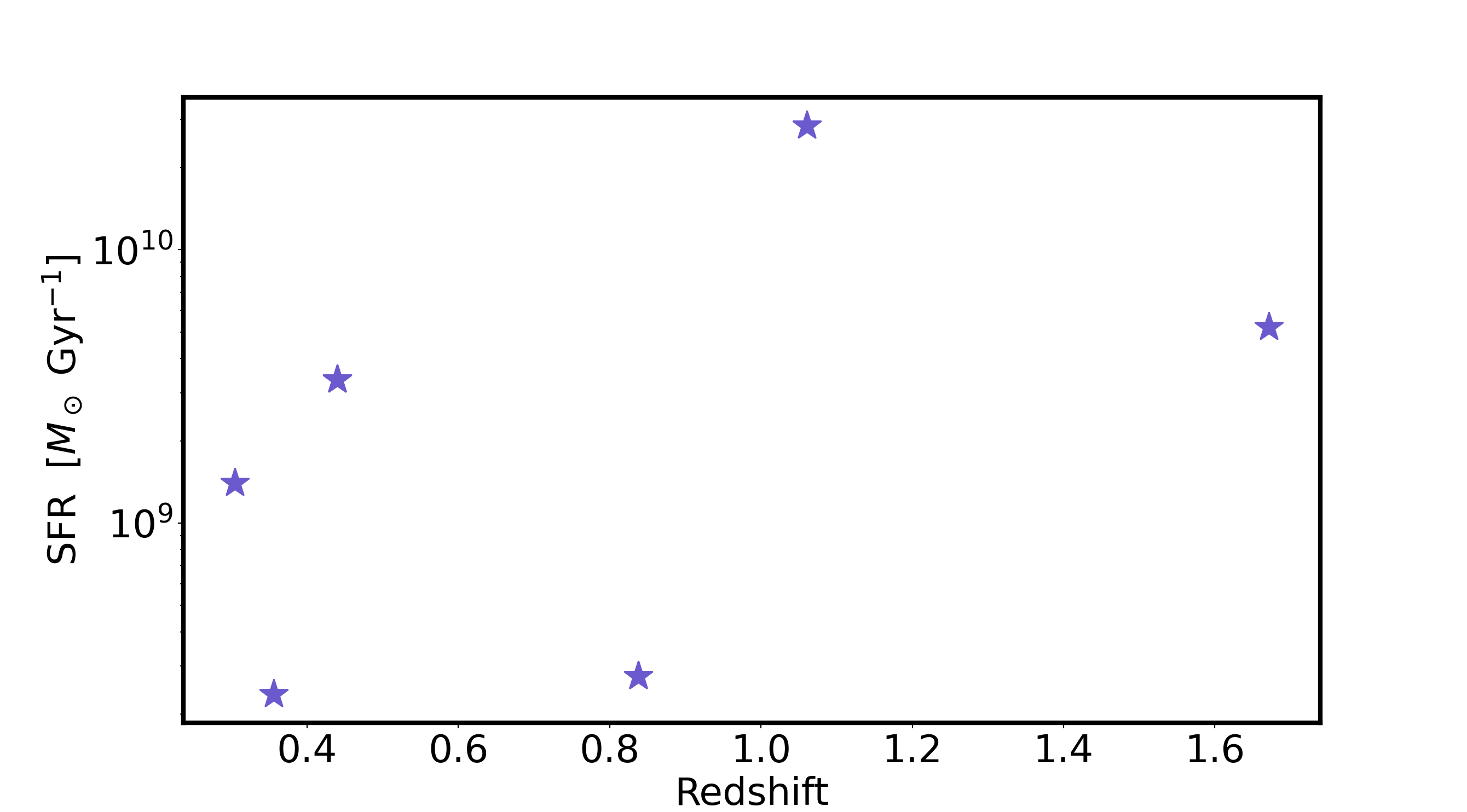}
     \includegraphics[width= 0.5\textwidth]{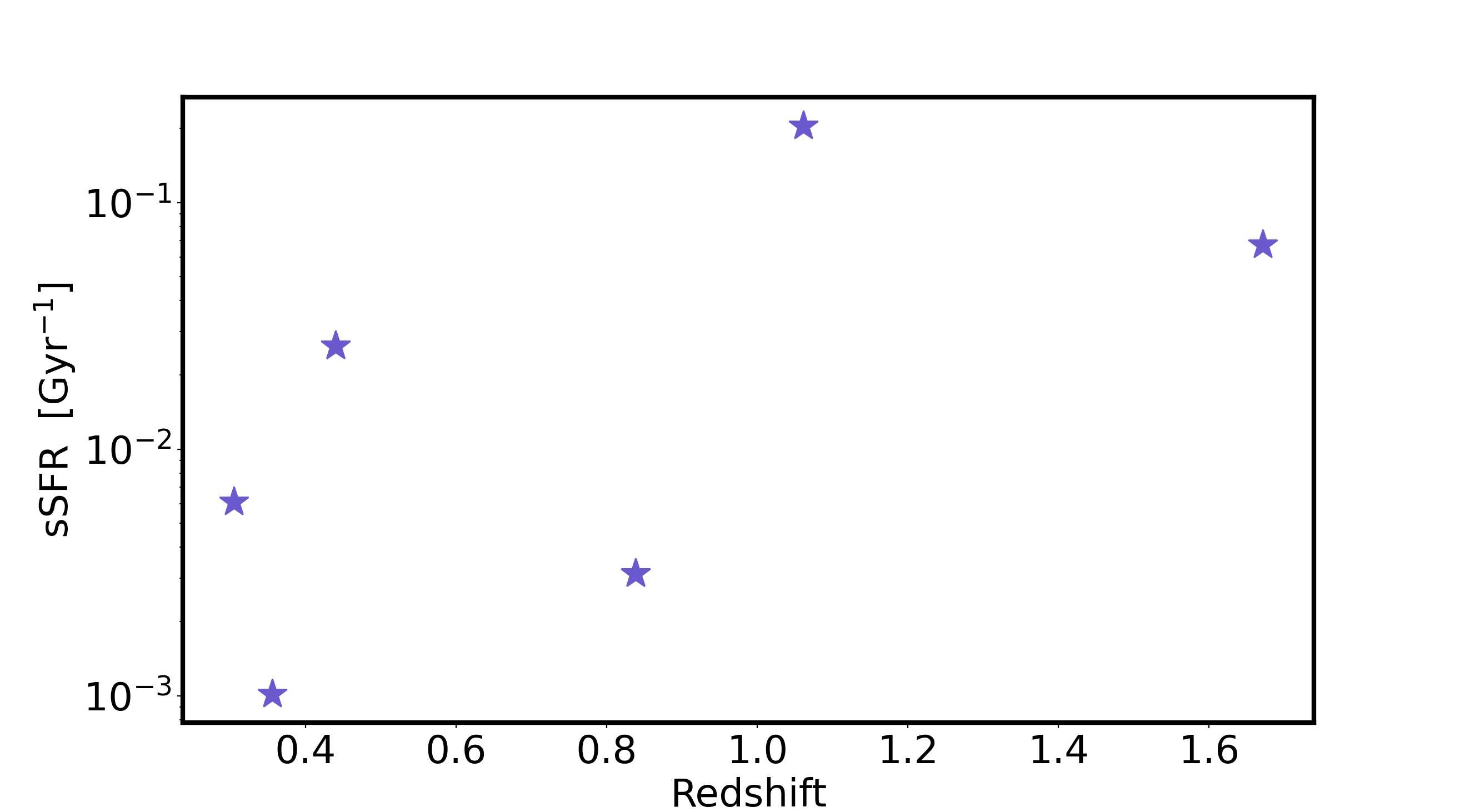}
    \caption{The global galactic properties of the galaxies hosting the merging SMBBH pairs identified as nano-Herz GW sources (i.e.~with chirp masses $M_c \geq 10^8$ M$_\odot$).  The panels show gas density (first panel), stellar mass (second panel), SFR (third panel), and sSFR (fourth panel) as a function of redshift.  All quantities are calculated inside a sphere with a 25 kpc radius around the galaxy center.}
    \label{fig:halo-z}
\end{figure}

\subsection{{Global galactic and host halo properties}}\label{sec:bhp2}

\begin{figure}
    \centering
    \includegraphics[width= 0.45\textwidth]{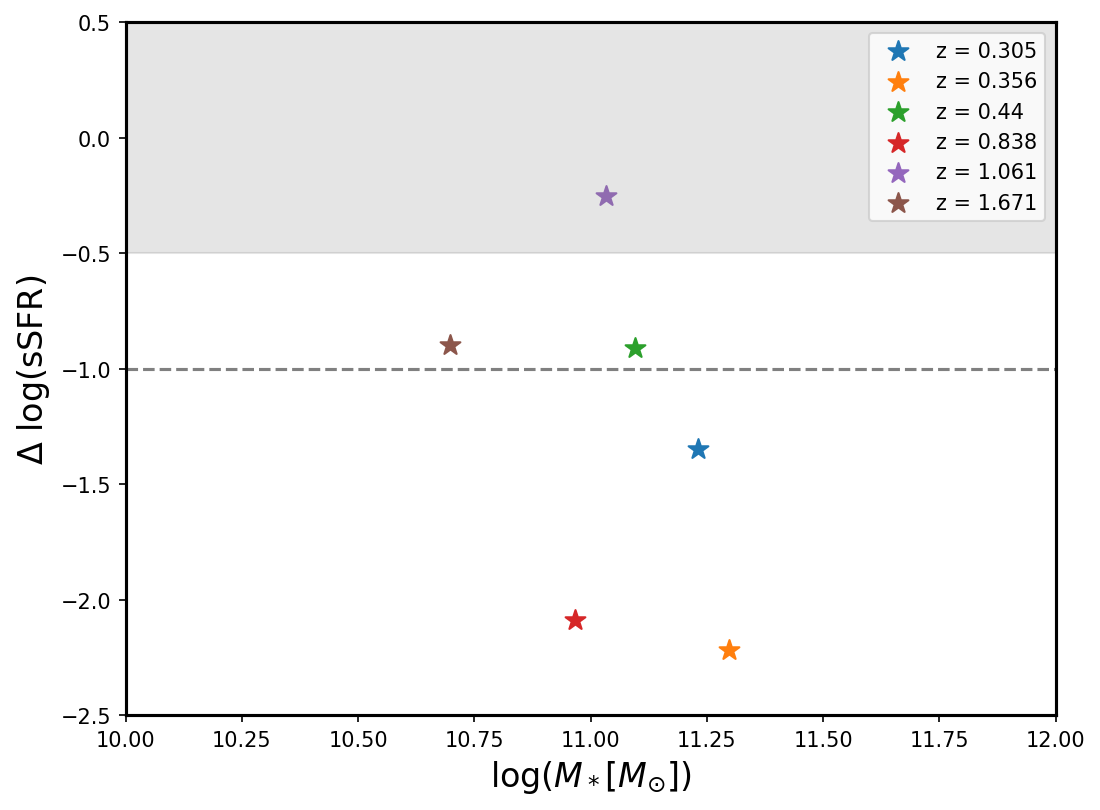}
    \caption{Categorising host galaxies of SMBBHs with $M_c > 10^8 M_\odot$ as star-forming or quenched following definition of \citet{genel2018sfquench}. Here we show $\Delta \rm log(sSFR) = log(sSFR_{galaxy}) - log(sSFR_{ridge})$ for these host galaxies. The dashed line, which marks 1 dex below the ridge for each redshift, serves as the quenched threshold. The shaded region indicates the ``main sequence". See text for detailed definitions.}
    \label{fig:sfquenched}
\end{figure}

\begin{figure*}
    \centering
    \includegraphics[width= 1.\textwidth]{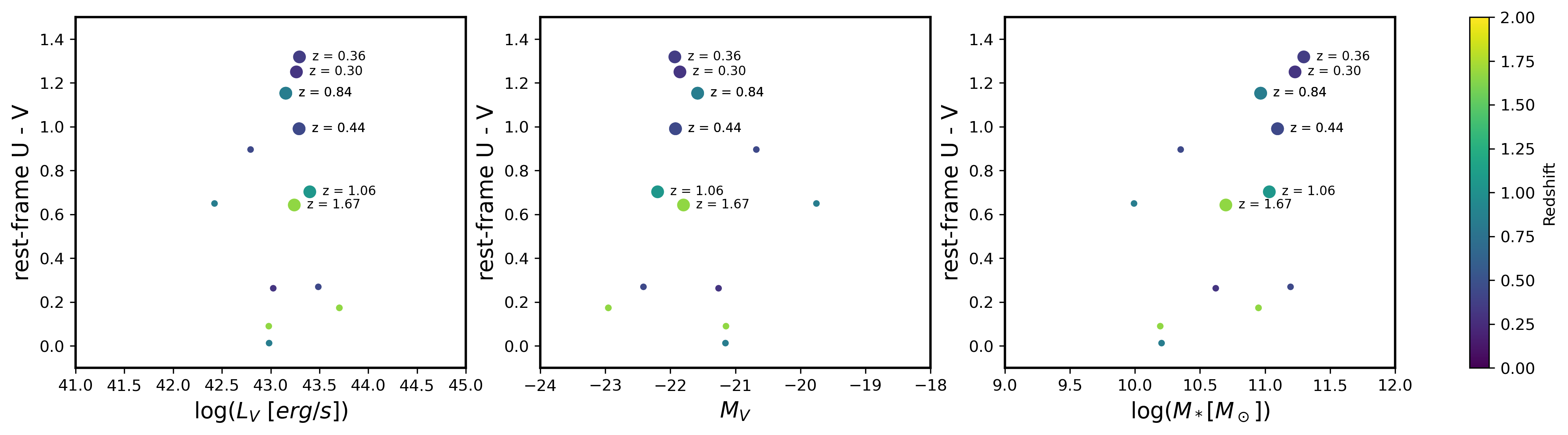}
    \caption{The rest-frame U-V vs rest-frame V-band luminosity ($L_V$) (first panel), vs rest-frame absolute magnitude ($M_V$) (second panel), and vs stellar mass $M_*$ 
    (last panel) of the host galaxies of the SMBHs for chirp mass $M_c \geq 10^8$ M$_\odot$ (represented by big circles) and $10^7$ M$_\odot$ $\leq M_c \leq 10^8$ M$_\odot$ (represented by small circles). The data points are color-coded as a function of redshift {, and the annotations show the exact redshift they are detected.}}
    \label{fig:cmd}
\end{figure*}

\begin{figure*}
    \centering
    \includegraphics[width= 1.\textwidth]{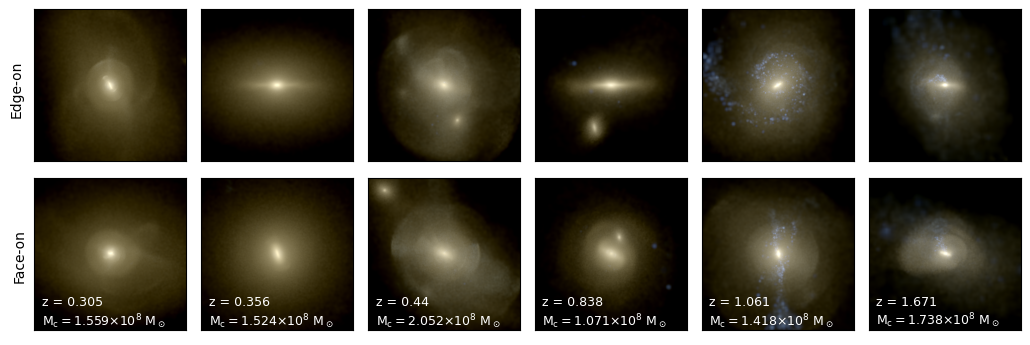}
    \caption{Multi-band composite image of the host galaxies for SMBBHs with chirp mass $M_c\geq 10^8$ M$_\odot$ is shown with edge-on (top) and face-on (bottom) views at redshifts they are detected at.}
    \label{fig:morpho}
\end{figure*}

\begin{figure*}
    \centering
    \includegraphics[width= 1.\textwidth]{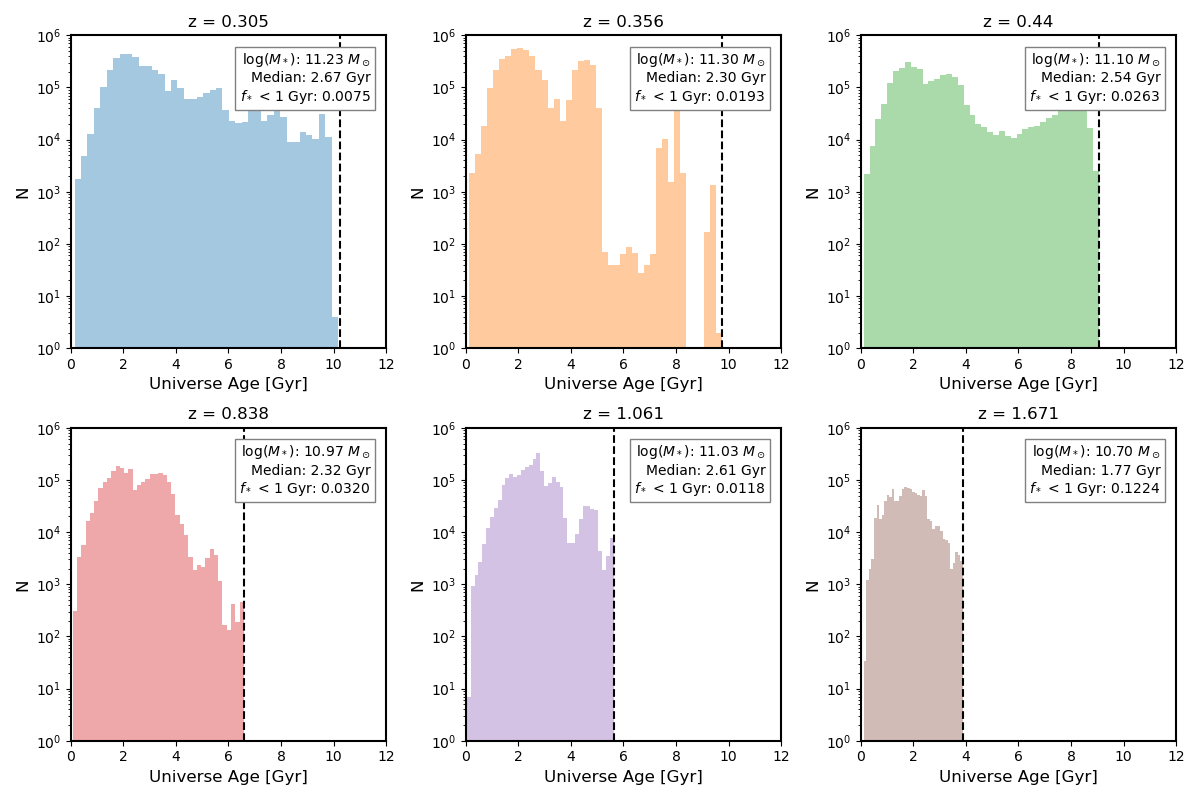}
    \caption{The age of the Universe when stars in the host galaxies of SMBBHs with a chirp mass $M_c\geq 10^8$ M$_\odot$ formed. The dashed line represents the age of the Universe at the time the SMBBHs are detected. The values in the top right box, from top to bottom, indicate the stellar mass within 25 kpc from the galaxy center, the median age of the stars, and the fraction of the stellar mass observed that is aged $\leq 1$ Gyr.}
    \label{startsform}
\end{figure*}

\begin{figure}
    \centering
    \includegraphics[width= 0.45\textwidth]{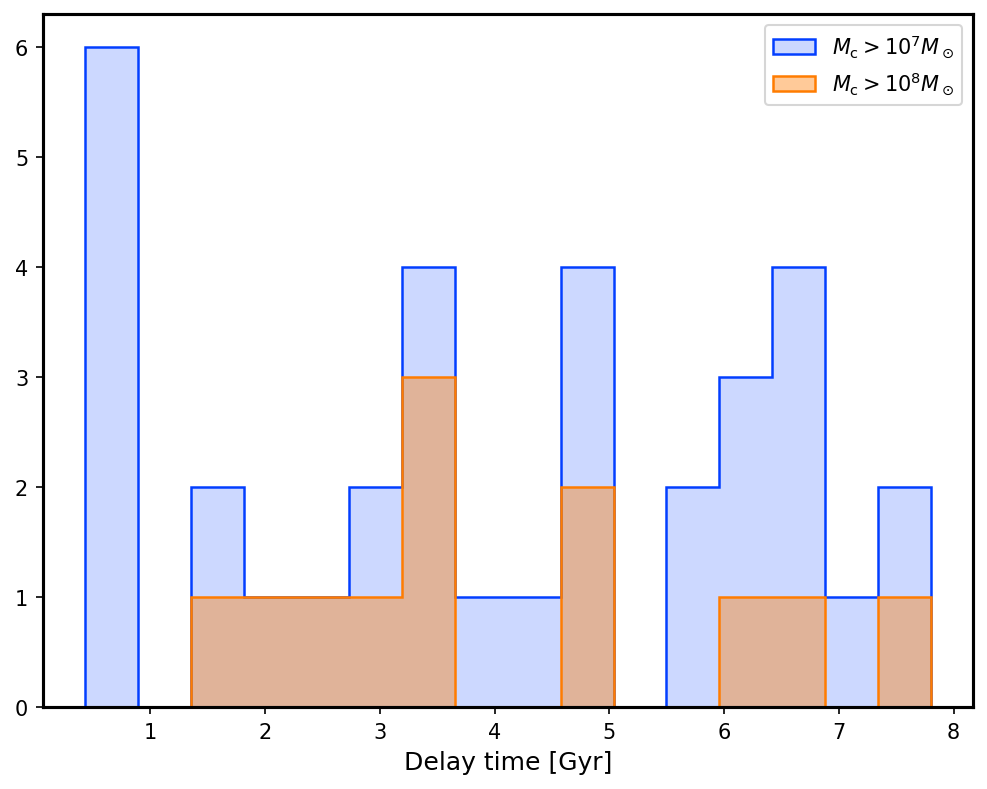}
    \caption{The amount of time taken by the SMBHs to grow the last $50\%$ of its mass it possesses at the time of the merger is shown for sources with chirp mass $M_{\rm c}\geq 10^7$ M$_{\odot}$ (in blue) and $M_{\rm c}\geq 10^8$ M$_{\odot}$ (in orange). The distribution indicates that the SMBHs with chirp mass $M_{\rm c}\geq 10^8$ M$_{\odot}$ need at least $10\%$ of the age of the Universe to grow, indicating these objects are likely to be host in old galaxies.}
    \label{fig:delaytime}
\end{figure}

\begin{figure*}
    \centering
    \includegraphics[width= \textwidth]{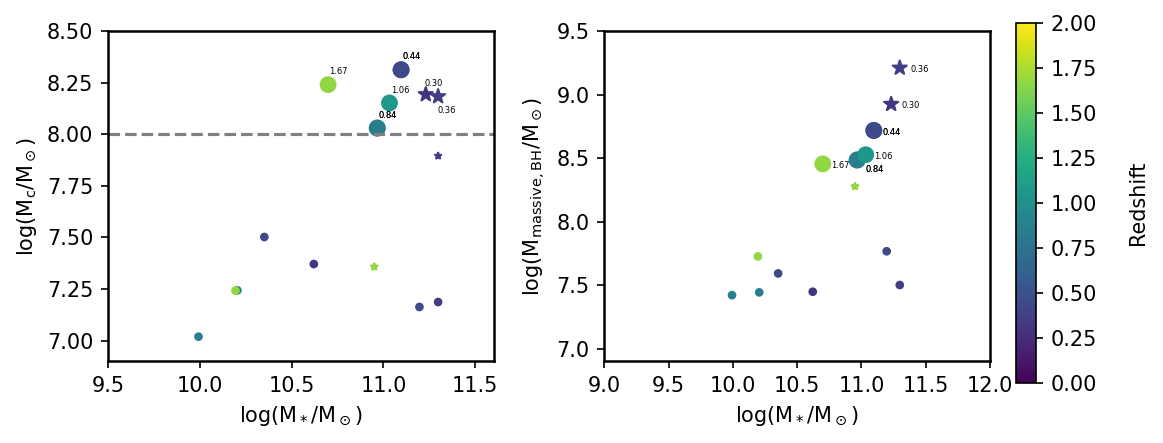}
    \caption{Left panel: The chirp mass of SMBBHs plotted against the stellar mass of their host galaxy. Right panel: Mass of the more massive black hole in SMBBHs versus the stellar mass of their host galaxy. SMBBHs with a mass ratio $q<0.1$ are represented by stars, while those with  $q>0.1$ are represented by circles. Large symbols correspond to SMBBHs with chirp mass  $M_c \geq 10^8$ M$_\odot$ and small symbols denote those with $10^7$ M$_\odot$ $\leq M_c \leq 10^8$ M$_\odot$. The data points are colored according to redshift {, and the annotations show the exact redshift they are detected at.}}
    \label{fig:chirpmass}
\end{figure*}

\begin{figure*}
    \centering
    \includegraphics[width= 0.32\textwidth]{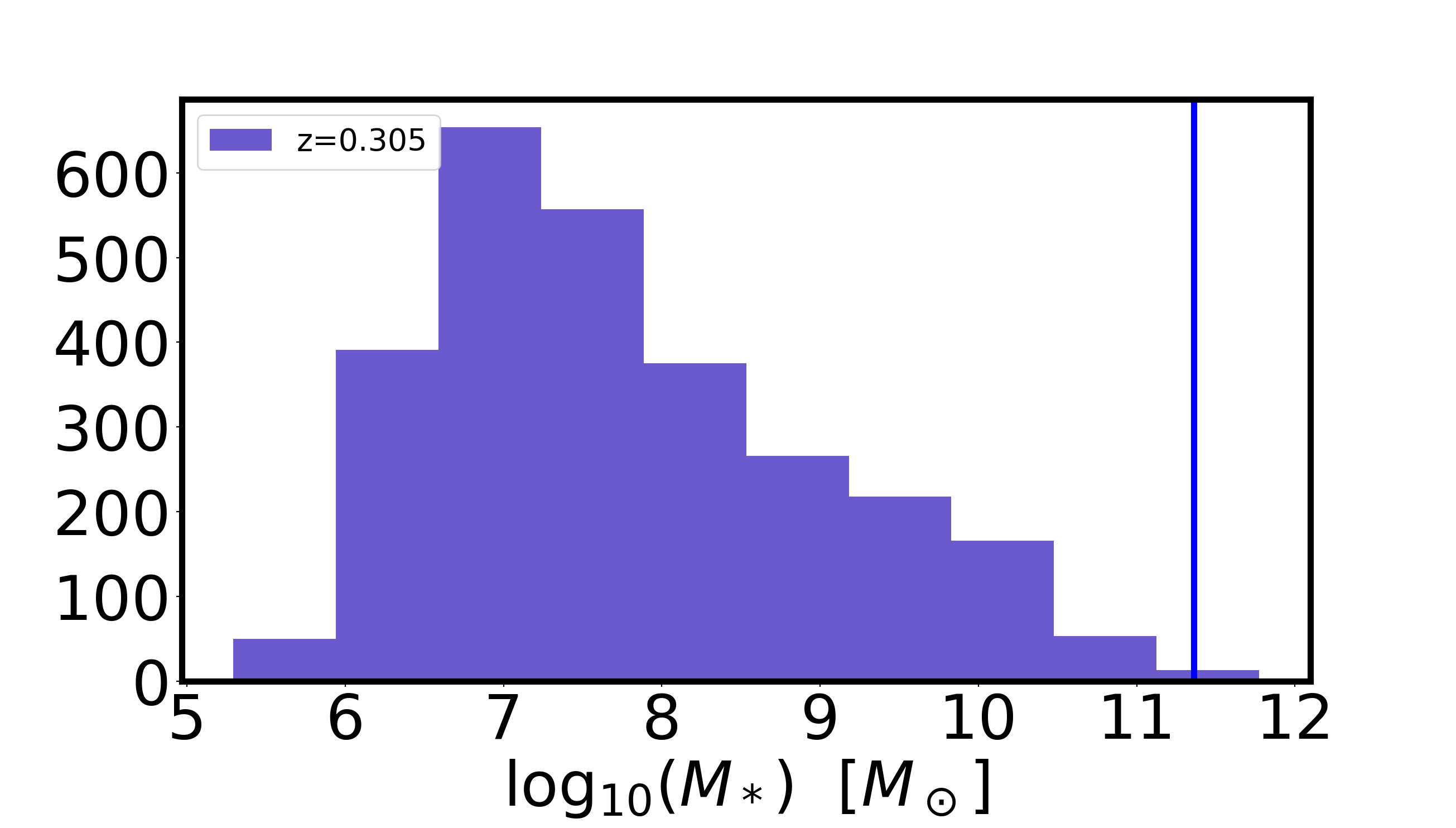}
    \includegraphics[width= 0.32\textwidth]{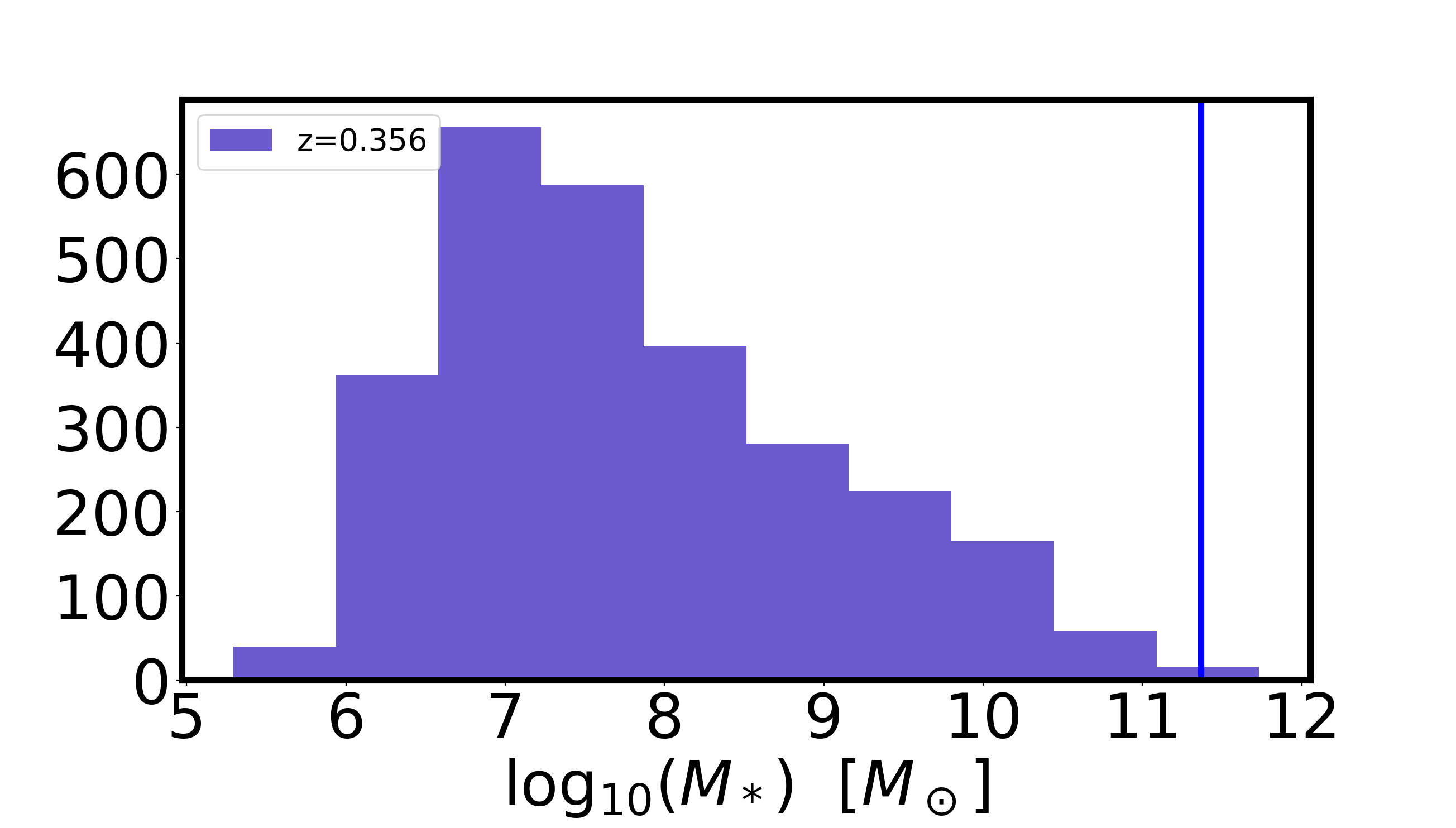}
    \includegraphics[width= 0.32\textwidth]{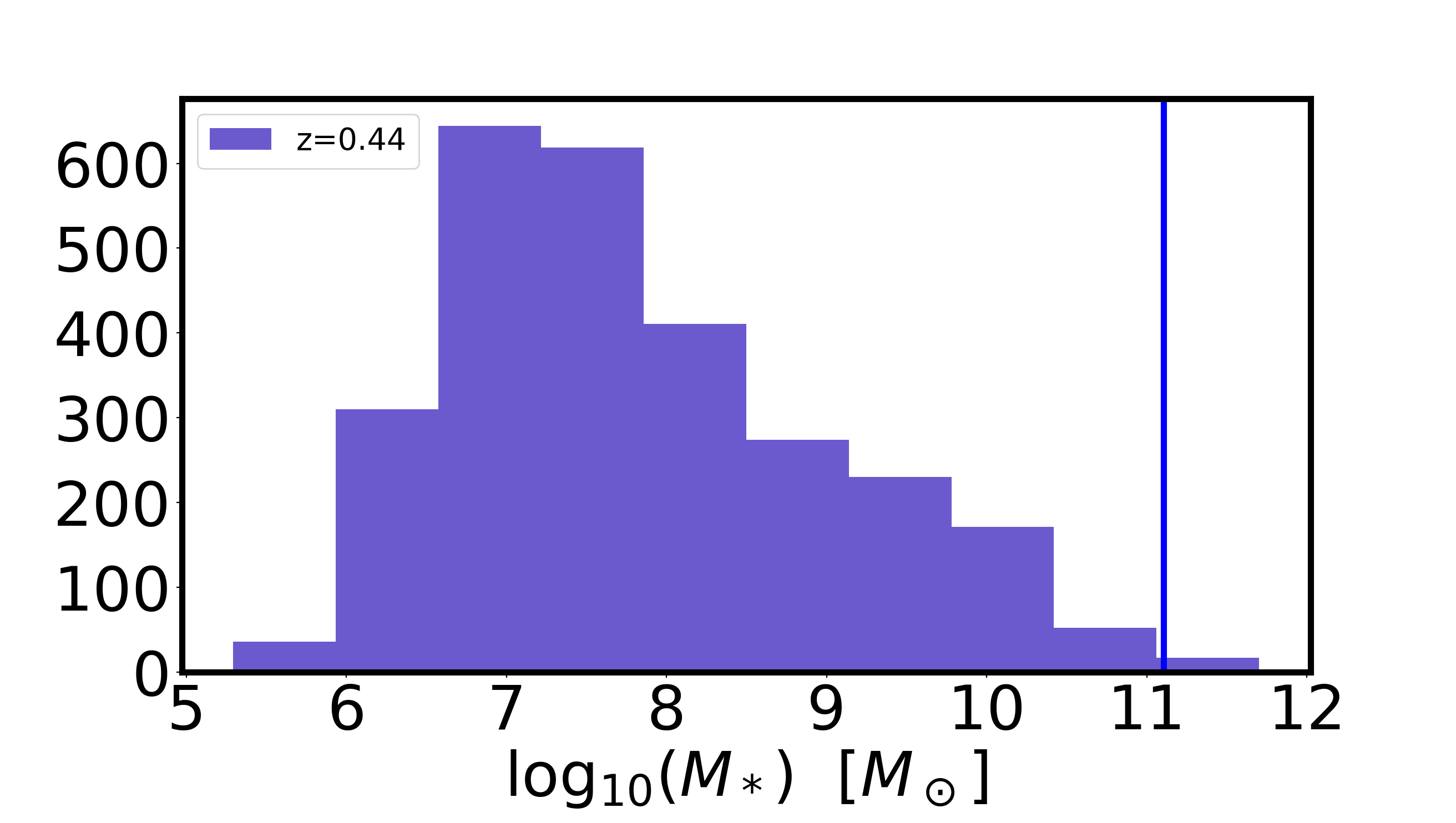}
    \includegraphics[width= 0.32\textwidth]{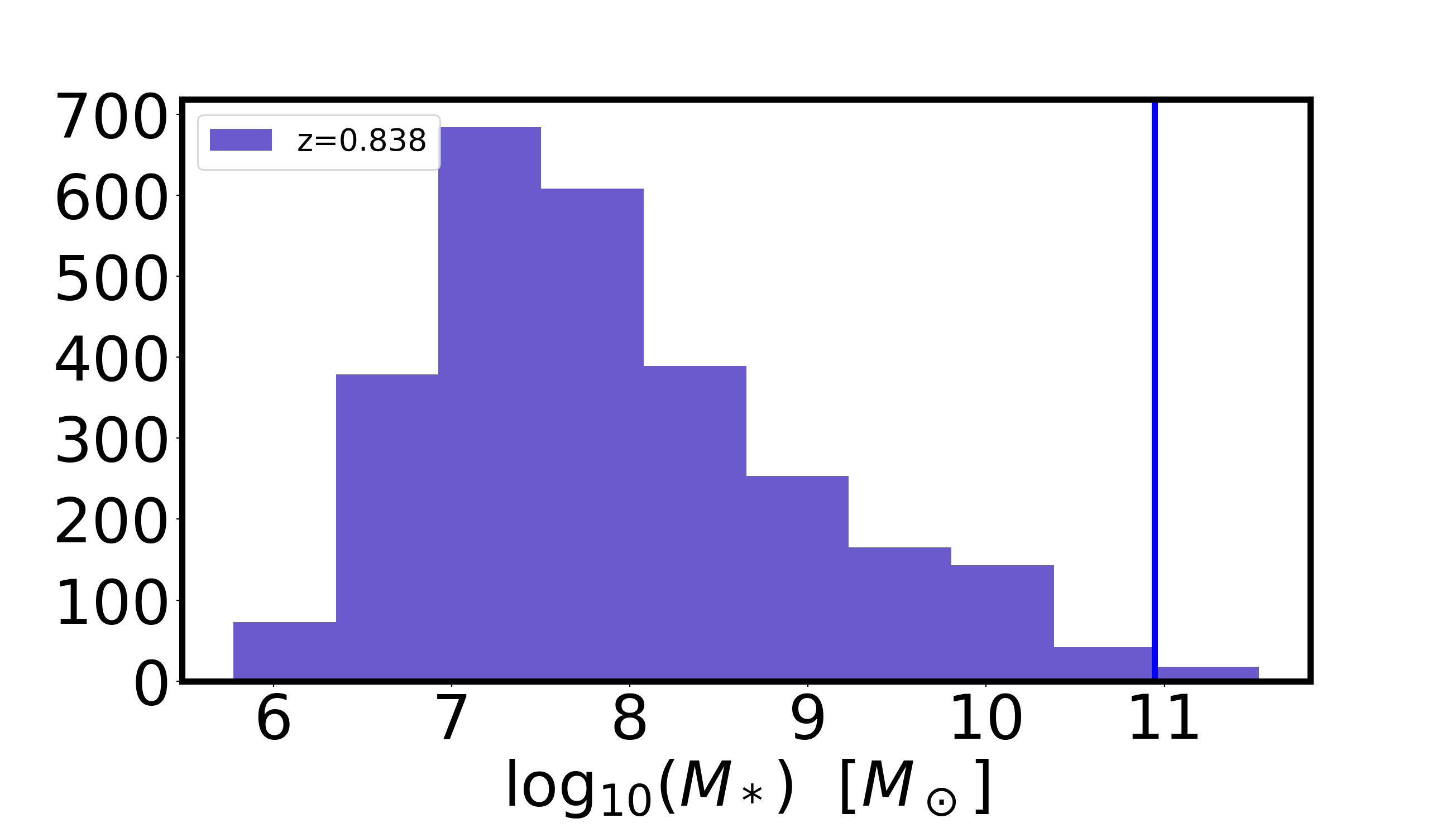}
    \includegraphics[width= 0.32\textwidth]{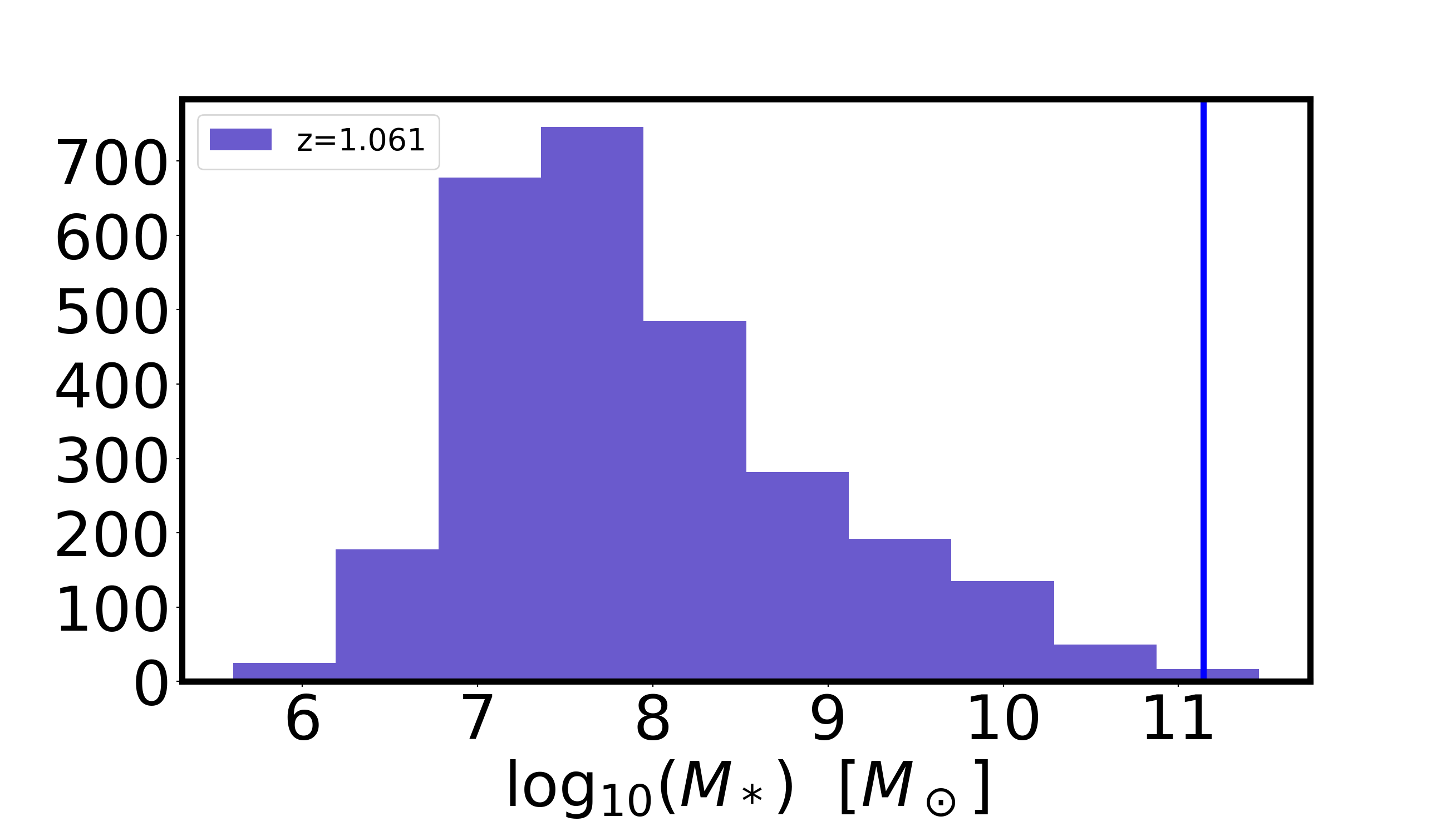}
    \includegraphics[width= 0.32\textwidth]{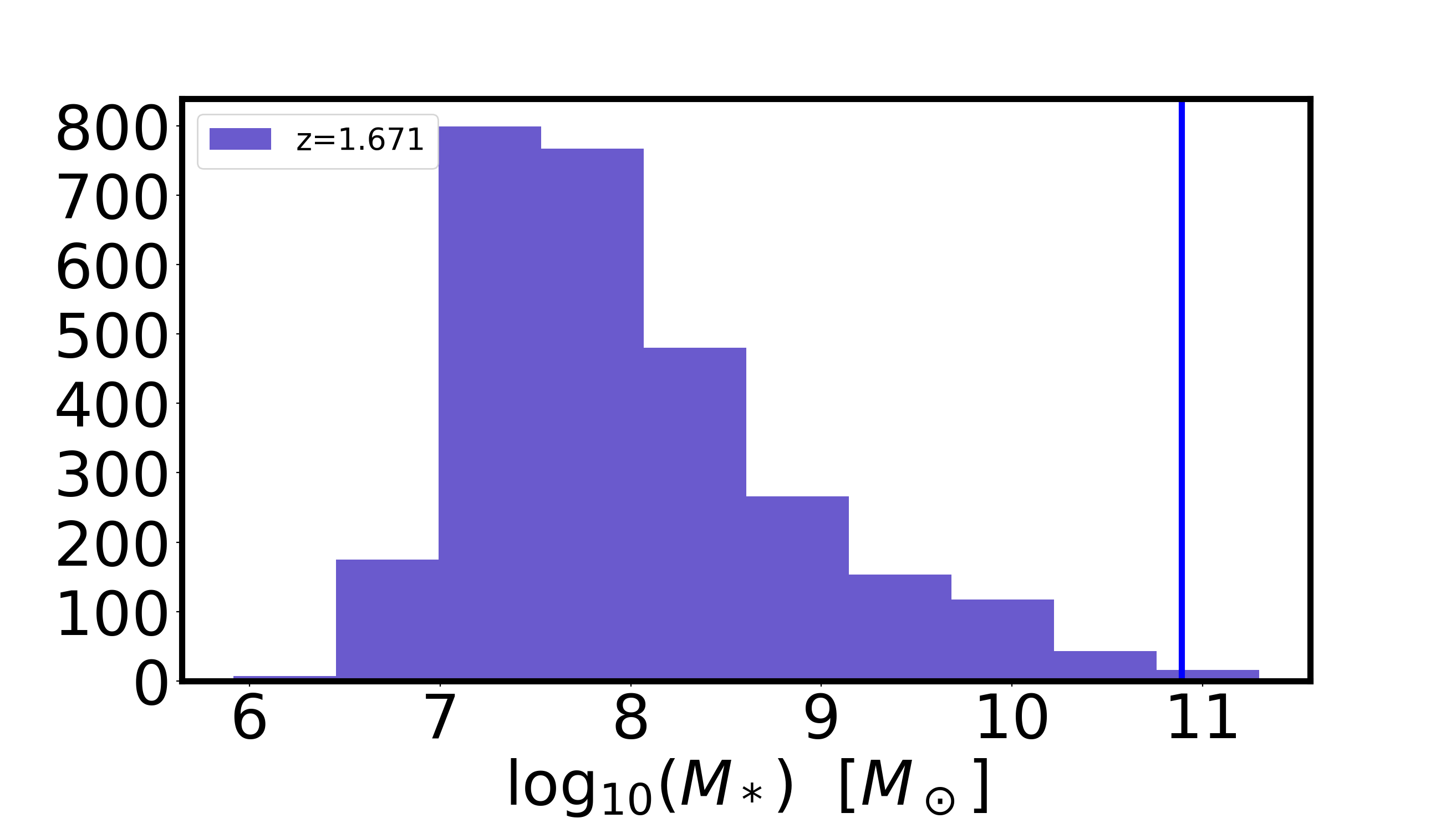}
    
    \hrule
    
     \includegraphics[width= 0.32\textwidth]{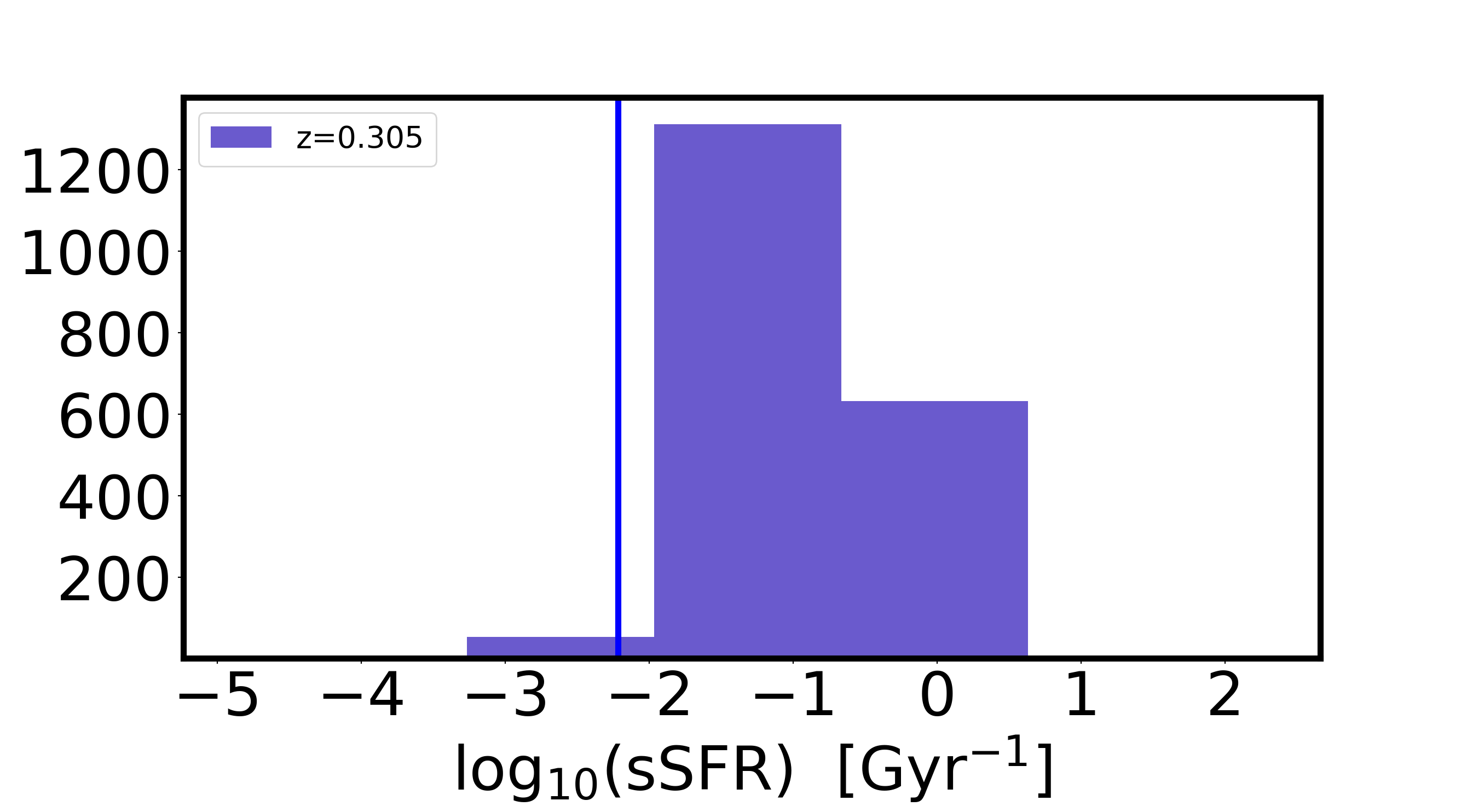}
     \includegraphics[width= 0.32\textwidth]{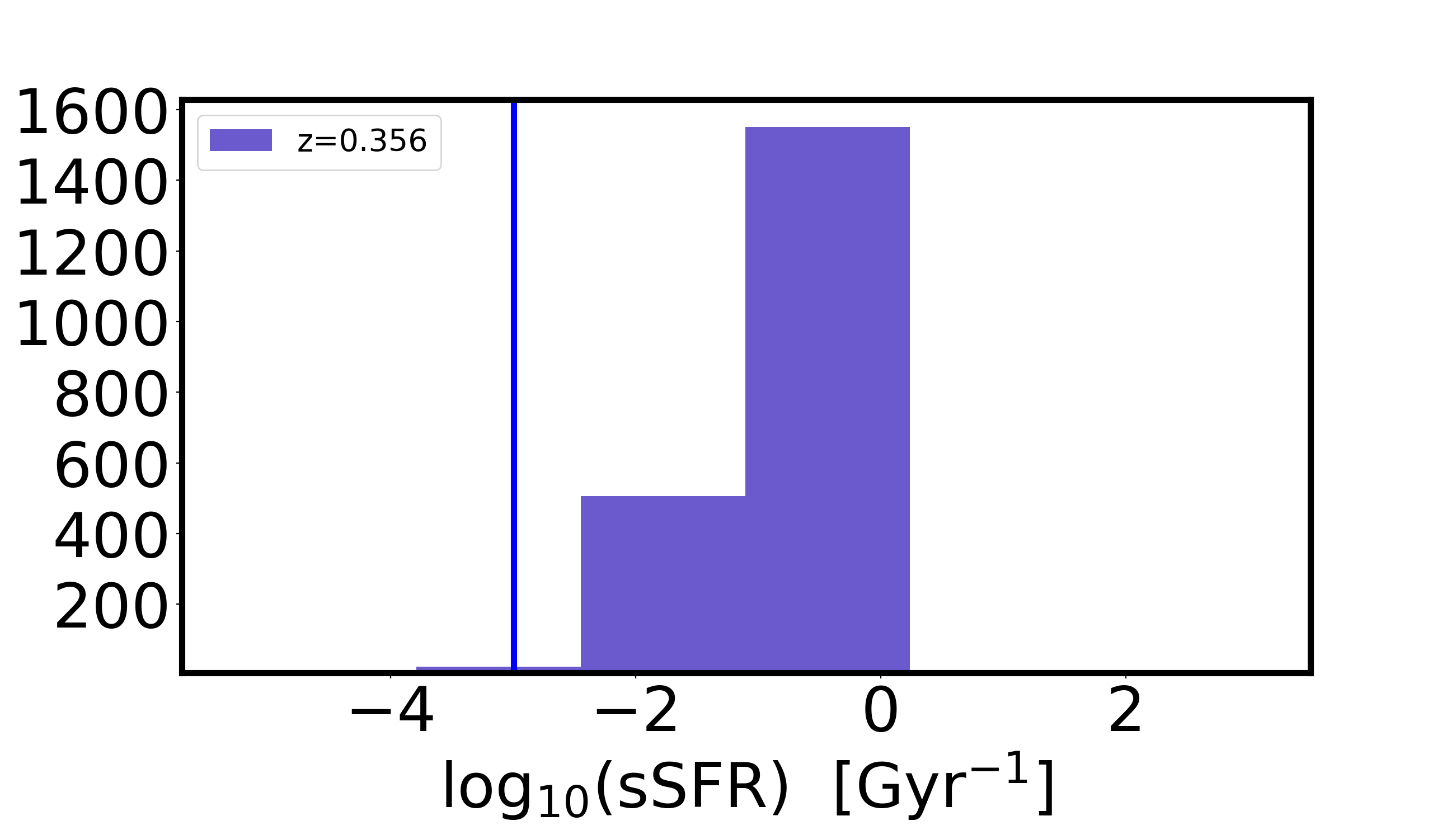}
     \includegraphics[width= 0.32\textwidth]{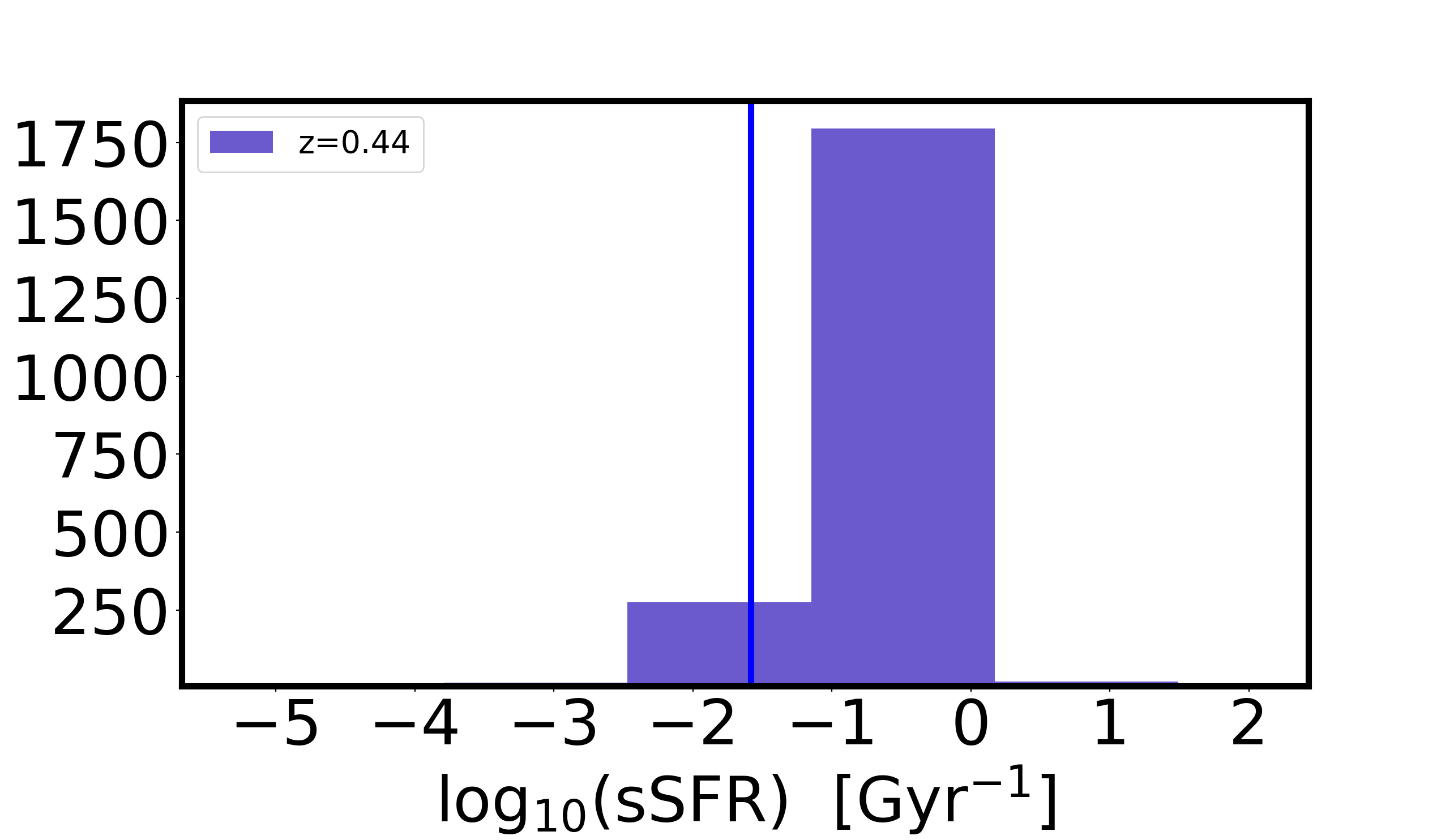}
     \includegraphics[width= 0.32\textwidth]{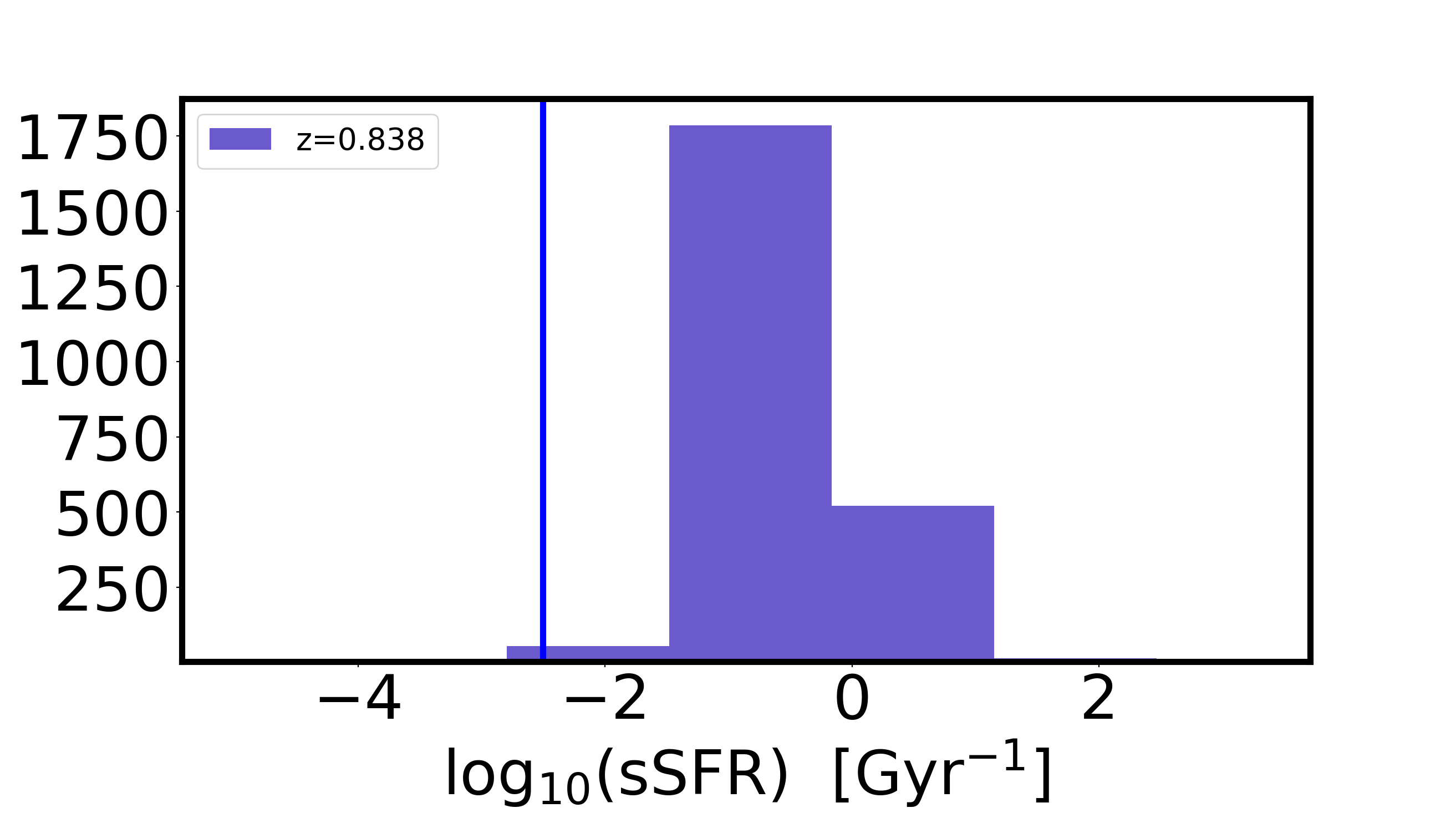}
     \includegraphics[width= 0.32\textwidth]{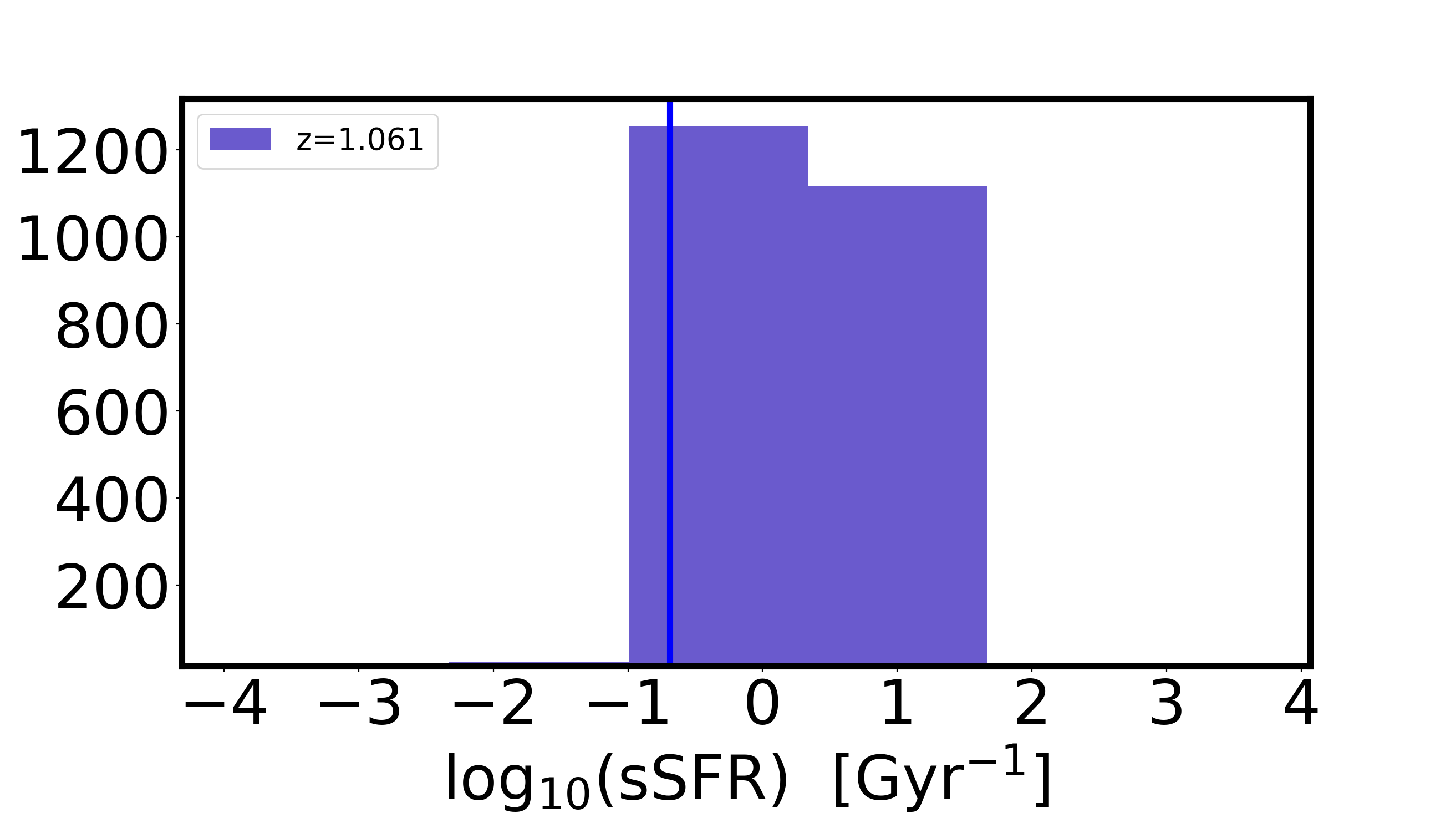}
     \includegraphics[width= 0.32\textwidth]{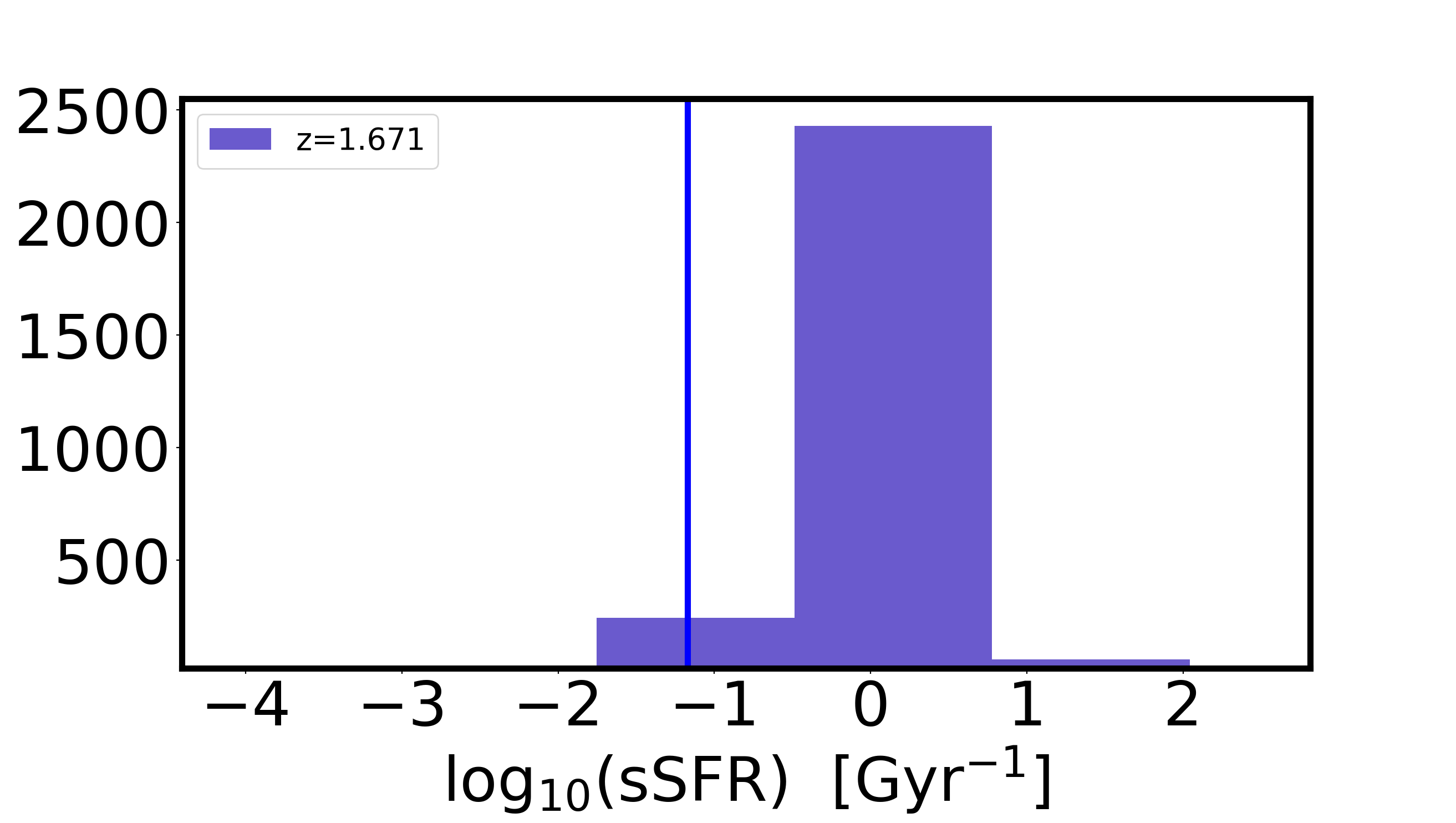}
     
     \hrule
     
    \includegraphics[width=0.32\textwidth]{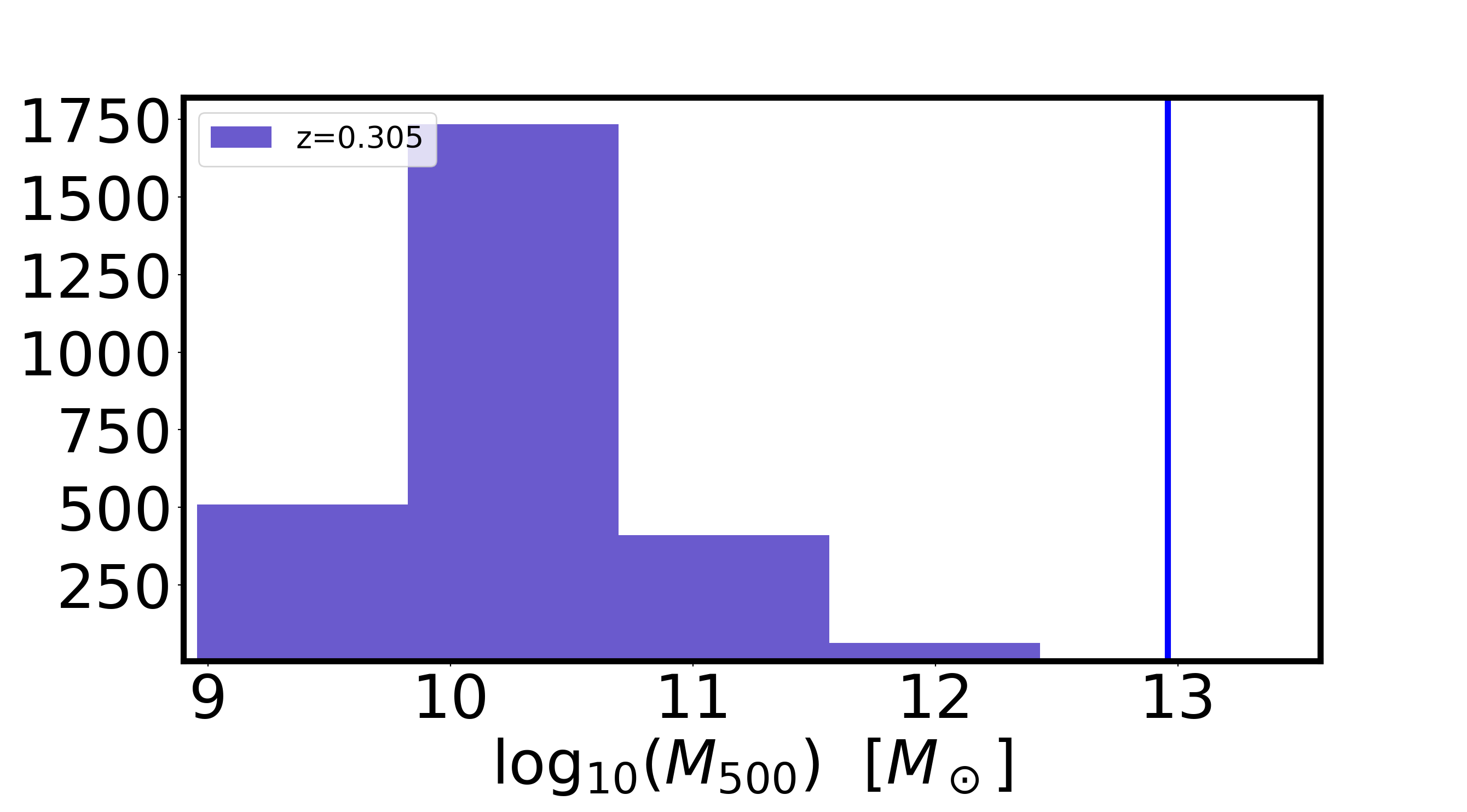}
    \includegraphics[width=0.32\textwidth]{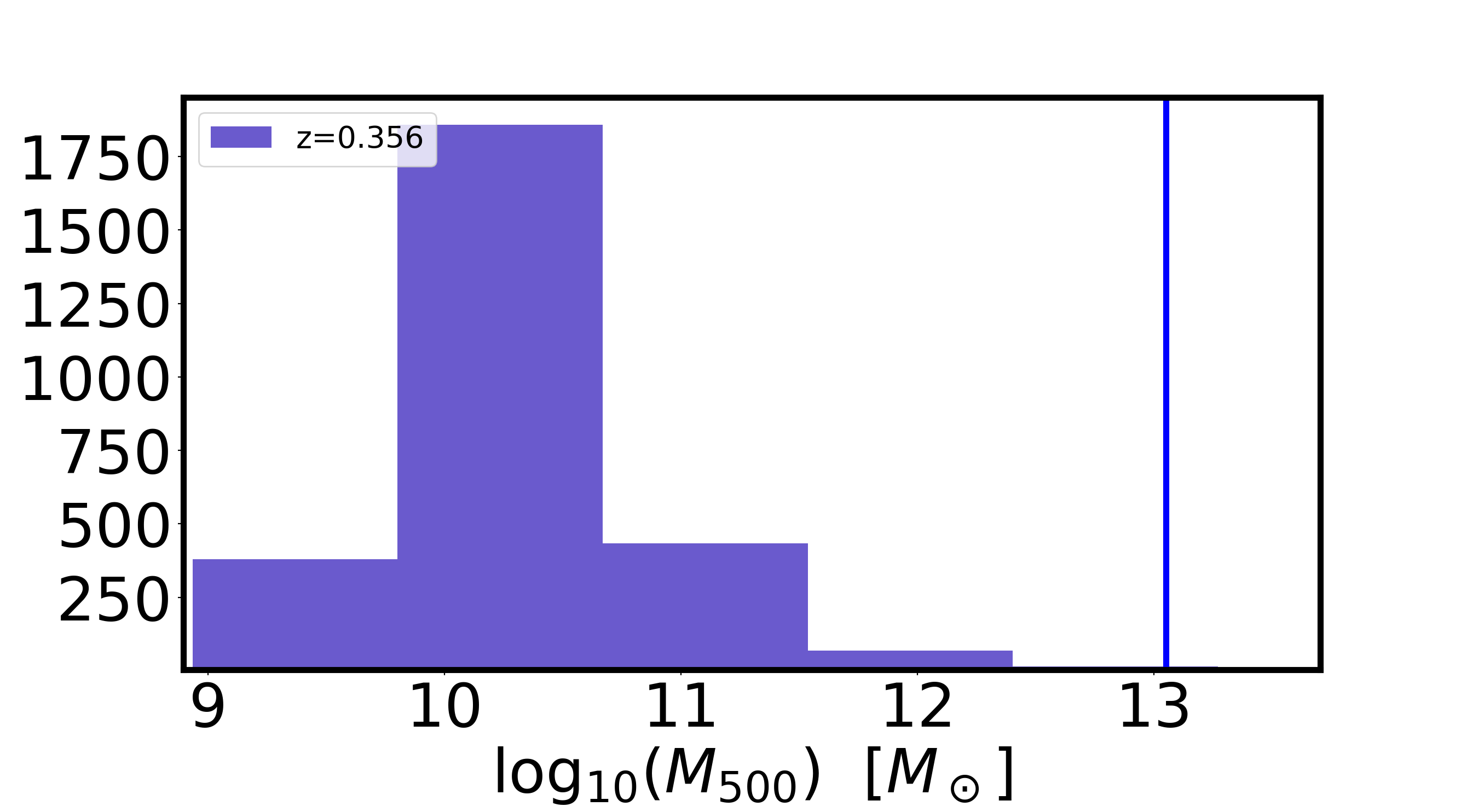}
    \includegraphics[width=0.32\textwidth]{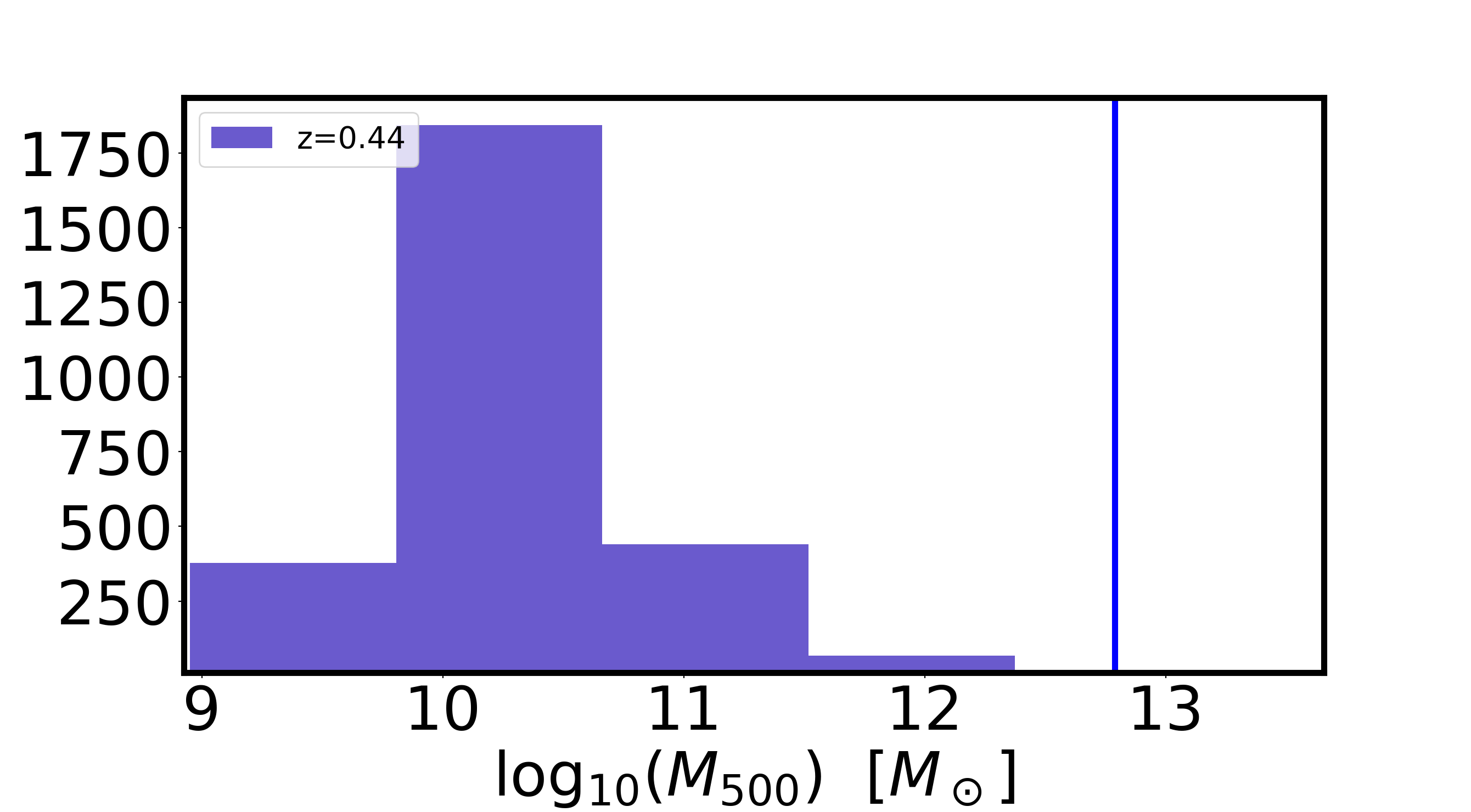}
    \includegraphics[width=0.32\textwidth]{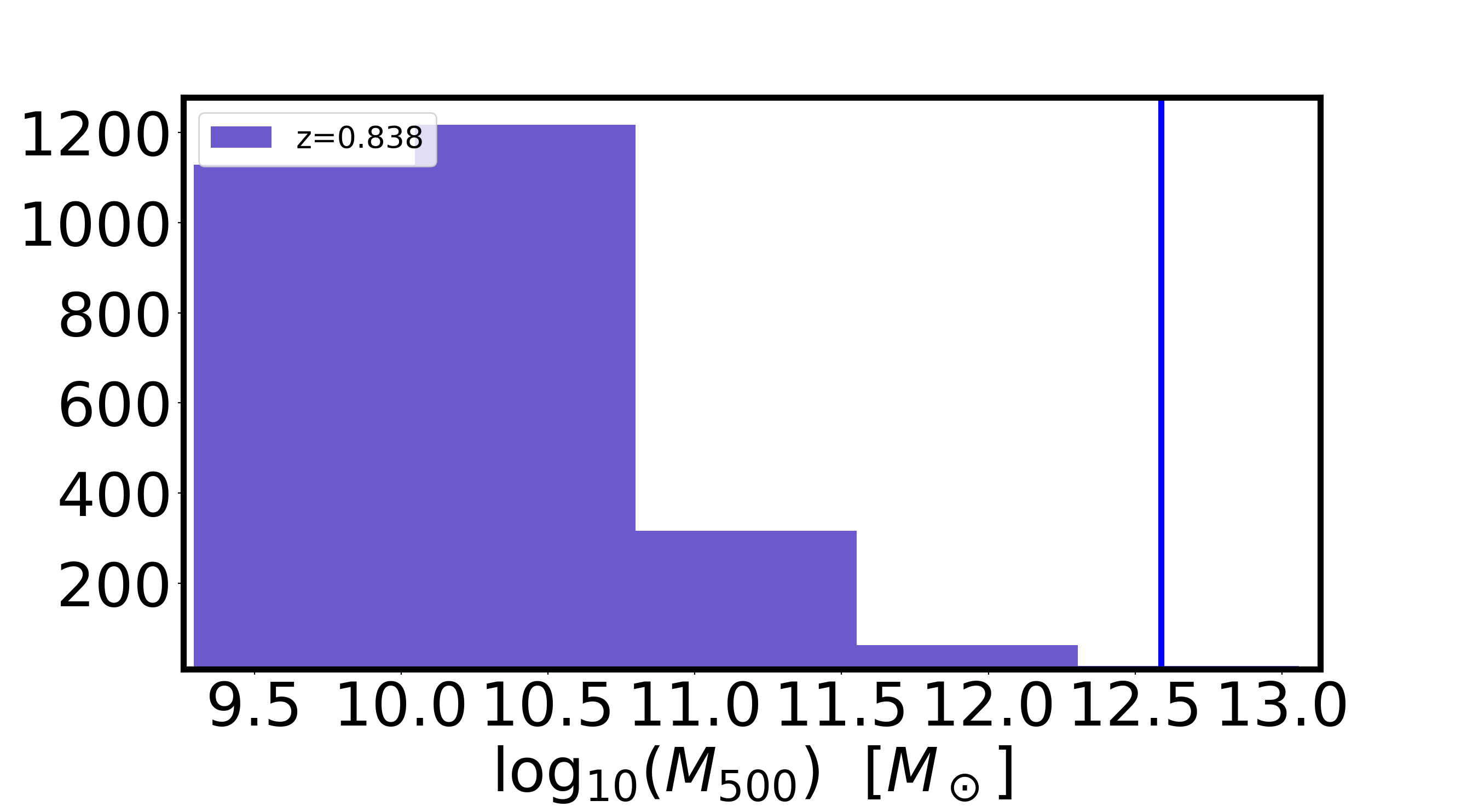}
    \includegraphics[width=0.32\textwidth]{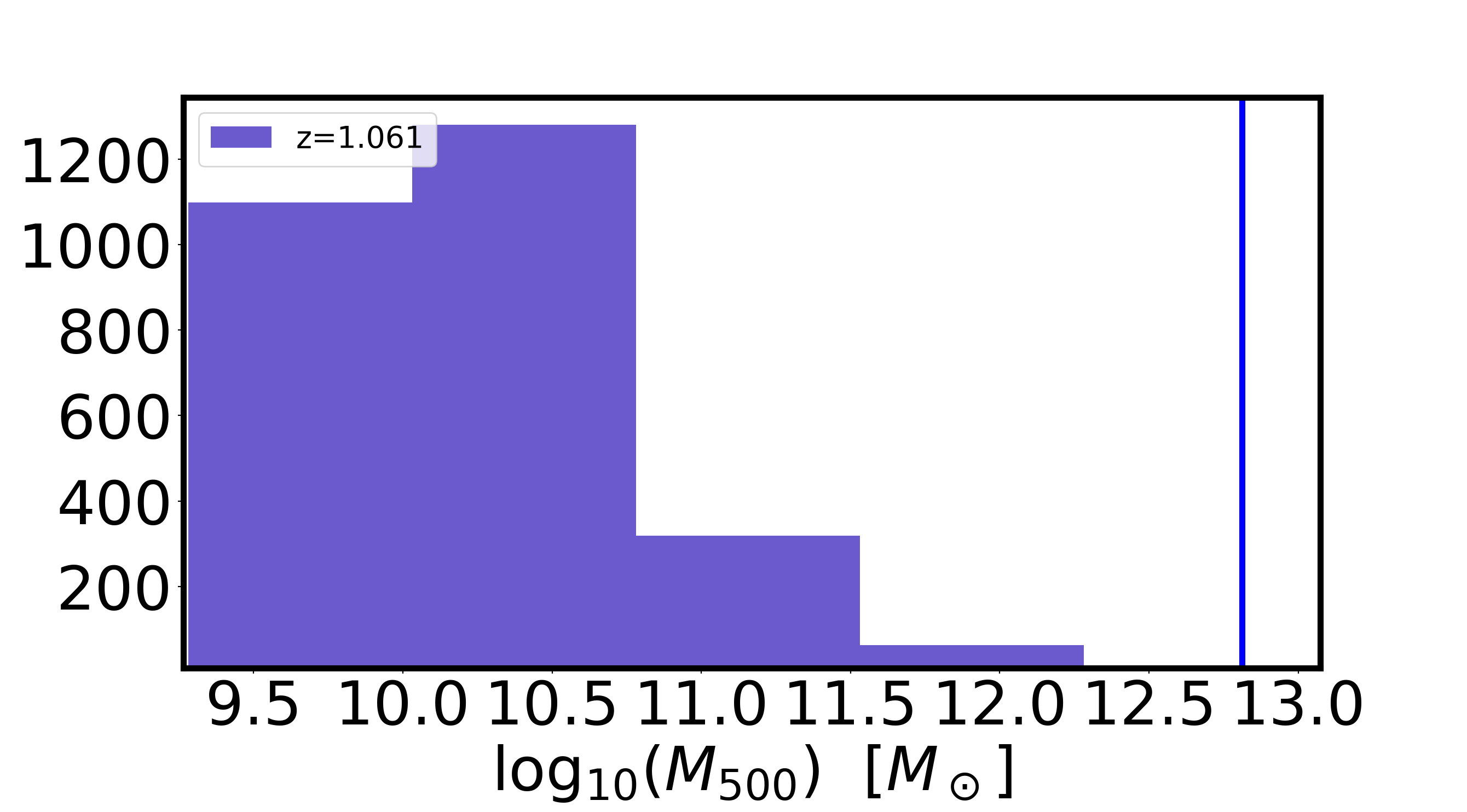}
    \includegraphics[width=0.32\textwidth]{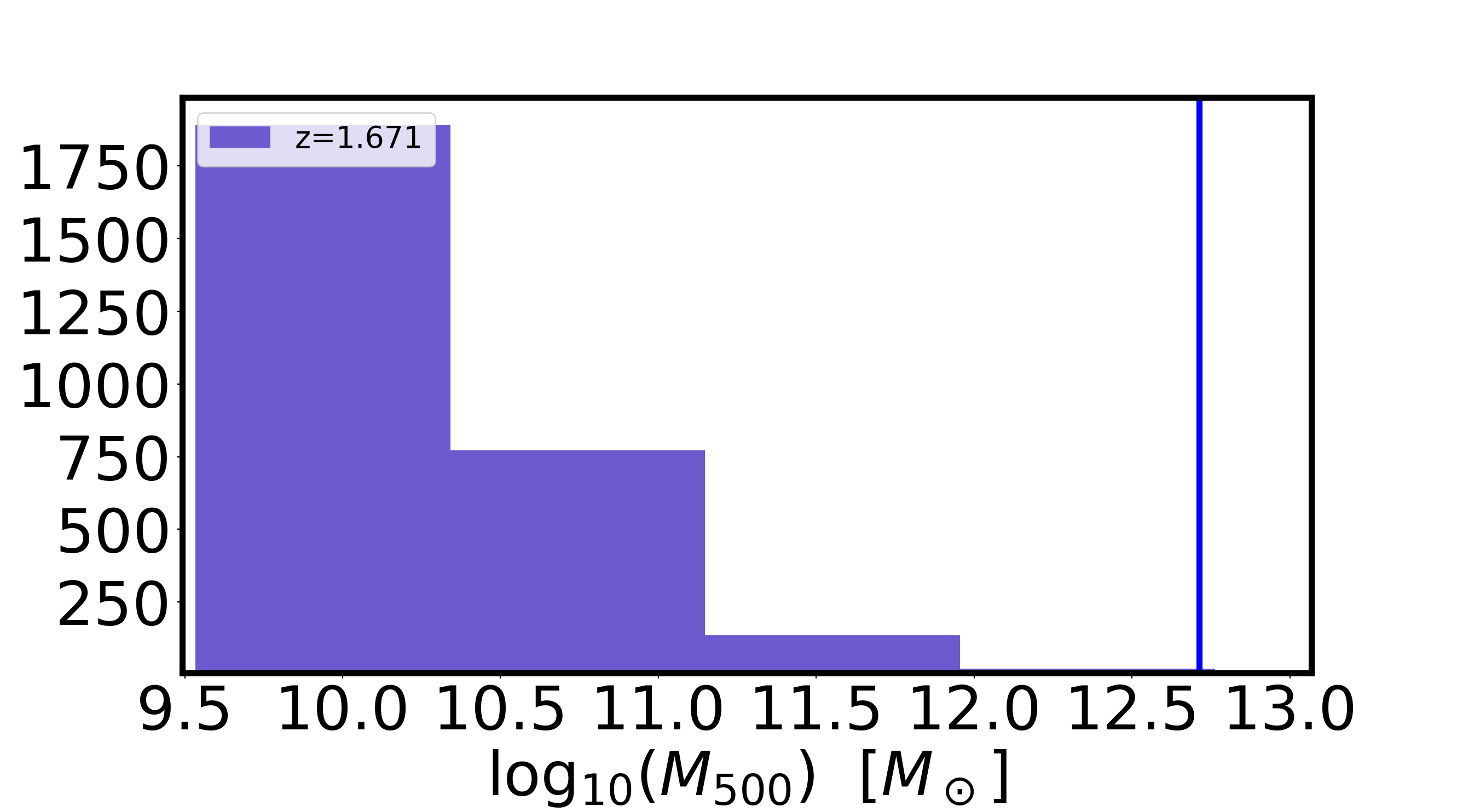}
    
   \caption{ vertical line in each panel shows properties of host galaxies and host halo of the merging SMBBH pairs identified as nano-Herz GW sources (i.e.~with chirp masses $M_c \geq 10^8$ M$_\odot$) at the corresponding redshift. The corresponding distribution for all galaxies or halos with $\rm M_{200} > 4.5 \times 10 ^ 9 M_\odot$ in the Romulus25 simulation at the same redshift with a background histogram. \textbf{First and second row:} display the stellar mass. \textbf{Third and fourth row:} present the sSFR. \textbf{Fifth and six row:} show M$_\mathrm{500}$ of the host halos. All properties are measured at the host galaxy's redshift. The stellar mass and sSFR are calculated within a 25 kpc sphere around the galaxy center. }
    
    \label{fig:halo-all}
\end{figure*}

\begin{figure}
    \centering
    \includegraphics[width= 0.45\textwidth]{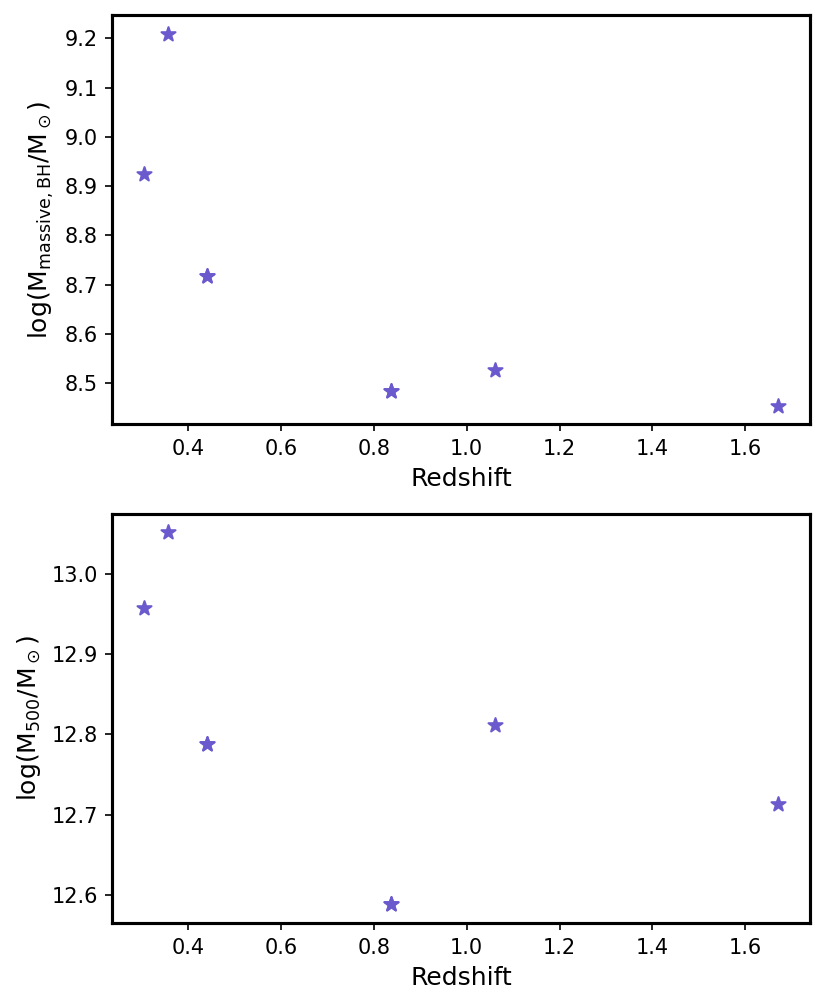}
    \caption{Top panel: Mass of most massive halo in SMBBH pairs identified as nano-Herz GW sources as a function of redshift. Bottom panel: The $\rm M_\mathrm{500}$ of host halos of nano-Herz GW sources vs. redshift.}
    \label{fig:halo-mass-z}
\end{figure}

In this section, we consider the global galactic characteristics of the hosts of nHz GW sources, delving into the observable properties of these host galaxies.  We also examine the properties of the halos of these host galaxies.

Figure \ref{fig:halo-z} illustrates the relation between galaxies' gas density ($\rho_{\rm gas}$), stellar mass ($\rm M_*$), star formation rate (SFR), and specific star formation rate (sSFR) vs redshift. Each of these quantities is calculated inside a 25 kpc sphere around the galaxy center. The first panel shows the galaxy gas density. No significant evolution with redshift is observed. On the other hand, the second and third panels clearly show a decrease in the stellar mass and an increase in SFR with redshift, respectively. Not surprisingly, this results in a rising sSFR with redshift, as shown in the lower right panel. Additionally, the stellar mass of the host galaxies suggests that at the redshifts of interest, these  galaxies are among the most massive systems in the {\sc Romulus} volume.  Based on analyses of \citet{jung2022massive} and \citet{saeedzadeh2023cgm}, we guess that these are the massive central galaxies in group-scale halos.   We will discuss this elaborately at a later part in this section. The specific star formation rate (sSFR) is frequently used to classify galaxies as star forming or quenched. We show the evolution of the sSFR with redshift for the host of the SMBBHs with $M_c \geq 10^8$ M$_\odot$ in the last panel in Fig. \ref{fig:halo-z}.

In this paper, we adopt the criteria from \citet{genel2018sfquench}, where they label a galaxy as ``main sequence'' if its sSFR is within $ \pm 0.5$ dex of the main sequence ridge and as ``quenched'' if its sSFR is 1 dex below the ridge. \citet{genel2018sfquench} give a relationship between sSFR and $M_*$ at few redshifts.  We linearly interpolate across these to determine the main sequence ridge sSFR at redshifts and stellar masses of our host galaxies. In Fig. \ref{fig:sfquenched}, we show $\Delta \rm log(sSFR) \equiv log(sSFR_{galaxy}) - log(sSFR_{ridge})$ for our host galaxies.  The shaded region shows the main sequence band and the dashed line corresponds to the threshold below which galaxies are classified as quenched.  Five of our six hosts either lie in the quenched territory or are on the border.  One galaxy, however, the $z=1.061$ host, falls within the star-forming main sequence band.  

 Next, we examine the rest-frame U-V colors of the host galaxies.   For completeness, we show (U-V)$_{\rm rest-frame}$ versus rest-frame V-band luminosity ($\rm L_V$), rest-frame absolute magnitude ($\rm M_V$), and stellar mass ($\rm M_*$). In these plots, large circles represent the host galaxies of nHz GW sources with $M_c \ge 10 ^ 8 
\rm M_\odot$, while small circles delineate hosts of nHz GW sources with chirp masses ranging between $10 ^ 7 M_\odot < M_c < 10 ^ 8 M_\odot$.   Generally, the latter tend to be bluer and less massive than the hosts of our nHz GW sources.  The four $z<1$ host galaxies are consistent with the quenched, early-type galaxies in Coma and Virgo clusters \citep{Renzini2006} as well as the SDSS and the CANDELS Multi-Cycle Treasury Survey \citep{bell2012}. {This aligns with the trends identified by \citet{Izquierdo-Villalba2023}, who shows that elliptical galaxies are significant hosts of massive binary black holes at $z<1$.}
The two $z > 1$ hosts, one of which is star-forming and the other quenched according to the \citet{genel2018sfquench} criterion, are both just slightly bluer than the population of quenched early-types at comparable redshifts in the CANDELS Multi-Cycle Treasury Survey \citep{bell2012}. The comparison to galaxies at the same redshifts is important since the stellar population in higher redshift galaxies is inherently younger and therefore bluer.

To gain better insight into why the two higher redshift exhibit bluer colors and to further clarify the nature of the host galaxies generally, we examine the images of the host galaxies.   These are shown in Fig. \ref{fig:morpho}. The top row shows the edge-on view of the galaxies and the bottom row shows the face-on view.  
{The stars are colored based on their magnitudes, determined by their age and metallicity. The magnitudes from the `i' band influence the red component of the image, `v' the green, and `u' the blue. These channels are then combined to produce a multiband composite image of the galaxy.\footnote{The galaxy images are generated using {\sc pynbody} package \citep{pontzen2013pynbody}.}}
Visually, \emph{all} of the galaxies appear to be early types, and the majority of the stars in these galaxies are old in the sense that 50\% had formed within the first 2.5-2.6 Gyrs after the Big Bang.   This is illustrated by the histograms in Fig. \ref{startsform}, which show the cosmic age (i.e.~the age of the Universe) when the stars in the galaxies formed.  Turning back to Fig. \ref{fig:morpho}, we see that nearly all of the galaxies appear to have experienced recent merger(s).  They all manifest features like stellar streams and shells \citep[eg][and references therein]{Fardal2007shells}.  Some of the mergers are gas rich, as in the case of the host galaxy at $z=1.06$ where one can clearly see an extended stream with ongoing star formation against a background of older stars. A less prominent stellar stream can also be seen in the host galaxy at z = 1.671. A detailed analysis of the morphology of host galaxy of merging binary black holes by \cite{Bardati2023} shows that the dominant morphological signature of SMBH mergers is the presence of a classical bulge that is also a sign of major mergers of these host galaxies.
The panels in Fig. \ref{startsform} provide additional information, including the fraction of stellar mass at the time of observation that has an age $\leq 1$ Gyr.  The latter varies from $<1\%$ to as much as $12\%$ in the highest redshift system.   This small fraction of very young stars is the likely explanation for the two $z>1$ galaxies' bluer colors.  As shown by \citet{pipino2009uvblue}, even a small fraction ($\sim 1\%$) of young stars ($\sim 0.1$ Gyr) can have a dramatic impact on UV-optical colors.

For the SMBBHs, we show the time taken by these sources to grow to masses above $M_c=10^7$ M$_\odot$ from their progenitor mass of $50\%$ of the source masses in Fig. \ref{fig:delaytime} for both the SMBHs in a SMBBH. The distribution of the delay time for the black holes with M$_c$ $\geq 10^7$ M$_\odot$ and M$_c$ $\geq 10^8$ M$_\odot$ is shown in blue and orange respectively in Figure \ref{fig:delaytime}. The shortest delay time observed for these black holes is 1.4 billion years, roughly 10\% of the age of the Universe for sources with M$_c$ $\geq 10^8$ M$_\odot$, which contribute significantly to the SGWB. This indicates that the number of mergers of the PTA sources is likely to be more towards low redshift than high redshift, and the corresponding properties of the host galaxies will be towards older galaxies.

In Fig. \ref{fig:chirpmass}, {left panel,} we show the correlation between the chirp mass of SMBBHs with the stellar mass {of their host galaxy. The right panel shows the mass of more massive black holes in the SMBBHs vs. the host galaxies' stellar mass.} Sources with black holes mass ratio $q < 0.1$\footnote{The mass ratio is defined as $q\equiv m_1/m_2$ with $m_2>m_1$.} are shown by stars and ones with $q>0.1$ are shown by circles. Large symbols correspond to SMBBHs with chirp mass  $M_c \geq 10^8$ M$_\odot$ and small symbols denote those with $10^7$ M$_\odot$ $\leq M_c \leq 10^8$ M$_\odot$.
The plot indicates that sources with the highest chirp mass are primarily present at low redshift and are hosted in galaxies with stellar mass greater than $10^{11}$ M$_\odot$. {Furthermore, the heaviest black holes are found in pairs with chirp masses $M_c \geq 10^8$ M$_\odot$, which are hosted by massive galaxies. 
These observations support the hypothesis we presented at the beginning of this section: that nHz GW sources predominantly inhabit massive, group-central galaxies. We will explore this further below.

In Fig. \ref{fig:halo-all}, we compare the global properties (i.e.~global stellar mass and sSFR) and the environment (i.e.~the halo mass and location therein) of the host galaxies of nHz GW sources with $M_c > 10^8 \rm M_\odot$ (vertical lines) against the properties of all galaxies in the simulations within halos with $\rm M_{200} > 4.5 \times 10 ^ 9 M_\odot$\footnote{This threshold is applied solely during the calculation of background histograms to save computational time, as we are not interested in very small halos.} (histograms).
{The plot supports the discussion we have presented above in the context of the colors and quenched/star-forming status of the host galaxies, that the sSFR of these galaxies places them in the low sSFR tail of the sSFR distribution of the galaxies in the {\sc Romulus25} simulation volume.} {From Fig. \ref{fig:halo-all} we deduce that our host galaxies are among the most massive galaxies in the {\sc Romulus25} simulation volume. That the host galaxies of the nHz GW sources reside in galaxies with high stellar mass and low sSFR making them a unique class of objects.   

This is further confirmed by comparing the host halo $\rm M_\mathrm{500}$ to all\footnote{All halos with $\rm M_\mathrm{200} > 4.5 \times 10^9$ } halos $\rm M_\mathrm{500}$ in the { \sc Romulus} simulation which is shown in the histogram, we note that the host halo masses are in the high-mass tail of the distribution (see the last two panels in Fig. \ref{fig:halo-all}).  The halos in which the host galaxies reside are group scale systems and based on the results of \citet{jung2022massive} and \citet{saeedzadeh2023cgm}, we expected -- and have subsequently confirmed -- that these galaxies are massive central group galaxies.   Collectively these findings strongly suggest that the nHz GW sources are hosted by massive early-type galaxies at the centers of groups and clusters.  \emph{However, we assert that the typical hosts of the nHz GW sources will be \underline{group-central} galaxies.}   For one, there are many more groups than clusters.  Moreover, the lower velocity dispersion of group satellites makes dynamical friction in group halos more efficient, and consequently, group environments are much more conducive to mergers, especially between the satellite and the central galaxies \citep[][and references therein]{OSullivan2017clogs-I,Oppenheimer2021GroupsSimReview}.

For completeness, in Fig. \ref{fig:halo-mass-z}, we present the mass of the more massive black hole in nHz GW sources and the host halo mass of these sources in the top and bottom panels, respectively. As expected, these plots indicate that as redshifts increase, the halo mass decreases, and the black hole mass in the pair also reduces.

In this section, we have elucidated the unique astrophysical properties of the host galaxies of the SMBBHs which contribute to the SGWB signal in comparison to all galaxies in the simulation. Our findings highlight that GW source hosts (with chirp mass $M_c\geq10^8$ M$_\odot$) predominantly reside in galaxies characterized by lower star formation, higher stellar mass, and higher halo masses compared to most counterparts at a given redshift. Specifically, these hosts are located within group-scale halo systems, identifying as massive central group galaxies. These host galaxies are early-type galaxies, displaying a distinct trend in the color-magnitude diagram across redshifts. 
These astrophysical properties inferred theoretically about the SMBBHs make it possible to correlate electromagnetic observations of the galaxies with the GW sources. Exploring such connections, coupled with comparisons to theoretical models, offers insights into the interplay between galaxy formation and black hole formation.

\section{Possible techniques to connect observations with theoretical models}\label{sec:connectiontoobs}

In the last two sections, we have discussed a scheme to connect the global astrophysical properties of the galaxies with the spectrum of the SGWB signal in the PTA band and have applied that to the \textsc{Romulus} simulation to understand the underlying theoretical correlation. The next interesting step forward is to connect this with the observations available from currently ongoing/upcoming surveys (see \citet{Burke-Spolaor:2018bvk} for review article). The observation of the GW signal from PTA observations in the nHz range can happen as (i) SGWB and (ii) GW signal from individual events. Both of these kind of observations can bring complementary information. 

\textit{SGWB:} The measurement using PTA observations provides a measurement of the spectrum of the SGWB signal. However, it is still unclear what are the properties of the host galaxies of the SMBBHs that contribute to the signal.  As we have shown in the previous section{, the simulations} show that the binaries are likely to form in galaxies with high stellar mass, high halo mass, and low SFR, and mostly early-type galaxies, that show signs of mergers in not too distant a past. We also showed that the host galaxies are central group galaxies. The host of the GW sources also shows a trend in the color-magnitude diagram as a function of redshifts. 

Based on these understandings, we can classify galaxies from electromagnetic observations based on their color, stellar mass, halo-mass, SFR, and galaxy type and can explore spatial cross-correlation of the galaxy distribution with the anisotropic SGWB signal \citep{Mingarelli:2013dsa, Hotinli:2019tpc,Sato-Polito:2023spo} and explore cross-correlation between the two quantities \citep{Mukherjee:2019oma,Yang:2020usq,Mukherjee:2020jxa,Yang:2023eqi}. 
A detailed paper on this formalism will be followed up in a companion paper. The cross-correlation of the SGWB signal with the galaxies of different types will be maximum for types of galaxies that are host of the GW sources. The exploration of the cross-correlation signal will give us an understanding of the population of the GW sources contributing to the background and we can estimate the occupation number of SMBHs. This will be useful in understanding the SGWB measurement in terms of the astrophysical properties of galaxies given in Eq. \eqref{eq:conn} based on observations. In future work, we will explore this aspect from the measurement of the SGWB signal and galaxies detected from optical and infrared surveys. 

\textit{Signal from individual events: } The measurement of the nHz GW signal from individual sources is likely to be possible from the future array of radio antennae such as Square Kilometer Array (SKA) \citep{Ellis:2013hna, Burke-Spolaor:2013aba, 2017NatAs...1..727K}. With such observations of individual GW signals, we can fit the astrophysical properties of the galaxies with the frequency dependence of the GW signal and fit the parameters on the occupation number and the signature of the environmental effects on the GW strain by directly comparing the properties of the host galaxy such as the gas density, stellar mass, halo mass, SFR, galaxy morphology, and color. Furthermore, an interesting avenue will be to perform a dedicated study of the hosts of the GW sources with high-resolution spectroscopic surveys to better understand its astrophysical properties. 

\section{Conclusion and Future Outlook}\label{sec:conc}
In this work, we explore the astrophysical properties of the host galaxies of the {SMBBHs} which can produce nano-Hertz SGWB using the {\sc Romulus25} cosmological simulation.
{\sc Romulus25} is capable of modeling the astrophysical properties of galaxies and its unique approach to seeding, accretion, and particularly the dynamics of SMBHs makes it especially well-suited for investigating SMBH/SMBBH-galaxy connections. Using this simulation, we have calculated the SGWB signal from the SMBBHs by modeling the environmental effects around the SMBHs. 

{We found that SMBBHs with chirp mass $M_c > 10^8 \rm M_\odot$ are primary source of the SGWB signal. In our simulation up to $z = 2$, we found six such sources resulting in a number density of $7.7 \times 10^{-6} \rm cMpc^{-3}$ consistent with the results from PTA studies \citep{mingarelli2017,casey2022,antoniadis2023second}. Although the {\sc Romulus25} is a cosmological simulation and not an idealized simulation specifically designed for SMBH-SMBH physics, it still successfully produces a correct number density. This highlights its potential for further studies in this domain. }

We {then continue by studying} the redshift evolution of the astrophysical properties of the host galaxies such as gas density, SFR, stellar mass, and halo mass, across redshifts. 
{These host galaxies are early-type galaxies, characterized predominantly by older star populations. They exhibit a distinct trend in the color-magnitude diagram across redshifts, which could be of particular interest to compare with observations.}

{Our analysis further reveals that, compared to their counterparts at similar redshifts, the host galaxies of nHz GW sources exhibit lower SFRs, greater stellar masses, and more substantial halo masses. Our findings collectively suggest that nHz GW sources are predominantly hosted by massive early-type galaxies at the centers of groups and clusters. However, we assert that the typical hosts for these GW sources are expected to be group-central galaxies. This is supported by two main factors: (i) groups are more common than clusters, and (ii) the lower velocity dispersion in groups leads to more effective dynamical friction, thereby increasing the likelihood of mergers, especially between satellite galaxies and the central galaxy of the group \citep{OSullivan2017clogs-I,Oppenheimer2021GroupsSimReview}.}

{It is important to note that our conclusions remain robust even if the seed mass in {\sc Romulus25} were set lower than $10^6 M_\odot$. As discussed in Section \ref{sec:bhp1}, the SGWB power spectrum signal in the nHz regime is predominantly due to SMBBHs with chirp masses greater than $10^8 M_\odot$. These systems are composed of individual SMBHs, each with a mass exceeding $8 \times 10^7 M_\odot$, which is significantly above our minimum SMBH mass. Thus, our findings regarding the properties of host galaxies of nHz GW sources remain valid. }

The theoretical connection of the host galaxy properties of the GW sources and the black hole masses indicates which kind of galaxies and their evolution are linked with the black hole merger. This theoretical connection shown in this work will be a guideline for us to explore the connections from GW observations in the nHz band and optical and infrared galaxy observations. By measuring the spatial cross-correlation between the anisotropic SGWB with galaxies as well as a targeted search of individual galaxies for the nHz GW events in the SKA era. 

The multi-messenger technique by exploring the connection between the astrophysical properties of host galaxies with the SMBHs and the strain of the GW signal from the coalescing SMBHs will make it possible to establish from observations how the SMBBHs evolution depends on the astrophysical properties of the galaxies. The occupation number of SMBHs in galaxies of different types will make it possible to test theoretical models using observations. In the future with the data from the ongoing International Pulsar Timing Array and {upcoming} Square Kilometer Array (SKA) \citep{Janssen:2014dka}, we will be able to make high-precision measurements of the nHz GW signal. { In} synergy with the galaxy surveys up to high redshifts such as Dark Energy Spectroscopic Instrument \citep{DESI:2016fyo}, Euclid \citep{2011arXiv1110.3193L}, Vera Rubin Observatory \citep{LSSTDarkEnergyScience:2012kar}, {and} Roman Telescope \citep{2019arXiv190205569A} we will make joint estimation of GW and galaxies to unveil the open question of formation of SMBHs and its connection with the galaxy evolution. 

\section*{Acknowledgement}

The work of SM is a part of the $\langle \texttt{data|theory}\rangle$ \texttt{Universe-Lab} which is supported by the TIFR and the Department of Atomic Energy, Government of India. 
VS and AB acknowledge support from the Natural Sciences and Engineering Research Council of Canada (NSERC) through its Discovery Grant program. AB acknowledges support from the Infosys Foundation via an endowed Infosys Visiting Chair Professorship at the Indian Institute of Science. MT was supported by an NSF Astronomy and Astrophysics Postdoctoral Fellowship under award AST-2001810. AB, TQ, and MT were partially supported by NSF award AST-1514868. 

The {\sc Romulus} simulation suite is part of the Blue Waters sustained-petascale computing project, which is supported by the National Science Foundation (via awards OCI-0725070, ACI-1238993, and OAC-1613674) and the state of Illinois. Blue Waters is a joint effort of the University of Illinois at Urbana-Champaign and its National Center for Supercomputing Applications. Resources supporting this work were also provided by the (a) NASA High-End Computing (HEC) Program through the NASA Advanced Supercomputing (NAS) Division at Ames Research Center; and (b) Extreme Science and Engineering Discovery Environment (XSEDE), supported by National Science Foundation grant number ACI-1548562. The analysis reported in this paper was enabled in part by WestGrid and Digital Research Alliance of Canada (alliancecan.ca) and on the cluster of $\langle \texttt{data|theory}\rangle$ \texttt{Universe-Lab} supported by DAE. Our analysis was performed using the Python programming language (Python Software Foundation, https://www.python.org). The following packages were used throughout the analysis: numpy (\citealt{harris2020array}), matplotlib (\citealt{hunter2007matplotlib}), Pynbody (\citealt{pontzen2013pynbody}), SciPy (\citealt{virtanen2020scipy}), and TANGOS (\citealt{pontzen2018tangos}).

Finally, VS and AB acknowledge the l{\fontencoding{T4}\selectfont
\M{e}}\'{k}$^{\rm w}${\fontencoding{T4}\selectfont\M{e}\m{n}\M{e}}n 
peoples on whose traditional territory the University of Victoria stands, and the Songhees, Equimalt and
\b{W}S\'{A}NE\'{C} peoples whose historical relationships with the land continue to this day.

\section*{Data Availability}

The data directly related to this article will be shared on reasonable request to the corresponding author. Galaxy database and particle data for {\sc Romulus} is available upon request from Michael Tremmel.



\bibliographystyle{mnras}
\bibliography{paper_draft_MNRAS} 


\bsp	
\label{lastpage}
\end{document}